\documentclass[useAMS,usenatbib,usegraphicx]{mn2e}
\usepackage{graphicx}
\usepackage{threeparttable}
\usepackage{amssymb}
\usepackage{subfigure}
\usepackage{subfig}
\usepackage{pifont}
\newcommand{\cmark}{\ding{51}}%
\newcommand{\xmark}{\ding{55}}
\newcommand{\araa}{AR\&A}
\newcommand{\mnras}{MNRAS}
\newcommand{\apj}{ApJ}
\newcommand{\aj}{AJ}
\newcommand{\apjl}{ApJL}
\newcommand{\apjs}{ApJS}

\newcommand{\nat}{Nature}

\title[{\sc{CARMA}} follow-up of candidate high-z, massive galaxy clusters]{{\sc{CARMA}} observations of massive {\it{Planck}}-discovered cluster candidates at $z\gtrsim 0.5$  associated with WISE overdensities:  strategy, observations and validation}

\author[Rodr{\'i}guez-Gonz{\'a}lvez, Muchovej \& Chary]{Carmen Rodr{\'i}guez-Gonz{\'a}lvez$^{1}$,
Stephen Muchovej$^{2}$,
Ranga Ram Chary$^{1}$\\
$^{1}${U.S. Planck Data Center, MS314-6, Pasadena, CA 91125, USA, E-mail: carmenrg@ipac.caltech.edu, rchary@ipac.caltech.edu}\\
$^{2}${Owens Valley Radio Observatory, California Institute of Technology, Big Pine, CA 93513, USA}
}

\begin{document}

\maketitle

\begin{abstract}
We present $1-2$\arcmin\ spatial resolution {\sc{CARMA}}-8 31-GHz
observations towards 19 unconfirmed {\it{Planck}} cluster candidates,
selected to have significant galaxy overdensities  from the {\sc{WISE}}
early data release and thought to be at $z\gtrsim 1$ from the {\sc{WISE}} colors
of the putative brightest cluster galaxy (BCG). We find a Sunyaev-Zeldovich (SZ) 
detection in the CARMA-8 data towards 9 candidate
clusters, where one detection is considered tentative. For each cluster
candidate we present CARMA-8 maps, a study of their radio-source
environment and we assess the reliability of the SZ detection. The CARMA
SZ detections appear to be SZ-bright, with the mean,
primary-beam-corrected peak flux density of the decrement being
$-2.9$\,mJy/beam with a standard deviation of 0.8, and are typically
offset from the {\it{Planck}} position by $\approx 80\arcsec$. Using archival
imaging data in the vicinity of the CARMA SZ centroids, 
we present evidence that one cluster matches Abell\,586---a
known $z\approx 0.2$ cluster; four candidate clusters are likely to have
$0.3 \lesssim z \lesssim 0.7$; and, for the remaining 4, the redshift
information is inconclusive. We also argue that the sensitivity
limits resulting from the cross-correlation between {\it Planck} and {\it WISE} makes it 
challenging to use our selection criterion to identify clusters at $z>1$.
\end{abstract}

\section{Introduction}
Galaxy clusters form over a Hubble time from rare, high-density peaks in the primordial density field on scales of a few Mpc. Their assembly via the hierarchical merging of smaller haloes from  $0 < z < 3$ straddles the period of dark energy domination and, on the largest scales, is driven primarily by gravitational physics with little effect from e.g., complex gas dynamics, feedback and stellar mass (see e.g.,  \citealt{bond1996} and \citealt{allen2011} for a review). As a result,
cluster abundance---the number of clusters per comoving volume per solid angle above a certain mass---as a function of redshift depends solely on the expansion history of the Universe and the growth of the initial fluctuations (\citealt{bardeen1986}, \citealt{myers1996}). 
In principle, measuring the evolution of the cluster mass function with redshift provides an independent probe 
for placing strong constraints on cosmological parameters (e.g., \citealt{bahcall1998}, \citealt{viana1996} and \citealt{voit2005}). 
However, the ability to extract precision cosmology from cluster surveys relies on 
 the selection function being well understood and the precise characterization of how a cluster observable translates to a cluster mass.

Lying on the exponential tail of the mass function, the most massive clusters are of particular interest, especially those at high redshifts, $z$, as they yield the largest differences between cosmologies.  
But, identifying such systems has been challenging, since they are inherently rare and since historically clusters have been detected via their optical flux or from the X-ray emission of the hot intracluster medium (ICM; observational methods that suffer from cosmological dimming). In recent years, significant progress in this quest has been made through Sunyaev-Zel'dovich (SZ) surveys like ACT \citep{marriage2011}, SPT \citep{williamson2011} and {\it{Planck}}\,\citep{ESZ}. When cosmic microwave background (CMB) photons traveling towards us traverse the hot ICM, many are inverse-Compton scattered by the hot electrons in the plasma producing a shift in the
blackbody spectrum of the CMB known as the SZ effect (\citealt{Sunyaev_1972}; see \citealt{carlstrom2002} for a review).
The total, or integrated SZ signal, $Y_{\rm{SZ}}$, has been shown to correlate tightly with mass, it is only weakly dependent on redshift at $z \gtrsim 0.3$ and has a weak bias to gas concentration (e.g, \citealt{motl2005}, \citealt{bonaldi2007} and \citealt{kay2012}).

The ACT and SPT SZ cluster surveys cover relatively small areas of sky ($\approx 500$ sq degs;\cite{hasselfield2013} and  
$\approx 2500$ sq degs;\cite{williamson2011}, respectively) at arcminute  resolution with the benefit of probing the low-to-medium mass end of the mass function. These surveys are complementary to the {\it{Planck}} mission (\citealt{tauber2010} \& \citealt{planck1}), which is designed to detect the most massive clusters over the entire sky. The latest {\it{Planck}} results based on 15.5 months of data contain measurements towards 1227 systems; 683 of these entries match a known cluster and, out of the 544 newly-discovered objects, 178 have been confirmed to be clusters through follow-up observations (\citealt{Planck2013}). One of the major drawbacks of {\it{Planck}} is its spatial resolution, which is between 5$\arcmin\ $ and $10\arcmin\ $ at the relevant frequencies, and contributes to the $\gtrsim 30\%$ errors on cluster parameters (\citealt{PlanckAMI}). Thus, to fully exploit the characteristics of {\it{Planck}} clusters to constrain cosmology, high resolution follow-up of the catalogued clusters is necessary.

For this project we have used the 8-element subarray of the Combined Array for Research in Millimeter-wave Astronomy, {\sc{CARMA}}-8, to follow up 19 unconfirmed {\it{Planck}}-discovered cluster candidates which we believed to be at $z\gtrsim1.0$ from their WISE colors. Previous high resolution SZ follow up of {\it{Planck}}-detected clusters and candidate clusters has been undertaken by \cite{muchovej2010}, \cite{hurley2011}, \cite{sayers2012}, \cite{PlanckAMI} and \cite{perrott2014}.

This work is divided into two articles. The first one, presented here, focuses on validating the sample candidate clusters. The second \citep{rodriguez2014b} aims to constrain cluster parameters by fitting models to the data in a Bayesian Monte-Carlo Markov Chain (MCMC) framework. The current article is organized as follows: details on the instruments and target selection are provided in Section \ref{sec:sample}. The CARMA data, including processing and main results, are presented in Section \ref{sec:data}. Validation work using ancillary datasets is described in Section \ref{sec:vali}. In Section \ref{sec:discussion} we discuss whether any of the cluster candidates without a CARMA-8 SZ detection are likely to be real. In Section \ref{sec:redshiftandvalid}, we (1) explore which of the selection criteria correlate best with detectability and (2) make use of WISE and SDSS to estimate or place constraints on the cluster redshift. The conclusions drawn from this study are provided in Section \ref{sec:conclusions}.

Throughout this work we present images where North is up and East is to the left. We use J2000 coordinates, as well as a $\Lambda$CDM cosmology with $\Omega_{\rm{m}}=0.3$, $\Omega_{\Lambda}=0.7$, $\Omega_{\rm{k}}=0$, $\Omega_{\rm{b}}=0.041$, $w_o=-1$, $w_{\rm{a}}=0$ and $\sigma_8=0.8$. $H_0$ is taken as 70\,km\,s$^{-1}$\,Mpc$^{-1}$.

\begin{table*}
\caption{Details on the CARMA-8 data. For simplicity and homogeneity in the cluster naming convention we use a shorthand ID for the targets. The PSZ (Union catalog) name (Planck Collaboration 2013 XXIX) is provided, where available$^{\dag}$. The RA and Dec coordinates correspond to the map centre of our CARMA-8 observations. For the short and long baseline data we provide the ellipse parameters for the synthesized beam and the visibility noise. Targets that have been detected in the CARMA-8 data have their ID highlighted. For P014, the SZ signal in the CARMA-8 data is considered tentative.}
 \label{tab:clusters}
 \tabcolsep=0.13cm
\begin{tabular}{lcccccccccc}
\hline \hline
Cluster  & Union Name  & RA & Dec  &  \multicolumn{3}{c}{Short Baselines (0-2k$\lambda$)}   &  \multicolumn{3}{c}{Long Baselines (2-8 k$\lambda$)}     \\
               &                          &       &           &                    Beam  & Position & $\sigma$ & Beam  & Position & $\sigma$\\
ID           &                          &       &           &                                 & Angle     &                   &             & Angle     &                  \\
               &                          & hh mm ss  & dd mm ss          &  $( \arcsec \times \arcsec )^a$ &  ($^\circ$) &   $(\rm{mJy/beam})^{\rm{b}}$  &  $(\arcsec \times \arcsec )^a$    &$(^\circ$) &    $(\rm{mJy/beam})^{\rm{b}}$                    \\ \hline

$\bf{P014}      $ & PSZ1G014.13+38.38 & 16 03 21.62 & 03 19 12.00         & 91.4 $\times$102.2 & 83.0    & 0.309         & 15.7$\times$19.5 & 44.9       & 0.324\\
P028     & PSZ1G028.66+50.16 & 15 40 10.15 & 17 54 25.14          & 123.8 $\times$129.9 & -37.4          & 0.433         & 17.4$\times$24.0 & 64.4       & 0.451\\
P031     & - & 15 27 37.83 & 20 40 44.28          & 92.6 $\times$234.9 & -84.5   & 0.727         & 16.5$\times$28.9 & 48.0       & 0.633\\
P049     &  - &  14 44 21.61 & 31 14 59.88        & 108.0 $\times$158.0 & -52.9          & 0.557         & 17.5$\times$25.0 & 85.1       & 0.572\\
P052     &  - & 21 19 02.42 & 00 33 00.00         & 87.4 $\times$107.5 & 53.1    & 0.368         & 14.7$\times$21.0 & 38.2       & 0.386\\
P057     & PSZ1G057.71+51.56 &15 48 34.13 & 36 07 53.86           & 119.7 $\times$133.9 & -41.8          & 0.451         & 16.7$\times$25.9 &  88.0      & 0.482\\
$\bf{P086}      $ & PSZ1G086.93+53.18 & 15 13 53.36 & 52 46 41.56         & 124.0 $\times$143.7 & 70.4   & 0.622         & 18.2$\times$24.4 & 54.0       & 0.599\\
P090     & PSZ1G090.82+44.13 &16 03 43.65 & 59 11 59.61           & 118.2 $\times$147.7 & -65.0          & 0.389         & 19.5$\times$24.4&  42.7       & 0.427\\
$\bf{P097} $&  - & 14 55 13.99 & 58 51 42.44      & 115.3 $\times$169.4 & -84.2          & 0.653         & 21.0$\times$25.0&  48.4       & 0.660\\
$\bf{P109}$ & PSZ1G109.88+27.94 & 18 23 00.19 & 78 21 52.19       & 112.3 $\times$185.0 & -86.4          & 0.562         & 23.3$\times$25.6 & -39.1      & 0.517\\
P121     & PSZ1G121.15+49.64 & 13 03 26.20 & 67 25 46.70          & 82.2 $\times$193.1 & 85.4    & 0.824         & 21.5$\times$23.4 & 89.9       & 0.681\\
P134     & PSZ1G134.59+53.41  & 11 51 21.62 & 62 21 00.18         & 106.5 $\times$164.3 &  80.0          & 0.590         & 20.1$\times$25.7 & -86.2      & 0.592\\
P138     & PSZ1G138.11+42.03 & 10 27 59.07 & 70 35 19.51          & 51.1 $\times$246.7 & 68.2    & 2.170         & 20.6$\times$26.1 & 75.0       & 0.982\\
$\bf{P170}$ & PSZ1G171.01+39.44  & 08 51 05.10 & 48 30 18.14      & 119.0 $\times$126.8 & -18.3          & 0.422         & 16.7$\times$24.5&  68.2       & 0.469\\
$\bf{P187}$ & PSZ1G187.53+21.92  & 07 32 18.01 & 31 38 39.03      & 104.0 $\times$145.3 & -60.9          & 0.411         & 16.7$\times$23.7&  63.0       & 0.412\\
$\bf{P190}      $& PSZ1G190.68+66.46 & 11 06 04.09 & 33 33 45.23          & 109.0 $\times$180.1&  -48.7          & 0.450         & 17.1$\times$26.3 &  -86.6     & 0.356\\
$\bf{P205}$ & PSZ1G205.85+73.77  & 11 38 13.47 & 27 55 05.62      & 117.8 $\times$130.1 & -35.9          & 0.385         & 16.9$\times$23.9 & 65.2       & 0.431\\
P264     & - &  10 44 48.19 & -17 31 53.90        & 102.9 $\times$124.9 & -12.3          & 0.476         & 16.8$\times$24.8 &  7.1       & 0.513\\
$\bf{P351}      $ & - &  15 04 04.90 & -6 07 15.25        & 96.5 $\times$109.6&  -39.9   & 0.355         & 17.2$\times$19.7 &  11.6      & 0.392\\ \hline

 \end{tabular}
 
 \begin{tablenotes}
 \scriptsize
 \item ($\dag$)  Since the cluster selection criteria, as well as the data for the cluster extraction, are different to those for the PSZ catalog, not all the clusters in this work have an official {\sc{Planck}} ID.
 \item (a) Synthesized beam Full Width Half Maximum (FWHM) (in arcsec) and position angle measured from North through East.
 \item (b) Achieved rms noise in corresponding maps.

  \end{tablenotes}
\end{table*} 

 \begin{figure*}
\begin{center}
\center{\includegraphics[width=10cm]{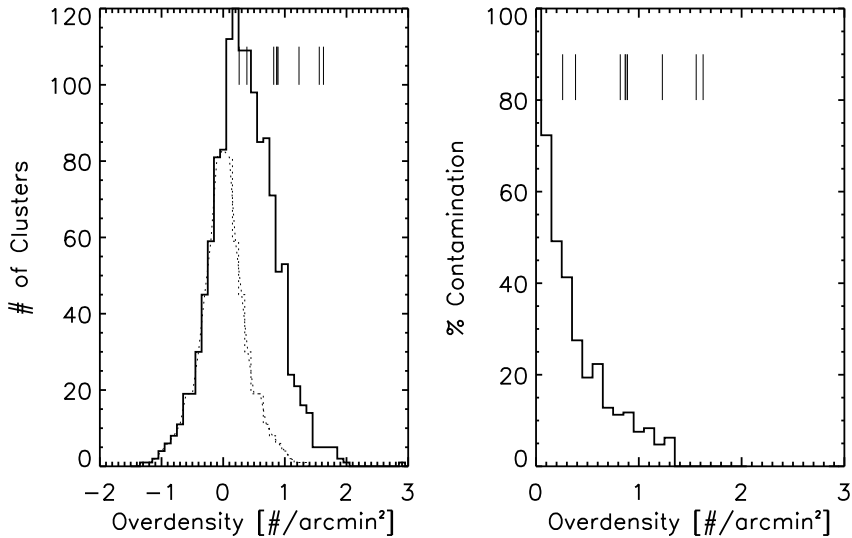}}
\vskip1cm
\caption{Left: the average {\sc{WISE}}-object overdensities at the location of {\it{Planck}} cluster candidates, where overdensities were calculated as the average density of sources within a radius of $4.75'$ minus the average density of sources in
an annulus with inner and outer radii of $4.75\arcmin\ $ and $7\arcmin\ $. Here, {\it{Planck}} clusters are all 1362 identified SZ-cluster candidates in the all-sky maps; this will include known clusters as well as candidate clusters discovered by {\it{Planck}} that are yet to be confirmed. The dotted line is the positive mirror-image of the negative side of the histogram. The vertical lines in both plots show the typical overdensities for our CARMA-8 detected cluster candidates. Right: percentage of clusters at a particular overdensity that do not have statistically significant
overdensities. The statistical significance is calculated as the ratio between the 
number of clusters at a particular overdensity and the number of clusters expected if 
the overdensity distribution followed a Gaussian centered on 0. Thus, at an overdensity of 0.75 galaxies/$\rm{arcmin}^2$, the percentage contamination is $<15\%$---the ratio between the dotted curve and the solid histogram in the left panel---and at overdensities of $\approx 1.25\,\rm{galaxies}/\rm{arcmin}^2$ (a typical WISE overdensity value for our WISE-Planck clusters) the contamination is $< 5\%$.}
\label{fig:rangawise}
\end{center}
\end{figure*}

\section{Cluster sample} 
\label{sec:sample}

\subsection{Instruments: CARMA and {\it{Planck}}}
The CARMA-8 telescope, previously known as the Sunyaev Zel'dovich Array (SZA), is an interferometer operating at 31\,GHz comprising eight antennas of 3.5\,m in diameter. Six of the antennas are arranged in a compact configuration (with baselines between $\approx 4-20$\,m or $\sim 0.4-2$\,k$\lambda$) to be sensitive to large-scale structure at $1-2$ arcminute-resolution and the two outliers, with baselines of $\approx 50$\,m or $2-10$\,k$\lambda$, provide the high resolution data ($\approx 20\arcsec$) to enable contaminating radio-point sources to be detected and removed accurately. It has a bandwidth of 8\,GHz divided into sixteen 500\,MHz channels, a 10.5$\arcmin$ FWHM primary beam and typical system temperatures of 40-50\,K.
Further details on the instrument can be found in \cite{muchovej2007}. Henceforth, we shall refer to the short-baseline data from the compact subarray ($0-2$\,k$\lambda$) as SB data and to the long-baseline data ($2-8$\,k$\lambda$) as LB data.

The {\it{Planck}} satellite (\citealt{tauber2010} \& \citealt{planck1}) is a third generation space-based mission to study the CMB and foregrounds to the CMB. It has mapped the entire sky at nine frequency bands from 30 to 857\,GHz,  with angular resolution of $33\arcmin\ $ to 5\arcmin, respectively. The bands where the SZ decrement is strongest, 100 and 143 GHz, have resolutions of 10\arcmin\ and 7\arcmin\ respectively.

\subsection{Sample selection}

The nineteen clusters comprising our sample are all candidate clusters detected in the {\it{Planck}} all-sky maps; they are listed in Table \ref{tab:clusters}. They were selected by cross-correlating WISE early data release\footnote{At the time when our sample of targets was selected the WISE early data release was the most up to date publicly available WISE data product.} and {\it{Planck}} catalogs of SZ candidate clusters. 
The {\it{Planck}} SZ catalog used for the primary
selection was an intermediate {\it{Planck}} data product known internally as DX7\footnote{\label{foot1}{\it{Planck}} data are collected and reduced in blocks of time. The DX7 all sky maps used in this analysis correspond to the reduction of {\it{Planck}} data collected from 12$^{\rm{th}}$ August 2009 to the 28$^{\rm{th}}$ of November 2010, which is the equivalent to 3 full all-sky surveys, using the v4.1 processing pipeline. The DX7 maps used in this work are part of an internal release amongst the {\it{Planck}} Collaboration members and, thus, is not a publicly available data product. The {\it{Planck}} Union catalog (PSZ) is based on more recent and refined processing of the data, including improved pointing and calibration.}. Candidate clusters were identified in the maps using a matched-filter component separation algorithm, MMF3\footnote{Three algorithms: MMF1 \citep{herranz2002}, MMF3 \citep{melin2006} and PwS (\citealt{carvalho2009} \& \citealt{carvalho2012}) have been used to identify cluster candidates from the {\it{Planck}} data (see \citealt{Planck2013}).} (\citealt{melin2006}). Henceforth, results derived from the MMF3 analysis of {\it{Planck}} DX7 data will be referred to as {\it{Planck}} results. 
We initially searched the WISE early data release \citep{wright2010} at the location of the ${\it{Planck}}$ cluster candidates and estimated a value for the overdensity of WISE objects.
The average density of WISE galaxies detected at 3.4 and 4.6$\micron$ was first calculated within $4.75\arcmin\ $  of the {\it{Planck}} position and, secondly, within an annulus with a $4.75\arcmin\ $ inner radius and a $7\arcmin\ $ outer radius\footnote{The choice of 4.75\arcmin as the inner radius is based on the {\it{Planck}} beam ($\approx 9.5\arcmin\ $ at 100\,GHz). Although the typical separations between {\it{Planck}} and X-ray cluster centroids from the early {\it{Planck}} SZ cluster catalog (\citealt{ESZ}) were found to typically be $\approx 2\arcmin\ $, we chose a larger radius to avoid introducing selection biases that could arise from picking systems with smaller offsets, which might be in a more relaxed dynamical state. However, from the CARMA-8 results, we find that we only detect systems within $\lesssim 2.5\arcmin$\ from the {\it{Planck}} position.}. The difference between these two density measurements yields a value for the overdensity of WISE objects in the vicinity of each {\it{Planck}} cluster candidate (see Figure \ref{fig:rangawise}). 

Coarse photometric redshifts were calculated from the [3.4]-[4.6] {\sc{WISE}} colors of the brightest red object within $2.5\arcmin\ $
\footnote{For some of the clusters in our sample the brightest red object in WISE used to obtain a photometric redshift estimate is $>2.5$\arcmin from the {\it{Planck}} position since earlier versions of the algorithm used to select the cluster sample did not impose such a tight constraint on the radial search.} of the {\it{Planck}} position fainter than 15.8 Vega mags at 3.4 microns (which corresponds to a 10$\rm{L}_{*}$ galaxy at $z\approx 1$\footnote{WISE is sensitive to a galaxy mass of $5\times10^{11}$\,$\rm{M}_{\odot}$ at $z\approx 1$. }). The purpose of the magnitude cut for our sample selection is to exclude contaminating foreground sources, while that of the distance cut is to maximize the likelihood of the red object being associated with the WISE overdensity (and, hence, the {\it{Planck}} candidate cluster).

\begin{table*}
\caption{Cluster-candidate information.  For the peak of the SZ decrement in the CARMA-8 SB {\sc{CLEAN}}ed, radio-source-subtracted maps, we provide the RA and Dec coordinates, the peak flux density after correcting for primary-beam (PB) attenuation and the distance to the map center (coincident with the Planck position). Using the PB-corrected peak SZ flux density$^{\dag}$ and the rms of the SB data, we calculate the SNR of the CARMA-8 detection. Also included are the SNRs in our analysis of {\it{Planck}} data and, where available, from the {\it{Planck}} Union catalog. The final column contains the IRAS estimate for the 100-micron intensity within 5\arcmin of the pointing centre for our observations. Not all candidate clusters in our sample have positions in the Union catalog, hence, for homogeneity, we centre the 100-micron statistics on the CARMA-8 pointing centre. However, the typical offsets between the Planck Union catalog positions and our CARMA-8 pointing centers, do not result in a significant change in the 100-micron emission. This 100-micron emission information can be retrieved from http://irsa.ipac.caltech.edu/applications/DUST/. Clusters that have been detected in the CARMA-8 data, even if marginally so, have their cluster ID written in bold font.}
 \label{tab:clusterpos}
\begin{tabular}{lcccccccccc}
\hline \hline
Cluster & RA                       & Dec                                    & PB-corrected Peak SZ       & Distance from      & SNR    & SNR   & SNR    &  100-micron Emission  \\
 & & & Flux  Density & Map Center  & CARMA-8  & {\it{Planck}} &   UNION\\
ID                   & hh mm ss.s             &  $^{\circ}$ \arcmin \arcsec           & mJy/beam       & \arcsec   & & & &    MJy/sr \\ \hline
{\bf{P014}} & 16 03 26.16         & +03 16 48.00         & -1.5 & 159  & 4.2 & 5.4 & 4.5 &  $5.32  \pm 0.35$\\
P028 & - & - & - & - & -   & 5.2        & 5.1&  $1.54 \pm 0.01$   \\
P031  & - & - & - & -    &&4.1  & -     &  $ 3.26 \pm  0.06 $  \\
P049  & - & - & - & -   & -&4.3 & - &      $0.67 \pm  0.04$   \\
P052  & - & - & - & -   &- &5.1 & -  &   $4.27  \pm 0.15$ \\
P057  & - & - & - & -   &-&5.3  & 4.6 &  $0.99 \pm 0.03$        \\      
{\bf{P086}} & 15 14 00.42        & +52 47 49.55                 & -3.4 & 94  & 5.1 & 5.1 & 4.6 &  $0.77 \pm 0.02$       \\
P090 & - & - & - & - & - &5.5  & 5.4 &  $0.59 \pm 0.02$         \\      
{\bf{P097}}& 14 55 23.79          & +58 52 18.42          & -3.0 & 84  & 4.4 & 4.8 &- &  $0.58 \pm 0.03$   \\
{\bf{P109}}&  18 23 08.14 & +78 23 04.18      & -4.2 & 76     & 7.3  & 5.6 & 5.3 &  $2.63 \pm 0.06$   \\        
P121& - & - & - & - &  - &5.0   & 5.6   & $ 0.93 \pm  0.11 $\\  
P134& - & - & - & - &  - &4.6   & 5.0 & $ 0.91 \pm  0.06$   \\
P138& - & - & - & - &  - &4.6   & 5.1 &  $2.03 \pm 0.04$        \\
{\bf{P170}}&  08 51 00.67 & +48 30 30.13  & -3.1 & 45  & 7.3 & 5.8 & 6.7&  $1.32 \pm 0.03$      \\
{\bf{P187}} &  07 32 21.15 & +31 38 11.02   & -2.4 & 49 & 5.8  &        6.0 & 6.1&  $ 2.55 \pm  0.08 $  \\
{\bf{P190}}& 11 06 08.09 &+33 34 00.22 &  -3.6 & 52  & 7.8 &  4.1 & 4.6 &  $ 1.06 \pm  0.05 $\\ 
{\bf{P205}}&  11 38 07.82 & +27 54 30.61  & -2.6 & 83 & 5.7 &  5.9 & 5.7  &  $1.10 \pm 0.01$   \\
P264& - &-  &-  &-  &  -  & 4.2 & -&   $1.61\pm  0.02 $ \\      
{\bf{P351}}&  15 03 59.21 & -06 06 30.25  & -2.1 & 96 &  5.6  & 3.8 & - &   $3.60 \pm 0.04$\\
\hline
\end{tabular}
\begin{tablenotes}
 \scriptsize
 \item ($\dag$)  To correct the measured peak flux density for the effects of primary-beam attenuation, we divide it by $\rm{exp}^{-r^2/(2\times\sigma^2)}$, where $r$ is the distance from the map center to the pixel with the peak flux density of the radio source and $\sigma = \rm{Beam
\,FWHM}/(2\times(2\times\rm{log_{\rm{e}}}(2))^{0.5})$ 
  \end{tablenotes}

\end{table*}

\begin{table*}
\caption{Information on the radio-source environment towards our candidate clusters obtained from the LB CARMA-8 data and NVSS. Column one contains the ID name for the cluster-candidate in whose field of view (FoV) the radio source lies in. Columns two and three contain the RA and Dec of the pixel with the peak flux density of the point source. At this position, and with this peak flux density as a starting point, the {\sc{Difmap}} task {\sc{Modelfit}} was used to obtain a best-fit peak flux density by chi-squared minimization. This best-fit value, after primary-beam correction, is given in the 5th column. The distance from the radio source to the map centre is provided in column 4. In column 6 we report, where available, the 1.4\,GHz NVSS (fitted) peak flux density for the LB-identified radio source.  The `fitted' flux density values in NVSS refer to the peak surface brightness.
We note that, unless stated otherwise, we will henceforth refer to flux density as the primary beam-corrected flux density. 
From the LB and NVSS peak flux densities, we calculate a spectral index between 1.4 and 31\,GHz, $\alpha$, where $S_{\nu} \propto \nu^{-\alpha}$, with $S$ denoting flux and $\nu$ frequency. It should be noted that the NVSS and CARMA data were taken years apart and variability could affect $\alpha$. The mean value for $\alpha$ is 0.72. Clusters that have been detected in the CARMA-8 data, even if marginally so, have their cluster ID written in bold font.}
 \label{tab:sourceinfo}
\begin{tabular}{lcccccccc}
\hline \hline
Cluster      & RA                      & Dec      &         Distance from   & 31-GHz Peak &  1.4-GHz Peak & $\alpha$ \\
                &                           &              & Map Centre                                              & Flux Density   & Flux Density & \\
ID               & hh mm ss.s       &  $^{\circ}$ \arcmin \arcsec                              & \arcsec                                      & mJy                          & mJy                        & 1.4/31\,GHz\\ \hline
{\bf{P014}}         & 16 03 18.90       & +03 16 44.00  & 152  & 6.3 & 120.8 & 0.95 \\
{\bf{P014}}         & 16 03 30.20        &  +03 26 32.00 & 458 & 9.4 & 96.8  & 0.75 &  \\
P028         & 15 40 13.24    & +17 56 33.14  &134 & 3.7 & 21.9 & 0.57\\
P028         & 15 40 21.92    & +17 52 45.12  & 196 & 3.2 & 77.7 & 1.02 \\
P031         & 15 27 30.42    & +20 41 32.27  & 113  &  3.4  & - & - \\
P049         & 14 44 27.84    & +31 13 15.87   & 133 & 10.0 & 62.6 & 0.59 \\
P052         & 21 18 49.08    & +00 33 28.00  & 202 & 5.6    & 129.4 & 1.01 \\
P052         & 21 19 05.08    & +00 32 40.00   & 44 & 2.5 & 56.3 & 1.00\\
P057         & 15 48 49.32     & +36 10 29.80   & 240 & 3.4 & 68.6 & 0.97\\
P057         &15 48 41.06      & +36 09 33.85 & 130 & 3.2 & 7.6 & 0.28\\
{\bf{P109}}         & 18 22 52.25     & +78 23 04.18  & 76 & 2.6 & - & - \\
P121    & 13 02 40.94     &+67 28 42.30  & 311 & 23.2 & 317.1 & 0.84\\
P138     & 10 26 37.40    & +70 32 50.37  & 432 & 150.3 & 135.1 & -0.03\\
{\bf{P170}}     & 08 51 14.78     & +48 37 06.11  & 421 & 10.8 & 119.9 & 0.78\\
{\bf{P187}}      &  07 32 20.21   & +31 41 19.03  & 161 & 3.1 & 12.7 & 0.46 \\
{\bf{P351}}      & 15 04 18.65 & -06 05 15.24  & 237 & 3.3  & 51.7 & 0.89 \\
\hline
 \end{tabular}
\end{table*}

\begin{table*}
\caption{Predicted 31-GHz flux densities after primary-beam attenuation for radio sources detected in NVSS  within 10\arcmin\ of our map centre but without a counterpart in the LB CARMA-8 data. These flux densities were calculated using the mean 1.4-31\,GHz spectral index (0.72) derived from radio sources detected in NVSS and our CARMA-8 data. We present only those sources whose estimated (attenuated) 31-GHz flux density is at least four times the noise in the LB CARMA-8 data. The expected SNR in the LB CARMA-8 data is provided in the last column. This was calculated using the LB rms and the expected 31-GHz beam-attenuated peak flux density. }
\label{tab:pred31source}
\begin{center}
\begin{tabular}{ccccccc}
\hline \hline
Cluster ID & RA & Dec & Distance & Predicted, Primary-beam-attenuated & SNR \\ 
   &                      &                      &                 &  31-GHz Peak Flux Density & \\
  & hh mm ss.s &   deg\,min\,sec  & \arcsec   & mJy & \\ \hline
P052 &21 18 48.31 &0 37 46.3& 356  & 2.4 & 6.1\\
P052 &21 19 25.44 &0 31 42.1 &354 & 3.3 & 8.6\\
P109 &18 22 40.96& 78 22 20.2 &64.6 & 4.0 & 7.7\\
P190 &11 05 53.78 &33 40 53.3 &447& 2.4 & 6.9\\
\hline
\end{tabular}
\end{center}
\label{default}
\end{table*}%

 We note that there is some scatter in the mid-infrared (MIR) color relation and some spread in the range of possible evolutionary tracks. Such that, in principle, lower redshift objects mimicking the $z>1$ WISE colors could lie within our sample. Though we do not expect to  have targeted many $z<0.5$ objects as it is likely that they would have been detected by other instruments and surveys, and a large fraction of our cluster candidates do not lie in close proximity to a confirmed cluster of sufficient mass (see Table \ref{tab:clustersfov}). Our goal is to eventually use spectroscopic data to obtain accurate redshift estimates. Objects are considered red if their flux in WISE channel 1 (3.4 microns) minus channel 2 (4.6 microns) $> -0.1$ in AB mags (0.5 in Vega); a method for preferentially identifying $z> 1$ objects (\citealt{papovich2008}). This MIR color criterion has been used by e.g., \cite{galametz2012} and \cite{brodwin2012}, who have also followed-up galaxy overdensities at $z=1.75$ with CARMA-8.

Upon identifying WISE galaxy overdensities at {\it{Planck}} candidate-cluster locations, we discarded systems with high ISM contamination, as measured in the IRAS 100\,$\mu$m-intensity band, as they are more likely to be spurious detections, and those at $\delta < -10^{\circ}$ in order to ensure sufficient $uv$-coverage of the CARMA-8 data. We assigned a figure of merit to the remaining cluster candidates based on several parameters: distance of the WISE brightest red object within 2.5\arcmin\ from the {\it{Planck}} candidate-cluster position, the SNR in {\it{Planck}}, and the magnitude of the WISE overdensity, and drew our final sample from the highest priority objects.

\section{CARMA-8 observations}
\label{sec:data}

\subsection{Data processing}
CARMA-8 observations were obtained towards 19 {\it{Planck}}-selected cluster candidates. Initially, data were collected for each cluster candidate for $\approx 4$\,hours; if a clear or tentative SZ decrement appeared in the maps, more data were gathered on that object to improve the SNR of the detection; otherwise, no more observations were queued for that particular system. This was an observing strategy designed to maximize the number of detections in the limited CARMA-8 time at our disposal ($\approx 110$\,hrs) and, as a result, there is significant spread in the rms of our final visibilities. The noise levels are provided in Table \ref{tab:clusters}, together with pointing centers, shorthand cluster IDs---which will be used throughout this work---and information on the beam, which is a measure of resolution.

Each cluster observation was interleaved every 15 minutes with data towards a bright, unresolved source to correct for variations in the instrumental gain. Absolute calibration was undertaken with observations of Mars. The raw data were exported in MIRIAD form and converted into a MATLAB format in order to be processed by the CARMA in-house data reduction software, which removes bad data points (e.g. visibilities that are shadowed, obtained during periods of sharp rises in system temperature or when the instrumental response changes unexpectedly or without calibrator data) and corrects for instrumental phase and amplitude variations; for further details see \cite{muchovej2007}. The output $uvfits$ file from the pipeline contains calibrated visibilities ($V$)---the response of the interferometer for a single baseline or the Fourier transform of the sky brightness distribution times the primary beam---which, for small FoVs like that of the CARMA-8 pointed observations, can be approximated by :

\begin{equation}
V(u,v) = \int \int A_N(l,m)I(l,m) \times \rm{exp}\left( {-2\pi j[ul+vm]} \right) dl dm,
\end{equation}
where $A_N$ is the normalized antenna beam pattern, $I(l,m)$ is the sky intensity distribution, $u$ and $v$ are the baseline lengths projected on to the plane of the sky, and $l$ and $m$ are the direction cosines measured with respect to the $(u,v)$ axes. The interferometer only measures some visibility values in the $uv$-plane. Hence, the array returns a sampled visibility function $S(u,v)V(u,v)$, where $S(u,v)$ is a function known as the {\it{synthesized beam}} that equals one for sampled visibilities and zero otherwise. By applying an inverse Fourier transform to the sampled visibilities measured by the array, the sky image (or {\it{dirty map}}) can be recovered. In order to reconstruct the sky brightness distribution from an incomplete visibility map, we use a deconvolution algorithm: {\sc{CLEAN}} \citep{hogbom1974}.

\subsection{Radio Source Contamination}

\begin{figure*}
\begin{center}
\includegraphics[width=15cm]{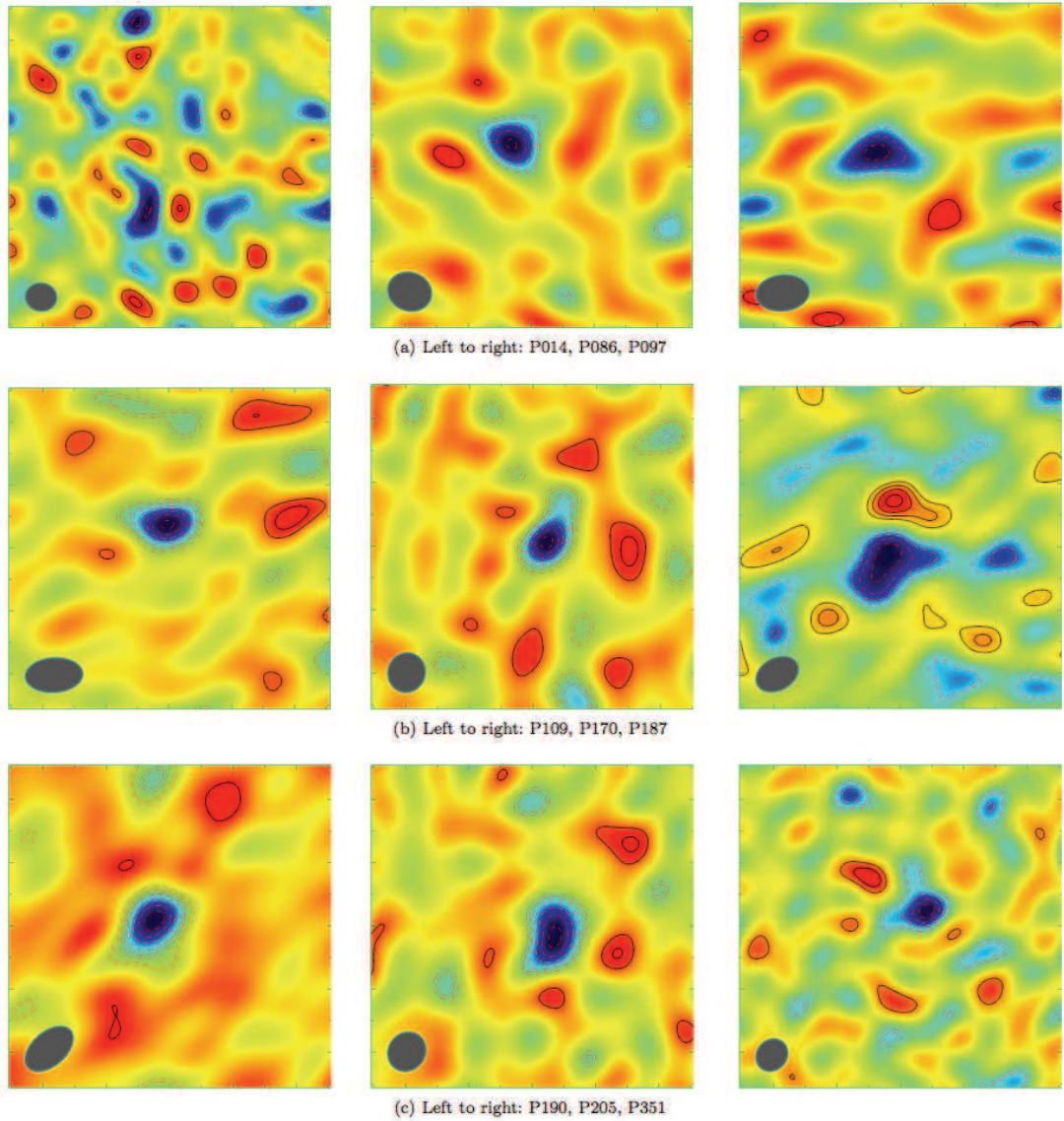}
\caption{
{\sc{CLEAN}}ed, naturally weighted maps of the SB data for CARMA-detected clusters after radio-source subtraction, where necessary. The dimensions are $1000\arcsec \times 1000\arcsec$ centered at the pointing center.  This is about twice the {\it{Planck}} beam FWHM (7-10\arcmin) relevant for SZ detection. Contours are scaled linearly starting from 2 to 10$\sigma$ , where $\sigma$ is the noise on the map 
(Table \ref{tab:clusters}), in integer multiples. Positive contours are shown in solid, black lines while negative contours are shown in dashed, red lines. The grey ellipse in the bottom left corner of each map is the synthesized beam.  No primary-beam correction has been applied. 
Table \ref{tab:clusterpos} contains the value for the measured peak SZ flux density and information on detected radio sources is provided in 
Table \ref{tab:sourceinfo}.
\label{fig:cleanmapsdet}}
\end{center}
\end{figure*}

In order to remove contaminating radio sources from the SB data, the LB CARMA-8 data were used to identify the location of the compact radio sources and provide an initial estimate of their peak flux density. This initial set of source parameters was fit directly to the LB visibility data and the best-fit parameters were determined using the {\sc{Difmap}} task {\sc{Modelfit}} (\citealt{shepherd1997}). Using these best-fit parameters, the contribution to the SB data from detected LB radio sources was removed. Radio source-subtracted, {\sc{CLEAN}ed}\footnote{{\sc{CLEAN}}ing was undertaken with a tight box around the cluster signal, where present, otherwise no box was used.} maps were produced, see Figure \ref{fig:cleanmapsdet} for CARMA-8 SZ-detected clusters, and Figure \ref{fig:CombiND} for candidate clusters without a CARMA-8 detection. Details on the SZ signal detected by CARMA and {\it{Planck}} are given in Table \ref{tab:clusterpos}. For those CARMA-8 data that required the contribution from LB-detected radio sources to be removed, the reader should note that the source-subtracted SB images represent the map with the most-likely source parameters and, thus, uncertainties in the source parameters are not reflected in the final map. Given that the cluster and source parameters can be degenerate, a quantitative analysis should fit for the cluster and radio-source contributions jointly. This analysis is undertaken in paper II of this work (Rodr\'{i}guez-Gonz\'{a}lvez et al. {\it{in prep}}).

Compact radio sources were detected in the LB CARMA-8 towards 12 cluster candidates; information on their properties is provided in Table \ref{tab:sourceinfo}. The LB-inferred radio-source environment towards candidate clusters with a CARMA-8 detection is, for the most part, benign, such that it has little or no effect on the SZ signal in the source-subtracted SB data. We find that this is not the case for a large fraction of the candidate clusters without a CARMA-8 detection, particularly for P031, P049, P052, P057, P121 and P138.

Noise levels in our LB data (on average $\approx 520$\,$\mu$Jy/beam) are comparable to typical noise levels in NVSS ($\approx 500 - 1000$\,$\mu$Jy/beam) but, since radio sources tend to be brighter at lower frequencies, NVSS often provides a more complete image of radio-emitting sources. 
Hence, we predict the 31-GHz radio-source environment by extrapolating NVSS (`fitted') peak flux densities for radio sources located within 10\arcmin of our map centre assuming a value for the spectral index, $\alpha$. Our choice of $\alpha$ is based on the average 1.4-to-31\,GHz spectral index derived from matching sources in NVSS and our LB CARMA-8 data (Table \ref{tab:sourceinfo}). In Table \ref{tab:pred31source} we present the position of those NVSS sources that, given our assumptions, we would expect to find at a significance of 4$\sigma$ or greater in the LB CARMA-8 data but, instead, remained undetected.  
A total of 4 radio sources satisfy these criteria. All but one of these sources (in the FoV of P109) are at $\gtrsim 6$\arcmin from the map center, with a predicted, attenuated 31-GHz peak flux density of $<3.5$\,mJy and, hence, should have a negligible impact on the recovered SZ signal. Undetected sources in P109 have the potential to have the most impact on the recovered SZ signal, with a source expected to have an attenuated 31-GHz peak flux density of 4.0\,mJy an arcminute away from the map centre. However, some of these sources could be varying, or could have flatter or inverted spectra, implying they could be brighter at 31 GHz than our predictions in Table \ref{tab:sourceinfo}. To estimate the contribution from radio sources below our detection threshold ($4\sigma \approx 1.5$\,mJy/Beam), we use a study by 
Murphy \& Chary (2014, in prep.), which stacks 30-GHz {\it{Planck}} data at the position of NVSS-detected radio sources of similar flux density. We expect to suffer from $\approx 90\mu{\rm{Jy/Beam}}$ of contamination from confused radio sources, which would account for $2-6\%$ of the peak SZ decrement of our clusters (Table \ref{tab:clusterpos}).

\subsection{CARMA SZ Detections}
\label{sec:notes}

Eight of the targeted 19 cluster candidates were detected with a peak beam-corrected SZ-flux-density with an SNR $\geq 4.4$ (Table \ref{tab:clusterpos})\footnote{The SNR for the CARMA SZ detections was calculated as the ratio of the peak decrement, after correcting for beam attenuation, and the rms of the SB data.}. We also found one tentative detection with an SNR of 4.2, which we deem tentative due to uncertainties in the data (see Appendix \ref{sec:appnotes}). 
Clusters have been classified as detections based on the inspection of the SB and LB CARMA-8 maps, taking into consideration ancillary data, which will be presented in detail in Section \ref{sec:anci}. For each cluster candidate we describe in Appendix \ref{sec:appnotes} the CARMA-8 data, including the SZ signal and the radio-source environment. 

The primary-beam-corrected peak flux density of the SZ decrements in the CARMA-8 data (after source subtraction, where necessary) ranges between -1.5 and -4.2\,mJy/beam, where P014---a marginal detection---has the smallest decrement. The SNR of the SZ signals in the CARMA data ranges between 4.4 (for P097) to 7.8 (P190). The peak of the CARMA-8 SZ signals is within 2\arcmin of the {\it{Planck}} position for all systems but for P014, which is $\approx 2.7\arcmin\ $ away (see Table \ref{tab:clusterpos}). 

\subsection{Potential Reasons for CARMA-8 Non-detections of the Remaining Candidate Clusters}

The data towards 10 cluster candidates did not show a significant decrement; though this does not provide conclusive evidence, in all cases, that the {\it{Planck}} cluster candidate is a spurious detection. A combination of factors could be responsible for the lack of an SZ signal in the CARMA-8 data (see Section \ref{sec:real} for comments on each of the undetected targets). These factors include, high noise levels in the SB data, confused radio sources (most relevant for observations with high rms in the LB data), poor subtraction of LB-detected radio sources and the cluster being significantly offset from the map centre, which results in strong beam attenuation. For example, a cluster with a relatively low peak-flux density of $-1.5$\,mJy/beam, at 3\arcmin (at 4\arcmin) from the map centre would have a flux density on the CARMA SB map of -1.2\,mJy/beam (-1.0\,mJy/beam). Given that the typical range of rms values in the CARMA-8 SB data range between $\approx400$ and 600\,$\mu$Jy/beam, such a signal would have, at best, an SNR $\approx 2-3$ ($1.7-2.6$) and, hence, would not be registered as a solid detection. This highlights one of the challenges of high-resolution interferometric follow-up of clusters detected
by a large beam experiment such as {\it Planck}.
Radio sources are often the dominant contaminant to the CARMA-8 observations. If they are not subtracted properly or they are not detected above the background noise---and, therefore, are not accounted for---they can fill and/or distort the SZ decrement. 

\begin{table*}
 \begin{center}
\caption{Qualitative summary of the information drawn from the main relevant data available towards those cluster candidates in our sample without a CARMA-8 SZ detection. Two tick marks indicate a strong presence of the emission of interest, while a single tick mark indicates a smaller contribution. Those systems for which the emission of interest is considered negligible are labelled with a cross mark. The SNRs given in the fourth column correspond to the most significant negative feature within $\approx 3\arcmin\ $ of the map centre. Those entries labelled with (I) do not appear in the Union catalog and the SNRs come from an intermediate {\it{Planck}} data product.
Based on this information, we include a comment on our own judgement on the likelihood of there being a real cluster in the neighborhood of the CARMA map center.}
 \label{tab:notesreal}
\begin{tabular}{lccccccc}
\hline \hline
Cluster ID &  \multicolumn{2}{c}{CARMA-8}                                                &  \multicolumn{2}{c}{{\it{Planck}}} & ROSAT & Real?\\
                   & Radio Contamination & SB Data                                             & SNR  & ISM$^{c}$ &                &  \\
                   \hline
P028         & \xmark &  $\,\,\,\,\,\,\,\,\,\,\,\,\,\,\,\,\,\,\,2\sigma$ ($\lesssim -0.9$\,mJy/beam)$^{a}$        &  5.1    & \cmark & \xmark & Unlikely \\
P031         & \cmark & $\,\,\,\,\,\,\,\,\,\,\,\,\,\,\,\,\,\,\,3\sigma$ ($\lesssim -0.9$\,mJy/beam)$^{b}$         &  4.1(I) & \cmark \cmark & \xmark & Unlikely\\
P049         & \cmark \cmark & $<2\sigma$ ($\lesssim -1$\,mJy/beam)    & 4.3 (I) & \xmark               & \cmark  & Likely \\
P052         & \cmark & $\,\,\,\,\,\,\,\,\,\,\,\,\,\,\,\,\,\,\,2\sigma$ ($\lesssim -0.7$\,mJy/beam)                & 5.1 (I)  & \cmark \cmark  & \xmark & Unlikely\\
P057         & \cmark \cmark & $<2\sigma$ ($\lesssim -1$\,mJy/beam)    & 4.6       & \xmark & \xmark & Unknown\\
P090         & \xmark & $\,\,\,\,\,\,\,\,\,\,\,\,\,\,\,\,\,2\sigma$ ($\lesssim -0.8$\,mJy/beam)                   & 5.4 & \xmark                  & \cmark & Likely \\     
P121         & \cmark \cmark & $\,\,\,\,\,\,\,\,<2\sigma$ ($\lesssim -1.6$\,mJy/beam) & 5.6 & \xmark  & \xmark  & Unknown \\
P134         & \xmark & $\,\,\,\,\,\,\,\,\,\,\,\,\,\,\,\,\,\,\,2\sigma$ ($\lesssim -0.7$\,mJy/beam)                & 5.0  & \xmark & \xmark & Unknown\\
P138         & \cmark \cmark & NA$^{d}$                                                             & 5.1  & \cmark & \cmark & Likely\\
P264         & \xmark & $\,\,\,\,\,\,\,\,\,\,\,\,\,\,\,\,2\sigma$ ($\lesssim -1$\,mJy/beam)$^{b}$             & 4.2 (I) & \cmark & \xmark & Unlikely\\
\hline
\end{tabular}
  \begin{tablenotes}
 \small
 \item(a) The map contains several other negative features of similar SNR.
 \item(b) The negative features coincide with features from interference or sidelobe patterns.
 \item(c) One (two) tick marks refer to 100 micron emission greater than 1\,MJy/sr (3\,MJy/sr).
 \item (d) This information is not meaningful for P138 given the poor $uv$ coverage and high rms of the CARMA-8 SB data towards this system.
  \end{tablenotes}
\end{center}
\end{table*}

\section{Validation}
\label{sec:vali}

We discuss the reliability of the cluster candidates in our sample by (1) searching for evidence for the presence of a cluster in catalogs in the literature and as excess emission in the ROSAT X-ray data and (2) by considering the effects of ISM contamination to the {\it{Planck}} data.

\subsection{Confirming the Presence of a Cluster in the CARMA-8 Data through Ancillary Datasets}
\label{sec:anci}

\paragraph*{Cluster Catalogs in the Literature: }

In Table \ref{tab:clustersfov} we provide details on known clusters found in the SIMBAD database \citep{wenger2000} and in the \cite{wen2012} SDSS-based catalog within 4\arcmin of the map center of our observations. At least one such match was found for thirteen of our clusters candidates. The CARMA-8 SZ signal for four of the systems in our sample---P170, P187, P190 and P205---coincides with the location of a registered overdensity of galaxies from Wen et al., with redshift estimates ranging from $\approx 0.2$ to $0.5$; though only P187 has spectroscopic confirmation.

Seven of our cluster candidates without a CARMA-8 SZ signal were found to have a known cluster (or cluster candidate) within 4\arcmin of the map center. In most cases, the lack of a CARMA-8 SZ signal towards these systems can be attributed to a combination of one or more of the following: challenging radio-source environment, high rms of the SB visibility data, primary-beam attenuation and insufficient ICM cluster mass (deduced from the X-ray data or estimated from the richness value). For further details on each CARMA-detected system see Appendix \ref{sec:appnotes}, and Section \ref{sec:real} for the remaining systems .

\paragraph*{ROSAT X-ray data: }

ROSAT X-ray images for each of our cluster candidates without (and with) a CARMA-8 SZ detection are provided in Appendix \ref{sec:ndetmaps} (and Appendix \ref{sec:rosat}). For the eight candidate clusters with a clear CARMA-8 detection, there is compelling evidence for X-ray cluster emission within a few arcminutes towards all of them, though the peak of the X-ray emission lies outside the CARMA SZ contours for two systems, P109 and P170. For P014---the candidate cluster with the tentative CARMA detection---there is no support for the presence of a cluster from the ROSAT image. Regarding the cluster candidates without a CARMA SZ detection, three seem to have evidence for extended X-ray emission in the ROSAT data: P049, P090 and P138.

\subsection{Investigating 100-micron ISM contamination in the {\it{Planck}} data}
{\it{Planck}} uses a multifrequency matched filter to detect both the SZ increment and the SZ decrement from a cluster. Due to strong ISM contamination in the upper frequencies (353, 545, 857\,GHz), the SZ increment may be biased high, resulting in a spurious detection, especially if the ISM emission falls on top of a CMB cold spot.
In Table \ref{tab:clusterpos} the 100-micron emission within $5\arcmin $ from the pointing centers of the CARMA-8 cluster observations is provided. The mean 100-micron emission is  $1.68 \pm 0.02$ \,MJy/sr towards clusters without a CARMA-8 detection and $2.10 \pm 0.01$\,MJy/sr for those with a clear detection. Hence, the presence of this foreground does not correlate strongly with detectability, at least for cluster candidates in our sample. The candidate cluster suffering from the highest ISM contamination is P014 with $5.3 \pm 0.4$\,MJy/sr. This system also shows the largest positional offset between the {\it{Planck}} and CARMA-8 data but the ISM contamination is not sufficient, in this case, to account alone for such a large discrepancy.

\section{Discussion}
\label{sec:discussion}

\subsection{Are any of the Candidate Clusters without a CARMA-8 Detection likely to be {\it{real}}?}
\label{sec:real}

We investigate the likelihood of the candidate clusters without a CARMA-8 detection being {\it{real}} clusters by considering all the relevant data available to us. That is, we consider the
ensemble of the CARMA-8 data (Table \ref{tab:clusters}), the NVSS catalog of radio sources (Table \ref{tab:pred31source}), the {\it{Planck}} data (Table \ref{tab:clusterpos}), contamination in the {\it{Planck}} data from the 100-micron emission (Table \ref{tab:clusterpos}), ancillary data (Section \ref{sec:anci}) including the ROSAT images (Appendix \ref{sec:ndetmaps} and \ref{sec:rosat}) and evidence in the literature for the presence of other clusters within 4\arcmin of the map center of the CARMA images (Table \ref{tab:clustersfov}). A summary of these results is given in Table \ref{tab:notesreal}.\\
\\
{\bf{P028}:} radio-source contamination in the CARMA-8 data is low. In the SB data there is a 2$\sigma$ negative feature $<2\arcmin\ $ away from the pointing centre but this is not an isolated 2$\sigma$ feature in the map. The SNR in {\it{Planck}} is quite high, 5.1, but could be affected by ISM emission ($\approx1.54$\,MJy/sr). No X-ray emission is seen in the ROSAT image. It seems unlikely that there is a single massive cluster associated with P028. But, since there are three low-mass (candidate) clusters cataloged in the literature within 3\arcmin\ of the CARMA-8 observations, with $0.07\lesssim z  \lesssim 0.4$, it could be that {\it{Planck}} picks up a combined SZ signal due to its large beam size.\\
{\bf{P031}:} has no decrement in the CARMA-8 data that could reliably be attributed to a low-SNR SZ signature, since the ones present in the map are likely to be a product of the inadequate subtraction of an extended radio source. The {\it{Planck}} data yields a low SNR cluster signal and suffers from high ISM contribution. There is no cluster detection in the ROSAT image. Thus, we conclude that, most likely, this is a spurious {\it{Planck}} cluster candidate.\\
{\bf{P049}:}  the challenging radio-source environment could be responsible for the lack of a CARMA-8 detection towards P049. The SNR in {\it{Planck}} is moderate, but contamination from ISM emission is low, which reduces uncertainties in the SNR value. The ROSAT image shows extended X-ray emission, with three point sources---a strong indication that there is a cluster towards P049.\\
{{\bf{P052}}:} a $3\sigma$ ($\approx -1.2$\,mJy/beam) feature is detected in the radio source-subtracted SB data within 2\arcmin\ of the pointing centre. The high level of radio-source contamination could be filling in partly an SZ decrement. This system appears in the {\it{Planck}} data with an SNR of 5.1, but these data suffer from high ISM emission. No X-ray cluster emission is seen in the ROSAT data. Overall, P052 does not seem a robust cluster candidate.\\
{{\bf{P057}}:} radio-source contamination in the CARMA data is substantial. The SNR in {\it{Planck}} is moderate and no cluster signal is seen in the ROSAT image.\\
{{\bf{P090}}:} there is a clear cluster detection in the ROSAT data. The x-ray counts in these data are higher over a similar area than for another cluster P187 (Abell\,586), which has an estimated mass within $r_{200}$---the radius at which the mean gas density is 200 times the critical density---of $\gtrsim 5$\,$h^{-1}_{100}M_{\circ}$ (\citealt{carmen2012}), according to SZ measurements by  the Arcminute Microkelvin Imager (\citealt{zwart2008}).  P090 has an SNR of 5.6 in the {\it{Planck}} data, which have negligible ISM contamination. Given all of this, it is surprising that, despite the seemingly benign radio-source environment and the low rms in the SB data, there is no clear SZ detection in the CARMA-8 data.\\
{\bf{P121}:} has a moderate SNR in the {\it{Planck}} Union catalog with low contamination from ISM emission. The CARMA-8 data is strongly affected by radio-source foregrounds and there is no sign of cluster-like X-ray emission in the ROSAT data.\\
{\bf{P134}:} has a {\it{Planck}} SNR of 5.0, which should not be greatly affected by ISM emission. There is no indication of the presence of a cluster in the ROSAT data. Regarding the CARMA results, radio contamination is negligible and, while the rms is high, $\approx 600$\,$\mu$Jy/beam, we  would expect to detect a massive ($\gtrsim -2.0$\,mJy/beam) cluster at $\gtrsim 3\sigma$ and yet not even a 2$\sigma$ negative feature is seen within 3\arcmin of the map center.\\
{\bf{P138}:} has an SNR of 5.1 in the {\it{Planck}} Union catalog, with little impact from ISM emission. This, together with the detection of extended X-ray emission in ROSAT, deem P138 a robust cluster candidate. The lack of a CARMA-8 detection is not surprising given the complications with radio-source foreground removal and the exceedingly high rms and poor $uv$ coverage due to the limited amount of data available towards this system. \\
{\bf{P264}:} is most likely a spurious cluster candidate. The {\it{Planck}} data suffer from significant ISM contamination and yield a low SNR for P264. The ROSAT image shows no evidence of a cluster nor do the CARMA-8 data, despite the reasonable rms and benign radio-source environment.\\

\section{Redshift Estimation}
\label{sec:redshiftandvalid}
The higher resolution CARMA-8 data showed that the putative BCG, on whose photometric redshift we relied on to select our $z\gtrsim1$ cluster candidates, did not lie within the CARMA-8 SZ contours for all but three of the systems: P097, P109 and P190, see Appendix \ref{app:multicol}. We now explore their plausible redshifts derived from photometric criterion. This is an independent approach, parallel to the XMM follow-up of {\it{Planck}} clusters, which has yielded redshifts of many {\it{Planck}}-selected clusters \citep{Planck2013}. We note that this analysis would be challenging to undertake without the CARMA-8 data since the location of the cluster would not be precisely known and, thereby, the galaxy counterparts might be difficult to identify.

\subsection{The {\sc{WISE}} Color Criterion}

In Figure \ref{fig:wiseccdet} (and \ref{fig:wiseccndet}) we explore the color-magnitude distribution of {\sc{WISE}} sources towards candidate clusters with (without) a CARMA detection within 2\arcmin of the CARMA ({\it{Planck}}) centroid. There does not appear to be any apparent difference in the number of red WISE galaxies for the CARMA detected and undetected clusters (Table \ref{tab:wiseinfo}).
We overlay the tracks for a passively evolving starburst of mass 10$^{12}\,M_{\odot}$, and an e-folding timescale of star formation of 100\,Myr using three formation redshifts for the burst. {\sc{WISE}} objects satisfying the MIR color criterion are shown as solid diamonds. 
The position of the brightest red object fainter than 15.8 mags at 3.4 microns is circled. The colors of this red object had suggested that P014, P170, P187, P190 and P205 would be at $z>1$ using
the $[3.4]-[4.6]>-0.1$ AB mag criterion. 
However, this object was found to lie outside the CARMA-8 SZ contours in all but three of the detected systems. With the CARMA-8 data in hand, we searched the WISE catalog for the
brightest red object within 2.5$\arcmin\ $ of the CARMA-8 SZ centroid; however, that only served to add P109 to the list of $z>1$ candidate clusters.

\subsubsection{Reliability of WISE colors to identify $z>1$ Clusters?}

\begin{figure*}
\begin{center}
\subfigure{
\centering
\includegraphics[width=5.7cm,clip=,angle=0.]{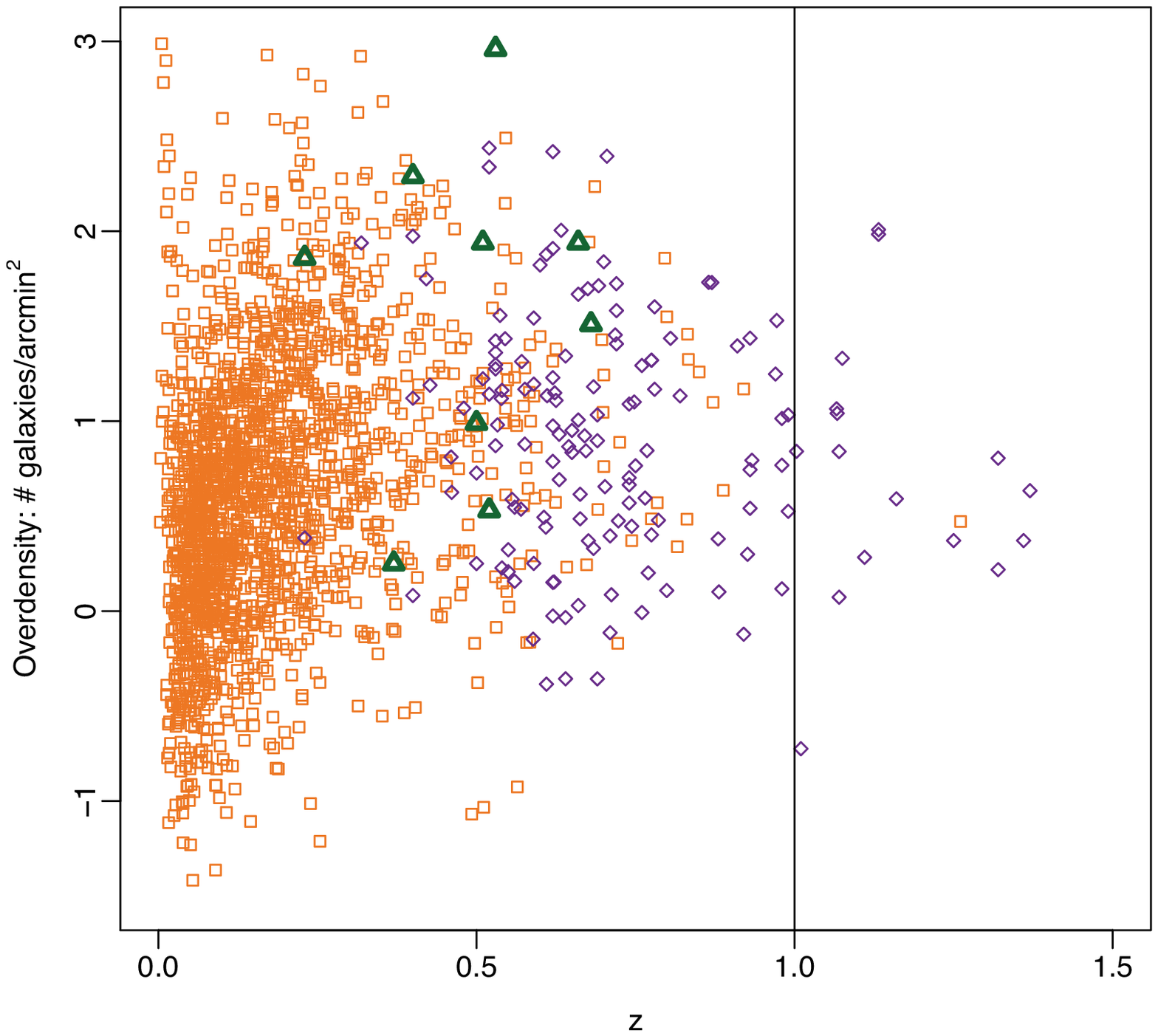}}\hspace{0.05em}%
\subfigure{
\centering
\includegraphics[width=5.7cm,clip=,angle=0.]{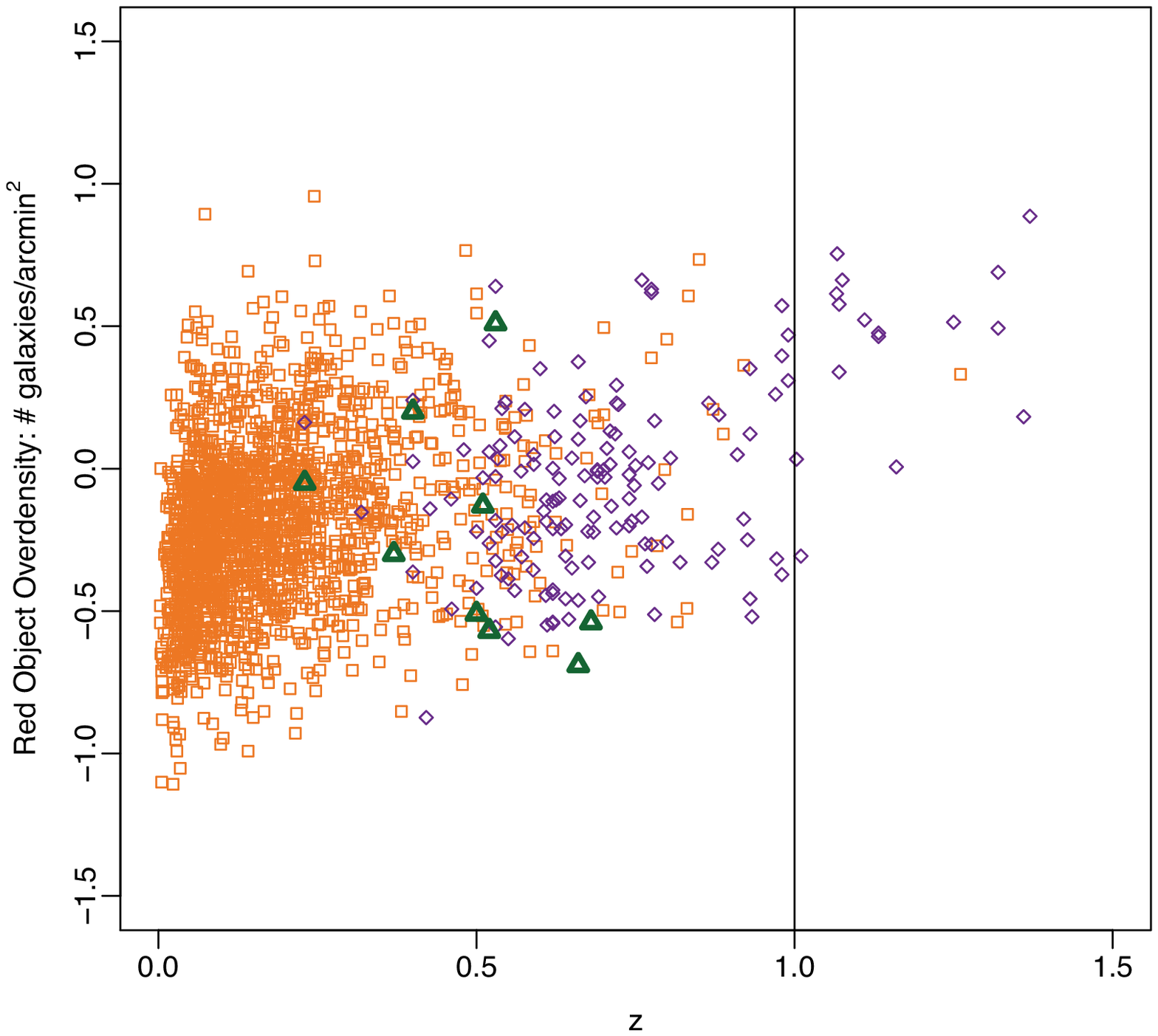}}\hspace{0.05em}%
\subfigure{
\centering
\includegraphics[width=5.7cm,clip=,angle=0.]{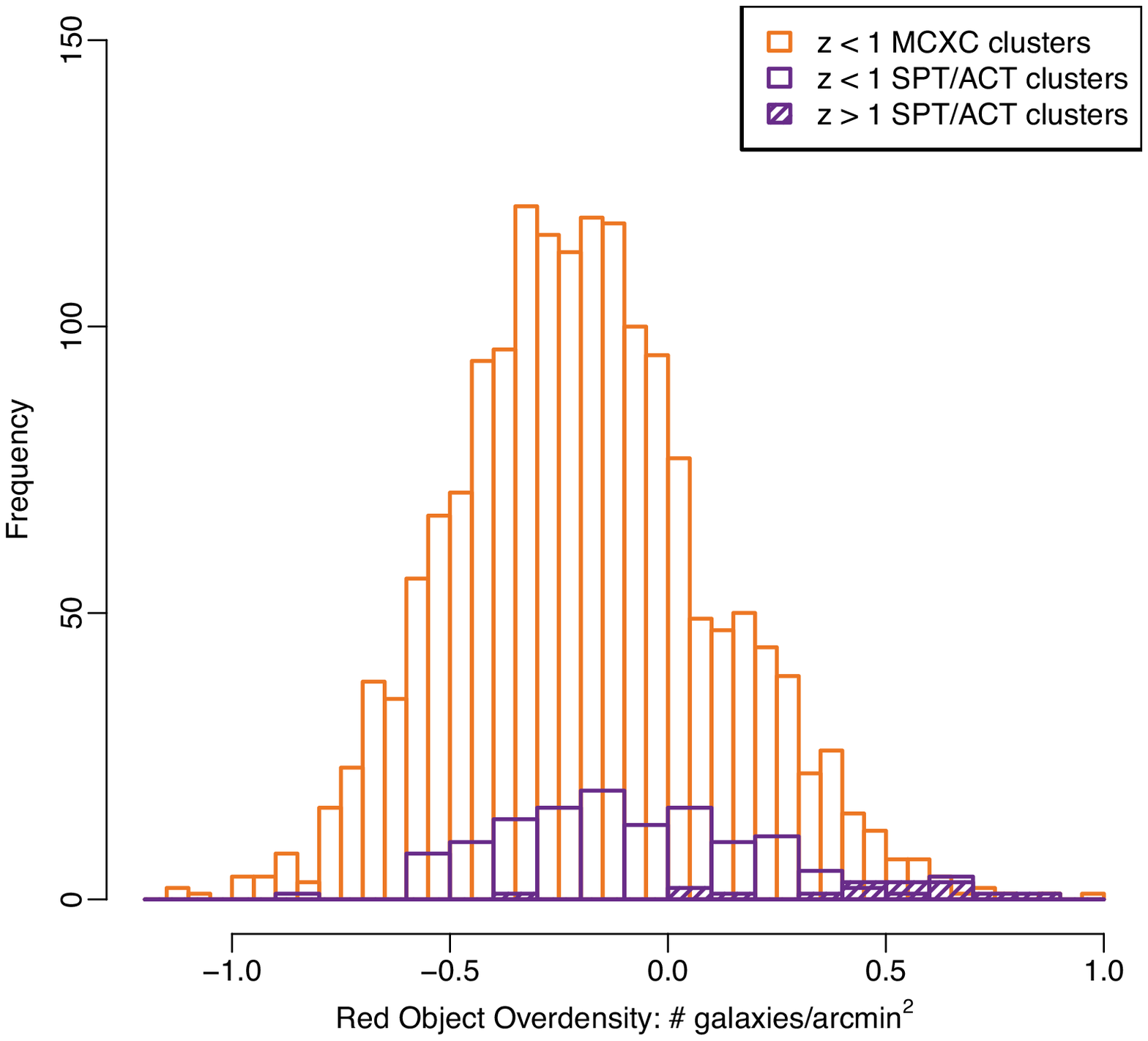}}
\caption{Left (and middle) panels: relation between redshift and the WISE all-object (red-object only) overdensity for (1) MCXC clusters (orange squares), (2) a selection of $z>0.5$ SPT and ACT-discovered clusters from \citet{williamson2011}, \citet{marriage2011}, \citet{reichardt2013} and \citet{hasselfield2013} (purple diamonds), whose redshifts can be spectroscopic or photometric and (3) {\it{Planck}} cluster candidates selected in this work with a CARMA-8 SZ detection (green triangles), where their redshifts are taken from Table \ref{tab:meansdssz}.  The overdensities have been calculated as the average density of WISE sources within a radius of $2\arcmin\ $ from the MCXC, SPT/ACT and CARMA centroids (as appropriate) minus the average density of sources in an annulus with inner and outer radii of $4.75\arcmin\ $ and $7\arcmin\ $. The vertical lines are included to highlight where $z=1$. These figures show that most of our cluster candidates are likely to be among the most massive known clusters at intermediate redshifts, which is not entirely surprising given that they were selected from {\it{Planck}} data, which are most sensitive to massive systems. The right panel shows histograms for the red-object WISE overdensities within 2$\arcmin\ $ for the MCXC clusters at $z<1$ (where only one cluster is at $z>1$) in orange and for the SPT/ACT clusters, where those at $z<1$ are shown in unfilled bins with a purple border and those at $z>1$ are shown in bins filled with diagonal purple lines. We do not include the cluster candidates from this work as the numbers are small.  While the overdensity of red objects is typically larger for clusters at $z>1$ than at lower redshifts, as one would expect from the application of the mid-IR criterion (e.g., \protect\citealt{papovich2008}), the contamination is high, such that selection of $z>1$ clusters from red-object WISE overdensities alone might not be very reliable. The histograms for the $z<1$ clusters in the SPT/ACT and MCXC samples are skewed towards negative red-object overdensity values, but this is likely due to uncertainties in the calculation for the overdensity and background estimation, though further investigation is out of the scope of this work.  
}
\label{fig:mcxcplck}
\end{center}
\end{figure*}

Selection of high redshift ($z\gtrsim1$) objects based on WISE or Spitzer colors has been discussed extensively in e.g., \citet{papovich2008}, \citet{gettings2012}, \citet{muzzin2013} and \citet{rettura2014}.  We explore further the reliability of the WISE mid-IR color criterion in Figure \ref{fig:mcxcplck} by looking at the relation between redshift and overdensity of WISE objects for a sample of clusters spanning a wide redshift range, $0.02 < z < 1.4$.
We use two overdensity measurements - one which considers all WISE objects (left panel) and another which only considers the red WISE objects i.e., those that satisfy $[3.4]-[4.6]>-0.1$\,AB mag (middle panel).
 The orange squares represent the 345 X-ray-detected clusters from the MCXC catalog (\citealt{pifa2010}). Clusters discovered by SPT and ACT (\citealt{williamson2011}, \citealt{marriage2011}, \citealt{reichardt2013} and \citealt{hasselfield2013}) with $z\gtrsim0.5$ (where some of these redshifts are photometric), are displayed in purple diamonds. {\it{Planck}} cluster candidates selected in this study with a CARMA-8 detection are represented by green triangles. For our cluster candidates, we use the SDSS-derived redshifts from Table \ref{tab:meansdssz}. The overdensities have been calculated as the average density of WISE sources within a radius of $2\arcmin\ $ from the MCXC, SPT/ACT and CARMA centroids (as appropriate) minus the average density of sources in an annulus with inner and outer radii of $4.75\arcmin\ $ and $7\arcmin\ $. The mean {\sc{WISE}}-all (and red-only) object overdensity for the MCXC clusters is 0.7 (-0.2)\,galaxies/\,arcmin$^2$, with a standard deviation of 0.8 (0.3); for the CARMA-8 detections it is 1.6 (-0.23)\,galaxies/\,arcmin$^2$, with a standard deviation of 0.9 (0.4) and for the SPT/ACT clusters it is 0.9 (-0.1)\,galaxies/\,arcmin$^2$, with a standard deviation of 0.6 (0.3) for systems at $z<1$ and 0.73 (0.4), with a standard deviation of 0.7 (0.3) for those at $z>1$. Although we do not know if all the {\sc{WISE}} objects are associated with the clusters, in a statistical sense, the average {\sc{WISE}}-object overdensities  for the different cluster sets suggests that our CARMA-8-detected clusters are likely to be massive systems. A strong correlation between WISE all object overdensity with redshift is not expected, though some selection effects could be manifesting themselves in the plot e.g., the highest redshift, X-ray-selected clusters are likely to be particularly massive, if they were serendipitous X-ray detections, and confusion in the WISE data is likely to have the highest impact at higher redshifts.
The red object criterion should be preferentially selecting $z\gtrsim1$ objects and, hence, we should expect a significant rise in the WISE red-object overdensity for systems at $z>1$. While we do indeed see a rise in Figure \ref{fig:mcxcplck}, middle panel, the histograms in the rightmost panel indicate that contamination levels are high. These contamination levels estimated from the positive-inversion of the negative side of the red-object overdensity histograms for $z<1$ systems suggest that relying exclusively on red-object WISE overdensities for selecting $z>1$ clusters can be risky.
We also find that there are some low redshift ($z<<1$) clusters that have large red-object overdensities. Again, this could be due to contamination. Otherwise, there may be an additional $z>1$ cluster within our 2\arcmin radius. There are also cases where $z>1$ background, unassociated objects could be producing an overdensity of red sources, in particular if a massive cluster is acting as gravitational lens, since such an effect would promote the detection of higher $z$ galaxies located behind the cluster that would otherwise remain undetected. It is also known that the $[3.4]-[4.6]>-0.1$ (AB mag) color selection of $z>1$ objects can sometimes misidentify low-redshift systems. Possible scenarios where an object might be falsely identified as red include, when the presence of an AGN leads to a rise in the galaxy SED at $\approx 3.4$\,microns or when the 3.3-micron polycyclic aromatic hydrocarbon (PAH) emission line in star-forming galaxies at a range of $z$ below 1 falls in the 4.6$\micron$-{\sc{WISE}} band (see e.g.,  \citealt{stern2012} and \citealt{assef2013} for methods on how to identify AGN in {\sc{WISE}} data). A targeted spectroscopic campaign in these cluster fields will be able to distinguish between these possibilities.

\subsection{The SDSS Criterion}
\subsubsection{Calibrating the SDSS photometric Redshift using a Subset of MCXC Clusters}
\label{sec:testmcxc}

We investigated the validity of using SDSS photometric redshifts, henceforth $z_{\rm{phot}}$, of galaxies in the vicinity of known clusters to estimate the cluster redshift. To do this, first we selected clusters in the MCXC catalog \citep{pifa2010} with $M_{500} \geq 2.5\times10^{14}M_{\rm{\odot}}$ and spectroscopic redshift $z_{\rm{spec}}\geq 0.5$,  in order to produce a cluster sample with similar properties to those expected for the {\it{Planck}}-discovered candidate clusters in this work. Secondly, we obtained $z_{\rm{phot}}$s from SDSS for all the galaxies within 1.5\arcmin\ and 1\arcmin\ of the MCXC cluster centroid. Histograms for the distribution of $z_{\rm{phot}}$s within these two radii for each of the selected MCXC clusters are shown in Figure \ref{fig:photozMCXC}. 

\begin{figure}
\begin{center}
\includegraphics[width=9.0cm,clip=,angle=0.]{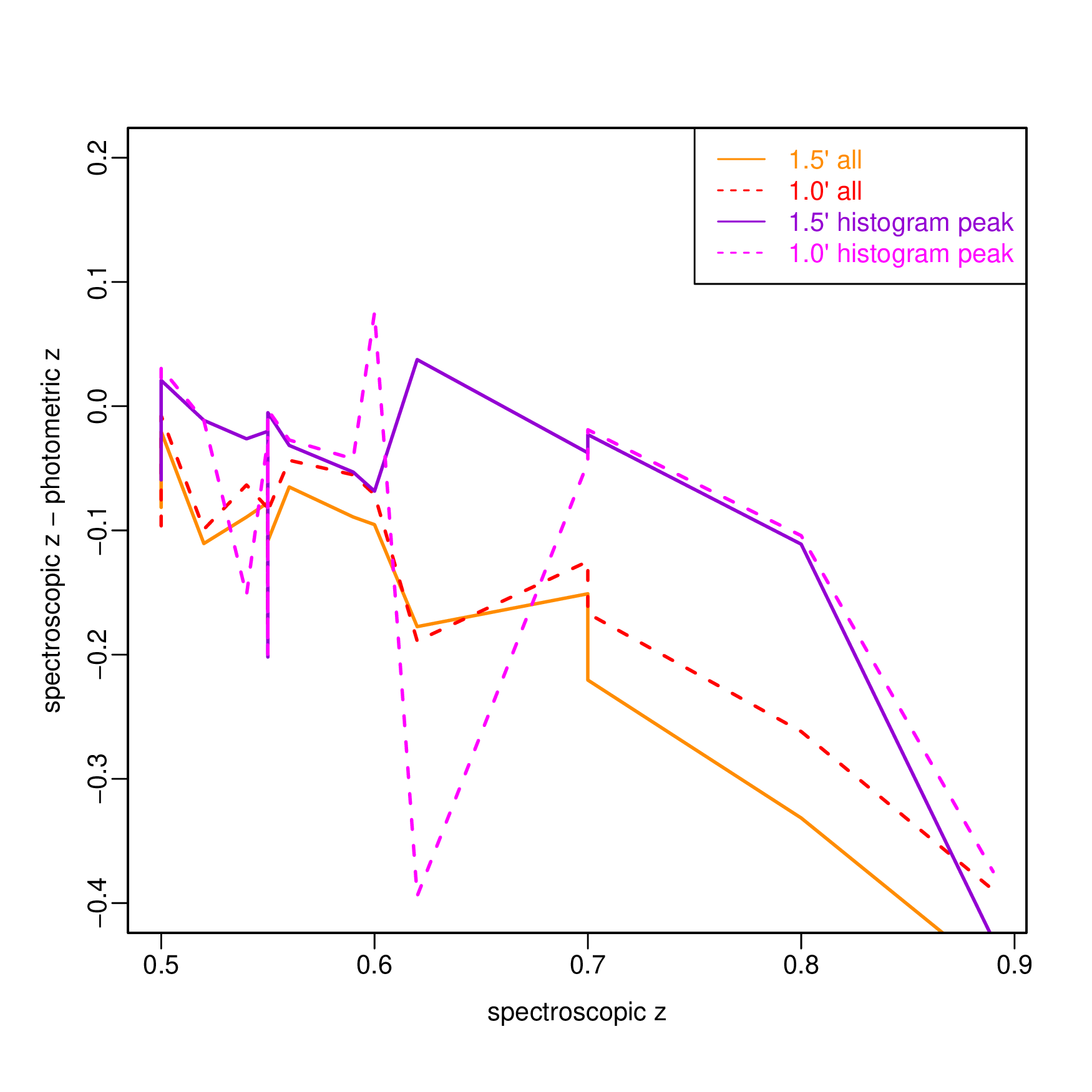}
\caption{The difference between the cluster spectroscopic  redshift and the photometric redshift estimates derived from two methods (see Section \ref{sec:testmcxc}) applied to SDSS objects with photometric information within 1.5\arcmin of the cluster X-ray centroid (from Piffaretti et al. 2011), in solid lines and within 1\arcmin , in dashed lines. The data points corresponding to each of the selected MCXC clusters  (see Section \ref{sec:testmcxc} for details) are joined by lines to depict trends with spectroscopic redshift more clearly. Typical errors on photometric redshift estimates for galaxies in the SDSS database are $\approx 0.1-0.15$. It is clear that, regardless of the method, the photometric redshift estimates from SDSS correlate poorly with the spectroscopic redshift beyond $z\gtrsim0.8$, due to the poor sensitivity in SDSS, though the data are scarce in this regime. On the other hand, there appears to be a reasonable correspondence up to $z=0.6$ to within $\approx 0.15$.}
\label{fig:diffz}
\end{center}
\end{figure}

We used two methods to estimate a value for the cluster $z_{\rm{phot}}$ and applied each method to the 1.5\arcmin and 1\arcmin$\,$ results; the results for each method together with the spectroscopic cluster redshift are provided in Table  \ref{tab:mcxcz}. For the first method (`All'), we took the mean of all the $z_{\rm{phot}}$ measurements; for the second method (`histogram peak'), we took the mean of the $z_{\rm{phot}}$s for objects lying in the histogram bin with the highest counts (Figure \ref{fig:photozMCXC}). 
The difference between the cluster $z_{\rm{spec}}$ and the $z_{\rm{phot}}$-estimate for each of the methods is plotted in Figure \ref{fig:diffz} as a function of $z_{\rm{phot}}$ for the MCXC clusters in Table \ref{tab:mcxcz}. Uncertainties are not included in this plot for clarity but typical errors in the galaxy $z_{\rm{phot}}$s in SDSS are  $\approx 0.1-0.15$. 

From Figure \ref{fig:diffz}, it can be deduced that, regardless of the method employed, beyond $z\approx 0.8$, the photometric estimates do not provide a good match to the spectroscopic redshift, likely due to the poor sensitivity of SDSS in this redshift regime, although this is based on two data points only. This figure also shows that typically the SDSS $z_{\rm{photo}}$ estimates are larger than the spectroscopic values. Closer inspection of the $z_{\rm{phot}}$ distribution from Figure \ref{fig:photozMCXC} shows that, for the $z \gtrsim 0.6$ systems, the histograms flatten out over a large portion of the redshift range. On other hand, until $z\approx 0.6$ the (mean, absolute) differences between the photometric and spectroscopic redshifts is reasonably small, $\lesssim 0.1$, and the histograms tend to have narrow peaks (with typical widths of $\approx 0.3$). The exception being RXJ1728.6+7041.
We find that the second method, when searching within a radius of 1.5\arcmin, provides the photometric redshifts closest to the spectroscopic value throughout most of the sampled redshift range; with the mean, absolute difference for the whole redshift range sampled being 0.08 and 0.06 for $z<0.6$. However, as appealing as this result might be, the number of clusters in this study are scarce and thus it cannot provide conclusive evidence as to whether the sole use of SDSS photometric redshifts can be used in a widespread fashion to estimate redshifts for clusters in the intermediate-to-high $z$ regime.

\begin{table*}
\caption{Spectroscopic redshifts ($z_{\rm{spec}}$) for clusters in the MCXC catalog (Piffaretti et al. 2011) with $M_{500}\geq 2.5\times10^{14}M_{\rm{sun}}$ and $z\geq 0.5$ with SDSS coverage. Photometric redshifts from objects in the SDSS database within 1.5\arcmin and 1\arcmin of the MCXC cluster position were used to estimate a photometric redshift for the cluster ($z_{\rm{phot}}$). We estimated $z_{\rm{phot}}$ in two ways: (1) using the mean photometric redshift for all objects within the radial search (ALL); (2) using the mean photometric redshift of objects within the radial search which fall within the peak bin of the histograms in Appendix \ref{sec:photoz} (Histogram Peak)$^{(a)}$. The average uncertainty in $z_{\rm{phot}}$ in the SDSS database for the relevant objects is typically between 0.1 and 0.15. Uncertainties are not included in the table since the mean SDSS $z_{\rm{phot}}$ error would be an underestimate. The last column contains photometric redshifts obtained from the analysis of Wen et al (2012) of SDSS DR6 data, where available. The $z_{\rm{phot}}$s in bold are the ones obtained from what was found to be the most reliable method.}
 \label{tab:mcxcz}
\renewcommand{\arraystretch}{1}
\begin{tabular}{ccc ccccccccccccc}
  \hline \hline
 \multicolumn{1}{l}{MCXC} & $z_{\rm{spec}}$ & \multicolumn{4}{c}{$z_{\rm{phot}}$ SDSS-DR9}& $z_{\rm{phot}}$ SDSS-DR6\\   
\multicolumn{1}{l}{Name} &                               & \multicolumn{2}{c}{All}  & \multicolumn{2}{c}{Histogram Peak}  & (Wen et al. 2012)\\
             &                               &  (1.5\arcmin) & (1\arcmin)   & (1.5\arcmin) &  (1\arcmin) & \\                                                   
  \hline
 
\multicolumn{1}{l}{BVH20072} & 0.50 & 0.42  & 0.40  & {\bf{0.36}}  & 0.52  & NA \\  
\multicolumn{1}{l}{MACSJ0911.2+1746} & 0.50 & 0.48 & 0.49  & {\bf{0.52}} & 0.53& NA\\
   \multicolumn{1}{l}{BVH2007198} & 0.52 & 0.41 & 0.42  & {\bf{0.51}} & 0.51  & 0.4960\\ 
  \multicolumn{1}{l}{MACSJ1423.8+2404} & 0.54 & 0.45  & 0.48   & {\bf{0.51}} & 0.39  & 0.5197\\ 
  \multicolumn{1}{l}{MACSJ0018.5+1626}  &  0.55  & 0.47   & 0.47  & {\bf{0.53}}  & 0.52  & 0.5602\\
  \multicolumn{1}{l}{ MACSJ1149.5+2223} & 0.55 & 0.46  & 0.49   & {\bf{0.53}} & 0.52  & 0.5603\\ 
  \multicolumn{1}{l}{ MS1241.5+1710} & 0.55 & 0.47 & 0.49   & {\bf{0.39}}  & 0.52 & 0.5572\\ 
 \multicolumn{1}{l}{  RXCJ1728.6+7041} & 0.55 & 0.44  & 0.37  & {\bf{0.35}}  & 0.35  & NA\\ 
  \multicolumn{1}{l}{ WARPJ0942.3+8111} & 0.55 & 0.44    & 0.47& {\bf{0.54}}  & 0.55  & NA\\ 
  \multicolumn{1}{l}{ BVH2007123} & 0.56 & 0.49 & 0.52    & {\bf{0.53}} & 0.53  & 0.5839\\ 
  \multicolumn{1}{l}{MACSJ2129.4-0741} & 0.59 & 0.50    & 0.52 & {\bf{0.54}}  & 0.55  & 0.6324\\ 
  \multicolumn{1}{l}{ BVH2007122} & 0.60 & 0.50 & 0.53   & {\bf{0.53}}  & 0.68  & 0.6055\\ 
  \multicolumn{1}{l}{BVH2007173} & 0.62 & 0.44 & 0.43 & {\bf{0.66}}  & 0.23  & 0.5470\\ 
  \multicolumn{1}{l}{BVH2007149} & 0.70 & 0.55  & 0.58   & {\bf{0.66}}  & 0.65  & NA\\ 
  \multicolumn{1}{l}{MACSJ0744.9+3927} & 0.70 & 0.48   & 0.49  & {\bf{0.68}}  & 0.68  & NA\\ 
  \multicolumn{1}{l}{WARPJ1350.8+6007} & 0.80 & 0.47 & 0.54   & {\bf{0.69}}  & 0.70 & NA\\ 
  \multicolumn{1}{l}{BVH2007154} & 0.89 & 0.43  & 0.50 & {\bf{0.46}}  & 0.52  & NA\\ 
   \hline
\end{tabular}
\begin{tablenotes}
 \scriptsize
 \item (a) When calculating the photometric redshift estimate using method 2 (i.e.. from the histogram peak) for histograms with multiple peaks, we took the mean value for all the objects in all those bins.
\end{tablenotes}
\end{table*}

\setcounter{subfigure}{0}

\subsubsection{Application of the SDSS photometric Redshifts to the CARMA-detected Clusters Candidates}

Histograms for the distribution of SDSS galaxy $z_{\rm{phot}}$s within 1 and 1.5\arcmin of the CARMA cluster centroids are given in Figure \ref{fig:photozP}. Comparing the number of SDSS objects found in the peak bin of the histograms for our CARMA-detected clusters and the selected MCXC clusters, shows that the former are very likely to be very massive systems since mass is expected
to be correlated with optical richness.
For four clusters, P170, P187, P190 and P205,  the histograms shows a distinctive narrow peak; for two clusters, P014 and P097, such a feature is not as prominent; and for P086 the distribution is bimodal while for P109 it is almost uniform out to $z=0.9$.
The mean $z_{\rm{phot}}$ estimates for the three methods outlined in the previous section are provided in Table \ref{tab:meansdssz}. 
The mean $z_{\rm{phot}}$ for all CARMA-detected systems from our sample using method three with a radial search of 1.5\arcmin---the preferred method from results in Section \ref{sec:testmcxc}---is $\approx 0.5$; results from this method are highlighted in Table \ref{tab:meansdssz}. 

We expect the histograms in Figures \ref{fig:photozMCXC} and  \ref{fig:photozP} to be biased towards low-redshift objects, since the sensitivity of SDSS drops as a function of redshift and the catalog is not expected to be very complete at $z\gtrsim 0.3$ (e.g., \citealt{montero2009}). Hence, 
for clusters like P014 and P187 with the $z_{\rm{phot}}$ distribution leaning towards $z\lesssim 0.4$, it is hard to rule out completely the possibility of a chance superposition of a low and a high redshift cluster without deep spectroscopic data. Though the lack of significant peaks in the $z_{\rm{phot}}$ distribution at higher redshifts,----where `significance' is measured in terms of total number of galaxies in the `peak' bin and relative to the other bins---is a strong indication that that is the likely, approximate redshift of the cluster.

To investigate the possibility of a higher-redshift cluster being associated with our cluster candidates, we found out how many of the SDSS objects towards each system had a matching {\sc{WISE}} object within 3\arcsec (FWHM of the {\sc{WISE}} short channels $\approx 6$\arcsec). If there was a high redshift cluster that was not seen in the relatively shallow SDSS data, there should be an overdensity of {\sc{WISE}} red objects without an SDSS counterpart\footnote{{\sc{WISE}} objects with two or more blended components or flagged as extended were removed.} (see Table \ref{tab:wisesdss}; note that for this overdensity calculation, unlike for earlier ones, no background galaxy contribution has been applied, since SDSS objects have been removed on the basis that they are likely to be low-$z$ interlopers). P170 has the largest overdensity of {\sc{WISE}},  red, unmatched objects and P187 the lowest, which provides further evidence that this cluster is indeed Abell 586 at $z \sim 0.2$. It is surprising that P109, a cluster that displayed an unconstrained SDSS galaxy-$z_{\rm{phot}}$ distribution compared to P187 (Figure \ref{fig:photozP}), has an overdensity  of these objects as low as 0.29\,arcmin$^{-2}$. One possibility is that the cluster is between $0.6 \lesssim z \lesssim 1$; then it might not have many objects that satisfy the MIR color criterion nor that are detected in SDSS. Alternately,  most of the galaxies of this cluster may be below the {\sc{WISE}} detection threshold, or confused by foreground sources within the large beams of the short-wavelength channels, which is more likely to happen in $z>>1$ systems.
 Thus, based on our analysis of SDSS photometric redshifts, the WISE MIR color criterion that we initially adopted appears to overestimate the redshifts of our detected clusters; however spectroscopic confirmation
 that we have undertaken, is required for a definitive confirmation.

\begin{table}
\caption{Photometric redshift $z_{\rm{phot}}$ estimates for the {\it{Planck}}-discovered, CARMA-detected clusters in our sample; note that P351 is not included due to lack of SDSS coverage. Photometric redshifts for objects found in the SDSS database within 1.5\arcmin\ and 1\arcmin\ of the CARMA cluster position were used to estimate a photometric redshift for the cluster. We estimated $z_{\rm{phot}}$ in two ways: (1) using the mean photometric redshift for all objects within the radial search (ALL); (2) using the mean photometric redshift of objects within the radial search which fall within the peak bin of the histograms in Appendix \ref{sec:photoz} (Histogram Peak). The average uncertainty in $z_{\rm{phot}}$ for the relevant objects in the SDSS database is typically between 0.1 and 0.15. We do not quote these errors since they would be an underestimate of the true error in the $z_{\rm{phot}}$ estimate. The $z_{\rm{phot}}$s in bold are the ones obtained from what was found to be the most reliable method.}
 \label{tab:meansdssz}
\begin{center}
\begin{tabular}{c cccccccccc}
  \hline \hline
Cluster & \multicolumn{4}{c}{$z_{\rm{phot}}$}\\   
ID                                & \multicolumn{2}{c}{All}   & \multicolumn{2}{c}{Histogram peak} \\
                                            &  (1.5\arcmin) & (1\arcmin)   & (1.5\arcmin) &  (1\arcmin)\\      
  \hline
    P014 & 0.38 & 0.39  & {\bf{0.37}} & 0.36 \\ 
   P086 & 0.54 & 0.55 &  {\bf{0.66}} & 0.65 \\ 
   P097 & 0.52  & 0.51  &  {\bf{0.68}} & 0.68 \\ 
   P109 & 0.48  & 0.48  &  {\bf{0.52}}& 0.38 \\ 
   P170 & 0.48  & 0.50&  {\bf{0.53}} & 0.53 \\ 
   P187 & 0.30  & 0.31 &  {\bf{0.23}} & 0.24 \\ 
   P190 & 0.46  & 0.45 &  {\bf{0.51}} & 0.51\\ 
   P205 & 0.46  & 0.44  &  {\bf{0.40}} & 0.39 \\ 
   \hline
\end{tabular}
\end{center}
\end{table}

\begin{table}
\caption{Information on the number of objects in the (allsky)  WISE and SDSS databases within 1.5\arcmin of the CARMA cluster centroid. In the fourth column, {\sc{WISE}} objects have an SDSS match if an SDSS object is found within 3\arcsec of the {\sc{WISE}} object. In the last column red {\sc{WISE}} objects are those objects that satisfy the MIR color criterion.}
\label{tab:wisesdss}
\begin{center}
\begin{tabular}{ccccc}
  \hline \hline
Cluster & $\#$ {\sc{WISE}} &  $\#$ SDSS  &  $\#$ {\sc{WISE}}  &  $\#$ Red {\sc{WISE}}   \\
ID                   &  Objects                                            &  Objects                                    & Objects  with no& Objects with no \\
                &                       &               &  SDSS Match         & SDSS Match \\
                &                       &               &                               & (overdensity /\arcmin$^2$) \\
\hline
P014& 34 &  76 &  16 & 8 (1.13) \\
P086 & 61 &  64 & 26 & 3 (0.42)\\
P097 & 62 & 87 &  26 &  9 (1.27)\\
P109 & 50 &   48 & 26 &  2 (0.28)\\
P170 & 56 &  70 & 26 & 13 (1.84) \\
P187& 43 &  74 & 23 &  1 (0.14)\\
P190& 55 &  99 &  15 &  8 (1.13)\\
P205& 58 &  94 & 9 &  7 (0.99)\\
P351 & 44&  NA &  44 &  10(NA)\\
\hline
\end{tabular}
\end{center}
\end{table}

\subsection{Photometric Redshift Summary}

\begin{itemize}

\item {\bf{P014}}: the redshift estimate for P014 is unclear. The histogram of SDSS $z_{\rm{phot}}$ is one of the least strongly peaked from our sample, peaking at $z\approx 0.4$, and the overdensity of red WISE objects without an SDSS counterpart is the third largest in our sample ($\approx 1.1/\rm{arcmin}^2$). It does not have an associated known cluster.

\item {\bf{P086}}: The distribution of SDSS $z_{\rm{phot}}$s has a peak redshift bin around $z\approx 0.65$ that contains almost 30 galaxies while the next highest-count redshift bin is at a much lower $z$, between 0.3 and 0.45 with $\approx 10$ galaxies less.  The overdensity of red {\sc{WISE}} objects with no matching SDSS source is very small  $\approx 0.4$ galaxies/arcmin$^2$. No known cluster is associated with P086. It is therefore likely that the redshift for P086 is around $0.65$.

\item {\bf{P097}}: The majority of the SDSS $z_{\rm{phot}}$s lie in two bins ranging from 0.45 and 0.75, which amount to a large number of objects (56). The overdensity of red WISE objects without an SDSS counterpart is the second highest in the sample with $\approx 1.3$ galaxies/arcmin$^2$. No known cluster lies within the CARMA SZ contours. The significant number of SDSS sources at intermediate redshifts, suggests that this cluster has a redshift of $\approx 0.6  \pm 0.15$. We now have spectroscopic data on this cluster which is being processed.

\item {\bf{P109}}: The redshift estimate for P109 is unclear. The SDSS $z_{\rm{phot}}$ histogram for this cluster is almost uniformly distributed, with few counts per bin, always below 12. Yet the number of unmatched red objects in {\sc{WISE}} is quite small, only  $\approx 0.28$ galaxies/arcmin$^2$ as is the overdensity of red objects in WISE. This system does not have an associated known cluster.

\item {\bf{P170}}: The distribution of SDSS $z_{\rm{phot}}$s has a distinct, strong peak around $z\approx 0.53$, which is in good agreement with the measured photometric redshift of an SDSS-identified cluster candidate system, WHLJ085058.7+483003, at $z_{\rm{phot}}=0.51$ \citep{wen2012} lying within the CARMA SZ contours. Though this is the most likely redshift for P170, it does have the largest overdensity of red {\sc{WISE}} objects without a match in SDSS ($\approx 1.84$ galaxies/arcmin$^2$) and the largest red-object overdensity in WISE of the entire sample, which suggests its at $z\gtrsim 1$.

\item {\bf{P187}}: The SDSS distribution of $z_{\rm{phot}}$s has a sharp peak at $\approx 0.2$. The centroid for the known cluster Abell 586 (at $z_{\rm{spec}}=0.17$) is only $\approx 15\arcsec$ away from the CARMA SZ peak. We found only two red {\sc{WISE}} objects without an SDSS counterpart. Hence, P187 is most definitely Abell 586 and there is no evidence for a line-of-sight cluster at a higher $z$.

\item {\bf{P190}}: This system has a clear peak in the histogram of SDSS $z_{\rm{phot}}$s, peaking at $\approx 0.51$. The counts in this `peak' bin ($\approx 45$) are the largest for all our of CARMA-detected clusters. The location of our SZ contours coincide with the position of an identified SDSS cluster candidate with a  $z_{\rm{phot}}=0.49$ (WHLJ11608.5+333340; \citealt{wen2012}). The overdensity of red {\sc{WISE}} galaxies with no SDSS match is of  $\approx 1.13$ galaxies/arcmin$^2$. Hence, we expect the redshift of P190 to be $\approx 0.5$.

\item {\bf{P205}}: The majority of the SDSS $z_{\rm{phot}}$s lie in two bins ranging from 0.3 and 0.6, which amount to a large number of objects ($\approx 80$). A known SDSS cluster candidate with $z_{\rm{phot}}=0.34$ (WHLJ113808.9+275431; \citealt{wen2012}) is located within the CARMA SZ contours. However, from the relation between the cluster mass and the r-band SDSS luminosity within $r_{200}$ from \cite{wen2012}, the $z=0.34$ cluster is expected to have $M_{200}=0.6\times 10^{14}M_{\circ}$. Such a low-mass cluster would not detectable in our CARMA data. The overdensity of red {\sc{WISE}} galaxies with no SDSS match is of  $\approx 1$ galaxy/arcmin$^2$. Hence, without much evidence for a higher redshift galaxy overdensity, we expect the redshift of P205 to be $0.34$.

\item {\bf{P351}}:  We do not estimate a redshift for P351 as there is no SDSS coverage towards this system.
\end{itemize}

\subsection{Improvements to the selection strategy for detecting high redshift, {\it{Planck}}-discovered cluster candidates with CARMA}
\label{sec:detect}

\begin{figure*}
\begin{center}
\centerline{\includegraphics[height=6.0cm,clip=,angle=0.]{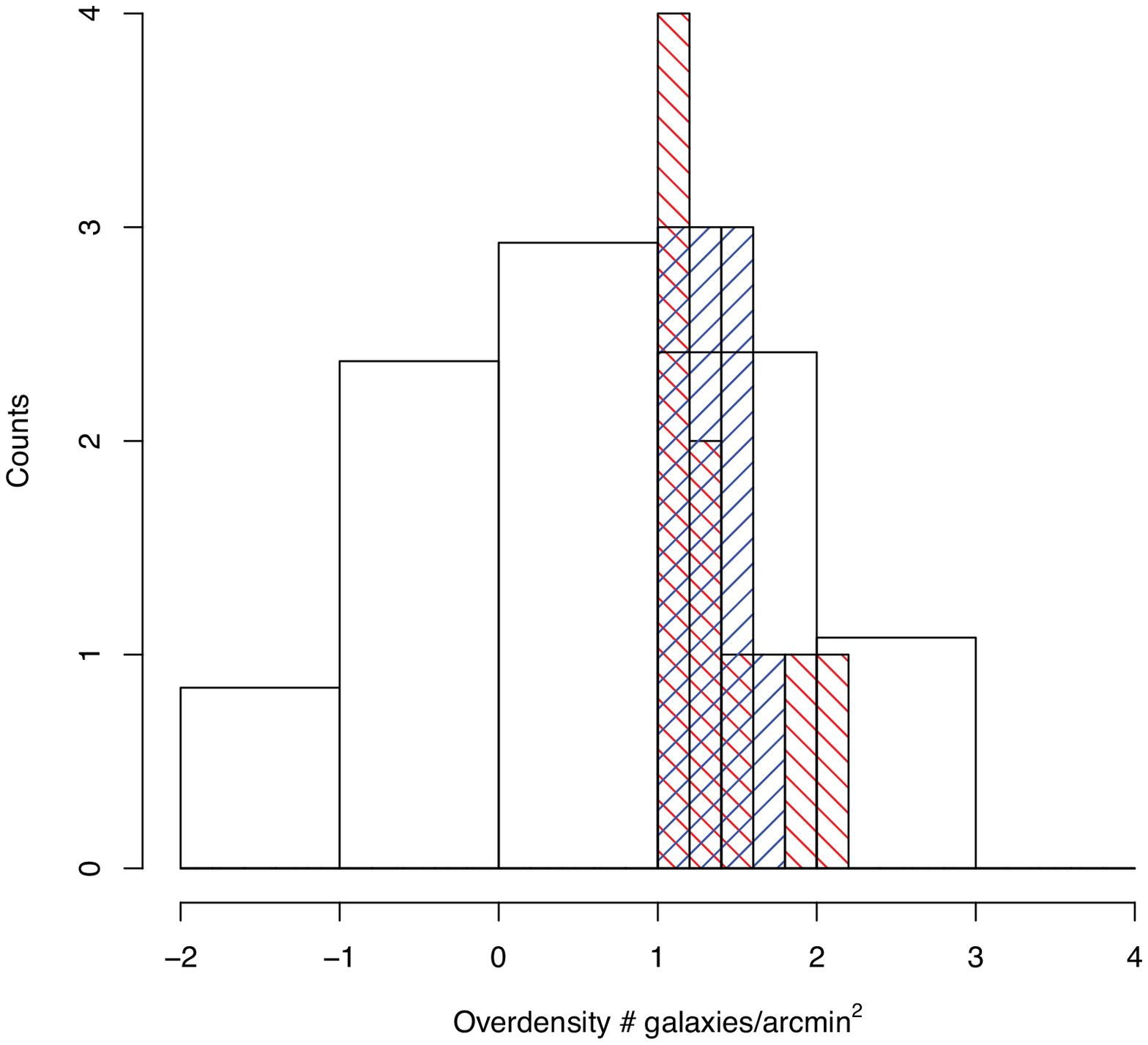}\qquad\includegraphics[height=6.0cm,clip=,angle=0.]{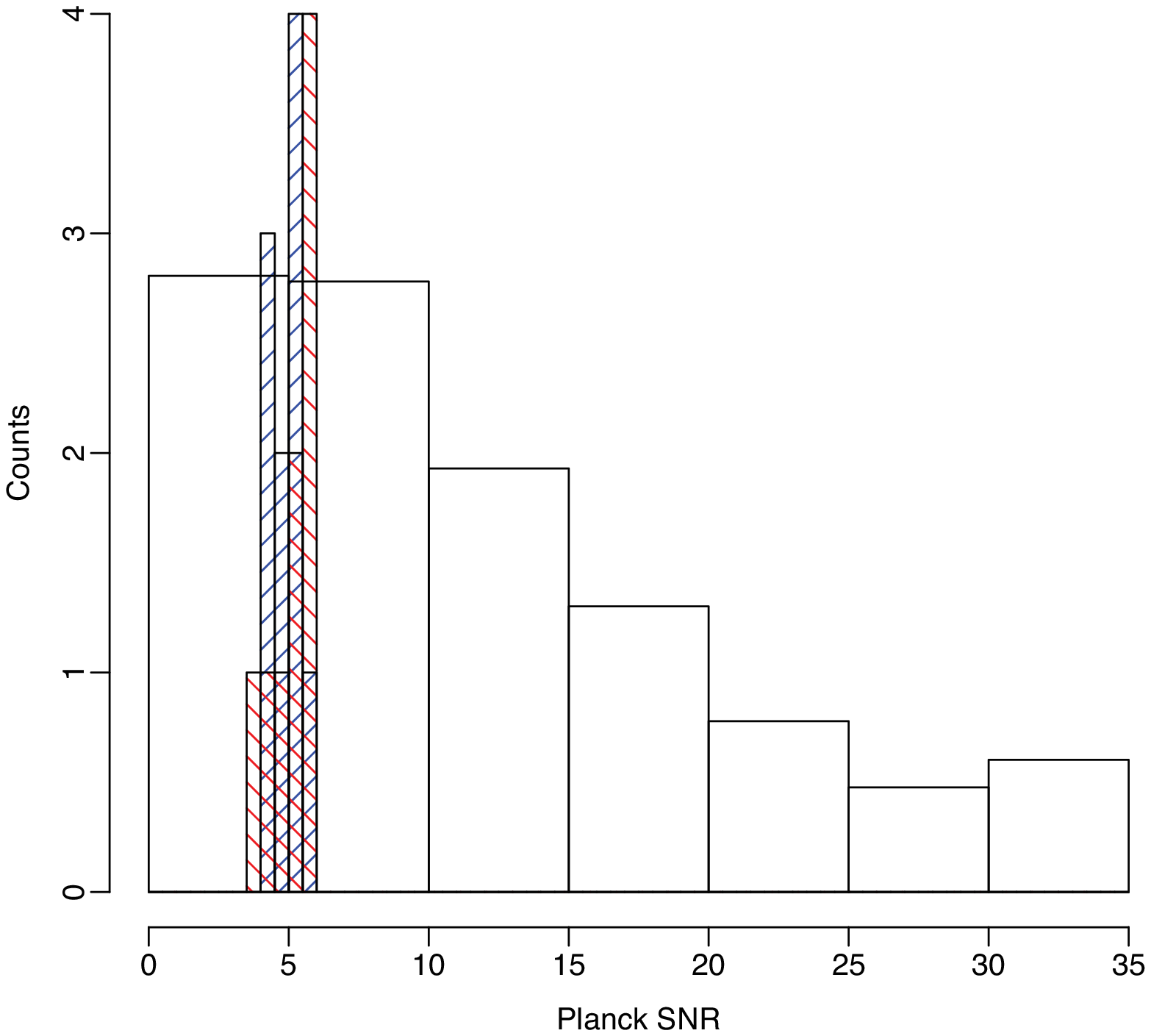}}
\caption{Left: Histogram of the WISE galaxy overdensities centered at the {\it{Planck}} cluster candidate position. Overdensities were calculated as the average density of galaxies within a radius of 4.75\arcmin minus the average density in an annulus with inner and outer radii of 4.75\arcmin and 7\arcmin. In white, plotted on a $log_{10}$ scale, are the overdensities for all {\it{Planck}} clusters. In red, diagonal lines with negative slope, plotted on a linear scale, are the 9 clusters with a clear or tentative CARMA-8 detection. While, in blue, diagonal lines with a positive slope, also on a linear scale, are the 10 cluster candidates without a CARMA-8 detection. No clear correlation can be drawn between WISE galaxy overdensity and CARMA-8 detectability for our cluster sample.  Right: Histogram of the {\it{Planck}} cluster SNRs. In white, plotted on a log10 scale, are the SNRs for all {\it{Planck}} clusters; in red, diagonal lines with negative slope (in blue, diagonal lines with a positive slope) plotted on a linear scale are shown clusters with a clear/tentative CARMA-8 detection (and without, 10 systems). Out of the 19 clusters in our sample  9 have a CARMA-8 detection and 10 do not. Clusters with an SZ detection in the CARMA-8 tend to have higher SNRs in the {\it{Planck}} data. }
\label{fig:WISEoverALL}
\end{center}
\end{figure*}

Our primary goal of this project was to attempt to identify the highest redshift, massive cluster by cross-correlating the {\it Planck} SZ-cluster catalogs and {\it WISE} galaxy catalogs.
The two main drivers in the selection of our clusters, which were found to have a large overdensity of sources in the WISE early data release, as well as  a bright red object satisfying the MIR color criterion, were:\\
\begin{itemize}
\item (1) the size of the {\sc{WISE}} overdensity
\item (2) the SNR in the {\it{Planck}} data. 
\end{itemize}
In  Figure \ref{fig:WISEoverALL}  we investigate if any of these two quantities correlate well with clusters with an SZ detection in the CARMA-8 data.
In the left panel we show the histogram for the distribution of WISE-object overdensities for all {\it{Planck}} cluster candidates, which has a mean value and standard deviation (s.d.) of 0.54 and 0.59. Overlaid are histograms of the {\sc{WISE}} overdensities for cluster candidates in our follow-up sample with and without CARMA SZ detections. The average overdensity (and s.d.) towards CARMA-8-detected systems is 0.95 (0.59), and 0.83 (0.43)  for those that were not detected. Hence, the {\sc{WISE}}-object overdensity centered at the {\it{Planck}} position, within a large (4.75\arcmin radius, see Section \ref{sec:sample}),  appears to only be a marginal way of selecting {\it{Planck}} candidates that are likely to be detected in a short ($\approx 5$ hr) CARMA-8 observation. Since the overdensity was initially calculated within a very large radius, due to the size of the {\it{Planck}} beam, but all but one of our  CARMA detections were within 2\arcmin, we re-calculated the WISE overdensity within 2\arcmin, see Table \ref{tab:wiseinfo}.  Narrowing the overdensity radius changed the over density values to 1.3$\pm$1 and 1.0$\pm$1.1 for the detected and undetected clusters. Applying the red color cut corresponding to the 
previously discussed MIR criterion to the WISE galaxies does not preferentially result in detectability (Table \ref{tab:wiseinfo}). It therefore appears that the density of WISE galaxies cannot be
used as a metric to preferentially detect Planck cluster candidates in ground-based SZ follow-up observations.

In the right panel of Figure \ref{fig:WISEoverALL} the {\it{Planck}} SNR has been plotted for all {\it{Planck}} cluster-candidates and, overlaid, are the {\it{Planck}} SNRs for clusters with a CARMA-8 detection and without. Clusters that have been detected in the CARMA-8 data tend to have the highest {\it{Planck}} SNRs, as would be expected, showing that {\it{Planck}} SNR does correlate well with detectability in our CARMA observations. 
Out of the clusters in our sample, 6 do not have a match in the Union catalog:  P031, P049, P052, P264, P097, P351. These 6 clusters have {\it{Planck}}  SNRs ranging from 3.8 to 5.1.  There are also 6 clusters: P028, P057, P090, P121, P134 and P138,  that appear in the Union catalog with SNRs between 4.6 and 5.6 but were not detected in the {\sc{CARMA}}-8 data. 
The next {\it Planck} SZ-catalog scheduled for released later in the year and based on the full mission data, will provide a more comprehensive picture of reliability as a function of SNR in the {\it Planck} catalogs. 

\section{Conclusions}
\label{sec:conclusions}

{\sc{CARMA}} 31\,GHz-data were collected towards 19 candidate clusters identified in {\it{Planck}} data, 
which were selected to have a significant ($1-2$ galaxies/\,arcmin$^2$) overdensity of objects in the {\sc{WISE}} early data release and a bright (but fainter than 15.8 mags at 3.4 microns), red ($[3.4]$ - $[4.6] > -0.1$\, AB mag) object---the putative BCG---within $\approx 2.5\arcmin\ $ from the {\it{Planck}} position. 
After removal of foreground radio sources identified in the long baseline CARMA-8 data, we detect eight clear ($\gtrsim 4.4\sigma$) SZ signals---most of which are associated with substantial extended X-ray emission in ROSAT---and one tentative signal in the short-baseline CARMA-8 data. 
The 8 clear SZ decrements are, on average, offset from the map center (and {\it{Planck}} position) by $\approx 1.2\arcmin\ $ 
and have peak primary-beam-corrected flux densities ranging from -2.1 to -4.2\,mJy; the tentative detection towards P014 is  offset by $\approx 160\arcsec$  and has a peak flux density of $-1.5$\,mJy. 

Out of the 10 {\it{Planck}} cluster-candidates without a {\sc{CARMA}} SZ signal, the likely causes for the lack of a robust detection towards 7 are contamination from radio sources and/or a high rms in
the CARMA data. For one, P028, while the radio-source environment appears to be benign at 31\,GHz, at 1.4\,GHz it does not; given the rms of the high resolution CARMA-8 data towards this system, undetected radio sources could be filling a possible decrement. The CARMA-8 and NVSS data suggest that P090 and P264 should have been detected. 
We also explore the level of 100-micron ISM contamination to the {\it{Planck}} data.
Based on this, together with results from CARMA and ROSAT, we conclude that, out of the cluster candidates without a CARMA detection, four are likely to be spurious and three are likely to be real, one of which is P090, as it has compelling, extended X-ray emission in ROSAT, comparable to that of Abell\,586 (P187), which has a measured SZ mass of $\approx 5 \times 10^{14}$\,$h^{-1}_{100}M_{\circ}$ (\citealt{carmen2012}) .

We find that, for our sample of objects, a CARMA-8 SZ detection is most likely to be obtained for candidate clusters with a higher SNR in {\it{Planck}} but not necessarily for those with a higher value of our estimate of WISE-object overdensity.

The high-resolution CARMA-8 data showed that, out of the SZ-detected systems, only three had the putative red BCG lying within the SZ contours. To shed light onto the possible photometric redshift of these systems, we produced histograms of the SDSS photometric redshifts of galaxies located within 1.5\arcmin of the CARMA SZ peak. We find that the histograms display a distinctive narrow peak for 4 systems and a less prominent peak for two more, while one system exhibited an almost uniform distribution and another a bi-modal distribution (SDSS data was missing for the remaining system).  The average SDSS photometric redshift for cluster candidates detected in CARMA was $\approx 0.5$. To test our methodology, we selected the 17 MCXC clusters with SDSS data, as well as $M_{500} \geq 2.5\times 10^{14}M_{\odot}$ and $z\geq 0.5$ (properties that we expect our sample of cluster candidates to share). We find that, out to $z\approx 0.6$, the difference in spectroscopic and photometric redshifts is $\lesssim 0.15$, for all but one cluster and that the method fails dramatically beyond $z\approx 0.8$, though the data are scarce in this regime. 

Given that the evidence suggests that most of our cluster candidates are not $z>1$ systems, we investigated the fidelity of the $[3.4]-[4.6]>-0.1$ (AB mag) color criterion for selecting such systems. We calculated the overdensity of red objects in WISE within 2\arcmin towards MCXC clusters and towards a selection of ACT and SPT-discovered systems at $z>0.5$. We found that, as expected, on average the $z<1$ clusters had smaller red-object overdensities than those at $z>1$. However, we also found that the contamination is high, such that using WISE red-object overdensities alone for the selection of $z>1$ clusters will likely result in the selection of many $z<1$ systems.

\section*{acknowledgements}

We thank the staff of the Owens Valley Radio observatory and CARMA for their outstanding support; in particular, we would like to thank John Carpenter. Support for CARMA construction was derived from the Gordon and Betty Moore Foundation, the Kenneth T. and Eileen L. Norris Foundation, the James S. McDonnell Foundation, the Associates of the California Institute of Technology, the University of Chicago, the states of California, Illinois, and Maryland, and the National Science Foundation. Ongoing CARMA development and operations are supported by the National Science Foundation under a cooperative agreement, and by the CARMA partner universities. The {\it{Planck}} results used in this study are based on observations obtained with the {\it{Planck}} satellite (http://www.esa.int/Planck), an ESA science mission with instruments and contributions directly funded by ESA Member States, NASA, and Canada. This publication makes use of data products from the Wide-field Infrared Survey Explorer, which is a joint project of the University of California, Los Angeles, and the Jet Propulsion Laboratory/California Institute of Technology, funded by the National Aeronautics and Space Administration. We acknowledge the use of NASA's SkyView facility (http://skyview.gsfc.nasa.gov) located at NASA Goddard Space Flight Center. We also wish to thank the referee for comments on this manuscript and Pedro Carvalho for useful discussions.

\begin{table}
\caption{Details on the overdensity of all (and red only) {\sc{WISE}} objects within 4.75\arcmin\ or within 2\arcmin\ of the {\it{Planck}} and CARMA SZ-decrement positions. These values were obtained using the ALLWISE data release, while the selection of our targets was made based on the WISE early data release. Our sample selection was based on the largest radius, $4.75\arcmin\ $, since only the low resolution {\it{Planck}} data was available at the time. The contribution from field galaxies was estimated within an annulus of 4.75\arcmin and 7\arcmin (see Section \ref{sec:sample}).}
 \label{tab:wiseinfo}
  \tabcolsep=0.13cm
\begin{tabular}{lcccccccc}
\hline \hline
 Cluster & \multicolumn{3}{c}{$\delta_{\rm{WISE,all}}$ ($\#$ objects/\,arcmin$^2$)} &   \multicolumn{3}{c}{$\delta_{\rm{WISE,red}}$ ($\#$ objects/\,arcmin$^2$)}\\
ID &   \multicolumn{2}{c}{At {\it{Planck}}} & At CARMA &   \multicolumn{2}{c}{At {\it{Planck}}} & At CARMA \\
 & $<4.75\arcmin$ & $< 2\arcmin$ & $< 2\arcmin$ & $ < 4.75\arcmin$ & $< 2\arcmin$ &$ < 2\arcmin $\\
 \hline
 {\bf P014} & -0.30 &  0.89 &  0.25&  -0.01&   0.44 & -0.30\\
 {\bf P086} &  0.24 &  0.25 &  1.94&  -0.45&  -0.83 & -0.69\\
 {\bf P097} &  0.87 &  1.58 &  1.51&  -0.11&  -0.54 & -0.54\\
 {\bf P109} &  0.99 &  0.97 &  0.53&   0.05&  -0.99 & -0.57\\
 {\bf P170} &  0.25 &  3.59 &  2.96&  -0.10&   0.83 &  0.51\\
 {\bf P187} &  0.95 &  1.30 &  1.86&  -0.02&   0.08 & -0.05\\
 {\bf P190} &  1.63 &  2.32 &  1.94&   0.13&  -0.11 & -0.13\\
 {\bf P205} &  1.22 &  1.87 &  2.29&   0.08&   0.70 &  0.20\\
 {\bf P351} &  1.07 &  0.71 &  0.99&   0.15&   0.18 & -0.51\\
 P028 &  0.29 &  0.35 &      &-0.37  &-0.01 &\\
 P031 &  0.89 &  2.43 &      & 0.20  & 0.21 &\\
 P049 &  1.24 &  1.47 &      &-0.33  & 1.05 &\\
 P052 &  0.74 &  0.30 &      &-0.02  & 0.38 &\\
 P057 &  1.67 &  0.74 &      & 0.01  & 0.43 &\\
 P090 &  1.44 &  0.12 &      & 0.04  &-0.62 &\\
 P121 &  0.41 &  1.31 &      & 0.10  & 0.22 &\\
 P134 &  0.68 & -0.75 &      &-0.04  & 0.74 &\\
 P138 &  0.77 &  2.33 &      &-0.40  &-0.55 &\\
 P264 &  0.95 &  2.63 &      & 0.02  & 0.39 &\\
 \hline
\end{tabular}
\end{table}

\appendix
\section{CARMA-8 SB MAPS, WISE Multi-Color Images and ROSAT Maps for {\it{Planck}} Cluster-Candidates from our Sample without a CARMA SZ Detection}
\label{sec:ndetmaps}

\begin{figure*}
\begin{center}
\centerline{\includegraphics[width=10.0cm,clip=,angle=0.]{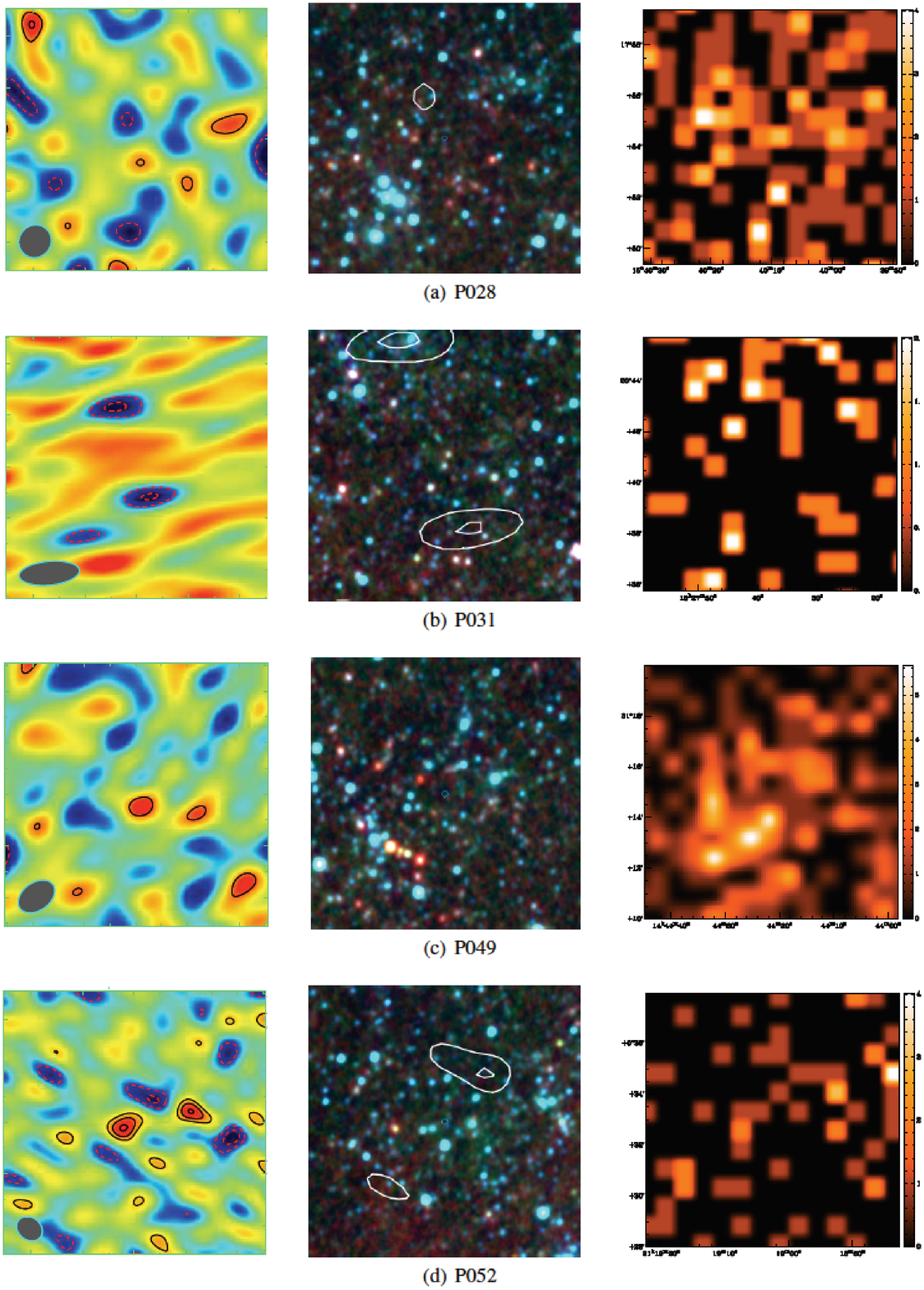}}
\caption{CARMA-8, WISE and ROSAT images for cluster candidates selected in this study without a CARMA-8 SZ detection. The left panel contains the $1000\arcsec \times 1000\arcsec$ {\sc{CLEAN}}ed CARMA-8 short-baseline maps centered at the pointing center after radio-source subtraction, where necessary. This is about twice the {\it{Planck}} beam FWHM (7-10\arcmin) relevant for SZ detection. Contours are scaled linearly starting from 2 to 10$\sigma$, where $\sigma$ is the noise on the map, in integer multiples. Positive contours are shown in solid, black lines while negative contours are shown in dashed, red lines. The grey ellipse in the bottom left corner of each map is the synthesized beam, a measure of resolution. No primary-beam correction has been applied. 
In the middle panel are the WISE multi-color $10\arcmin\times10\arcmin$ plots with the CARMA-8 negative contours overlaid. Channel 1 (3.4 microns) is shown in blue , channel 2 (4.6 microns) in green and channel 3 (12 microns) in red. The blue circles are at the pointing centre of the CARMA observations.  The right panel depicts ROSAT X-ray images smoothed by a Gaussian function with a radius of 5$\arcmin\ $. The images are centered at the CARMA pointing center, are $10\arcmin \times 10\arcmin\ $ and the pixel unit is counts. The FITS files were taken from SkyView \citep{McGlynn1998}. }
\label{fig:CombiND}
\end{center}
\end{figure*}
\clearpage
\newpage
\begin{figure*}
\begin{center}
\centerline{\includegraphics[width=10.0cm,clip=,angle=0.]{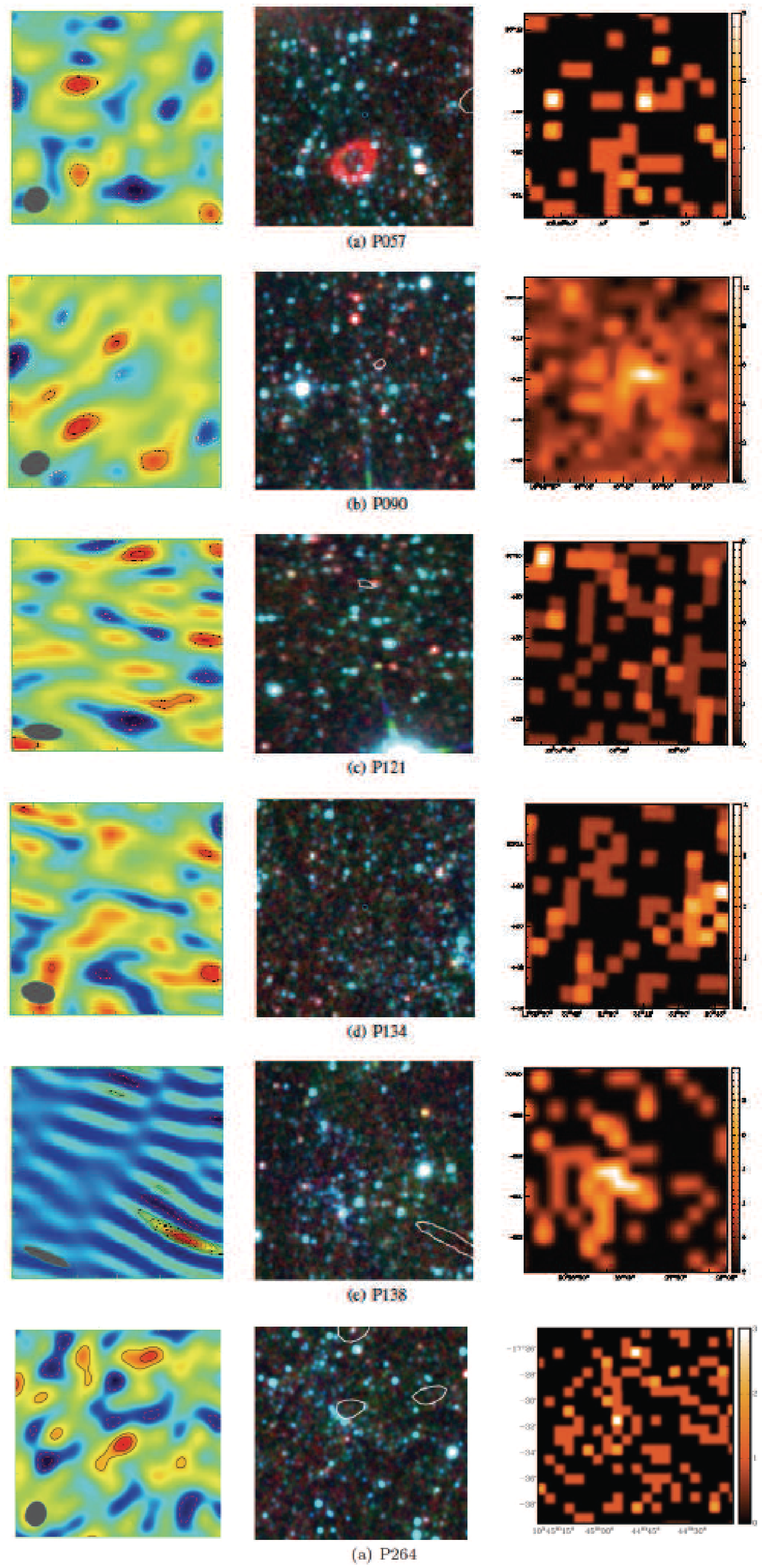}}
\end{center}
\end{figure*}
\clearpage

\newpage
\section{Notes on Individual CARMA-8-detected {\it{Planck}}  Cluster Candidates}
\label{sec:appnotes}

{\bf{P014:}}\\
Two radio sources were detected in the LB data, both of which coincide with an NVSS source. There are four more NVSS sources but they are unlikely to contaminate an SZ signal, given their location and 1.4\,GHz fitted flux densities. Despite one of the LB-detected radio sources lying $\approx 2.5\arcmin\ $ away from the map centre with a peak, PB-corrected flux density of $6.3$\,mJy, the source subtracted LB map is consistent with noise fluctuations, indicating that the removal of these two sources worked well. The SB map after source subtraction has a $4.2\sigma$ negative feature $\approx 2.7\arcmin\ $ from the map center. This is the most significant negative feature on the map and is likely to be a low-SNR SZ detection. This cluster does, however, have the least negative, primary-beam corrected flux density, $-1.5$\,mJy/Beam, which could be due to the 6.3\,mJy source located $\approx 2'$ away partially filling the decrement. The decrement is elongated in the N-S direction with respect to the synthesized beam. In the {\it{Planck}} Union catalog it has an SNR of 4.5. On the other hand, there is no X-ray cluster signal in ROSAT towards this system, which would suggest that this is indeed a high redshift object or, rather, a spurious detection.\\
{\bf{P086:}}\\
No radio sources were detected in the LB data and only two sources were identified in NVSS at $\lesssim 5$\,mJy at $\gtrsim 7$\arcmin . The cluster decrement is clearly detected at the $5.1\sigma$ level. In the inner, high SNR regions, the projected SZ signal towards this cluster appears to be close to circular. This system has an SNR of 4.6 in the {\it{Planck}} Union catalog, a peak decrement of -3.4\.mJy/beam and has substantial X-ray emission, as seen in the ROSAT image, Appendix \ref{sec:rosat}, whose peak lies within the CARMA-8 SZ contours. \\
{\bf{P097:}}\\
No radio sources were detected in the LB data. Six sources were identified in NVSS, none of which coincide with likely low SNR radio sources in the LB data. The brightest NVSS sources have peak flux densities of $\approx 16$\,mJy and are at least 9\arcmin from the map center, thus, we do not expect them to contaminate the $4.4\sigma$ {\sc{CARMA}} SZ detection. The detected SZ signal has a highly non-circular, extended shape with a peak flux density of -3.0\,mJy/beam. The ROSAT image shows X-ray emission within the CARMA-8 SZ contours with two peaks of $\approx 4$ counts each. A tentative filament of X-ray emission extends $\approx 5\arcmin\ $ South with a peak of 4 counts. P097 is not included in the Union catalog but has an SNR of 4.8 in our analysis of the {\it{Planck}} data.\\
{\bf{P109:}}\\
One radio source with a peak flux density of 2.6\,mJy was detected in the LB data, but it does not correspond to any of the six sources in the NVSS catalog. The LB-detected radio source is coincident with the SZ signal and could be affecting the 7.3$\sigma$ detection, whose peak SZ flux density is -4.2\,mJy/beam. From the maps, the cluster appears to be extended in the NE and S directions. The ROSAT image reveals the presence of X-ray emission, at the level of 3 to 4 counts, that could be associated with the CARMA-8 SZ signal. This X-ray emission extends South for $\approx 2\arcmin\ $, where it peaks with 5 counts. P109 is included in the Union catalog with an SNR of 5.3.\\
{\bf{P170:}}\\
There are four identified NVSS radio sources, one of which is located at the position of an LB-detected source. After removing the radio source, the LB map looks noise-like and a clear
$7.3\sigma$ SZ decrement with a peak flux density of -3.1\,mJy/beam is seen in the SB map. The cluster decrement is elongated with respect to the synthesized beam in the NW direction. In the ROSAT images not much X-ray emission can be seen, with a peak of 2 counts lying just outside the edge of the 2 sigma CARMA SZ contour. With a Union SNR of 6.7, this is the cluster in our sample with the highest SNR in the Union catalog.\\
{\bf{P187:}}\\
There is a high density of sources in NVSS towards this cluster candidate with 10 cataloged sources. One of these sources is located next to an LB-detected source that appears to be slightly extended in the LB data. Despite this, the source is removed well from the LB data, which are consistent with noise. A clear $5.8\sigma$ SZ detection is revealed in the {\sc{CLEAN}}ed SB maps. The decrement is one of the largest in our sample extending over $\approx$4\arcmin and branches out in several directions, primarily towards the SE, suggestive of a dynamically disturbed system. A known cluster, Abell 586, at $z=0.171$ lies $\approx 50\arcsec$ away from the map center. The peak of the X-ray emission towards Abell 586 is offset from the SZ peak in our observations by $\approx 110\arcsec$, to the SW. The X-ray image of Abell 586 is circular and compact ($\approx 1\arcmin\ $ in radius) such that there is only a small overlap between the {\sc{CARMA}} and X-ray signals. Abell 586 has been observed at arcminute resolution, at 16\,GHz with AMI (\citealt{carmen2012}), where it had a peak flux of -1.3\,mJy/beam coincident with the X-ray position and was clearly extended, with signal over $\approx 8\arcmin\ $. The AMI SZ signal is fairly circular around a radius of $\approx 2\arcmin\ $ from the SZ/X-ray peaks, where it is barely resolved, but is distinctly elongated in the SW direction at larger radii. The synthesized beam is approximately circular with a radius of $\approx 1.5\arcmin\ $. 
The higher-resolution  {\sc{CARMA}} data overlaps with the AMI signal to the East of the AMI SZ peak. The AMI signal does not show the same SE elongation in the signal in this region as the {\sc{CARMA}} data but this difference is likely to be instrumental rather than astrophysical, due to the poorer AMI resolution and the tilt in the {\sc{CARMA}} beam. What is interesting is that the SZ signal in the {\sc{CARMA}} data shows no extension in the SW direction, where a large fraction of the AMI SZ flux lies. This `missing' signal from the {\sc{CARMA}} data has a measured signal of almost -1\,mJy/beam in the AMI data; since the SZ flux is stronger at 31\,GHz than it is at 16\,GHz, we would expect to see this signal in the {\sc{CARMA}} data at $\gtrsim 3\sigma$. The AMI and {\sc{CARMA}} SZ peaks are almost equidistant from the map centre, just in a different direction, and, hence, the lack of signal in the SW region should not be due to a sensitivity issue. Most likely, the CARMA-8 data has resolved out part of the signal AMI detects on the larger scales. In the ROSAT image there is a distinctive excess of X-ray emission whose peak coincides with the CARMA SZ peak. The X-ray emission drops fast towards the SW, except for a lobe stretching out $\approx 1\arcmin\ $, but remains significant towards the SW with 5 counts even beyond the 2$\sigma$ CARMA SZ contour. P187 has an SNR of 6.1 in the {\it{Planck}} Union catalog. \\
{\bf{P190:}}\\
There are 6 NVSS radio sources within 10\arcmin of the map center, one $41$\arcsec away from the map center with a flux of 9.8\,mJy and another offset by 447\arcsec with a flux of 80.8\,mJy. No sources are seen in the LB data and, thus, no source subtraction was undertaken. A strong SZ decrement is seen in the SB data at almost $10\sigma$, with a peak flux density of -3.6\,mJy/beam. The ROSAT image shows some concentrated X-ray emission peaking at $\approx 4$ counts towards the edge of the CARMA-8 SZ contours. P190 has an SNR of 4.6 in the {\it{Planck}} Union catalog.\\
{\bf{P205:}}\\
The SZ decrement is clearly detected towards P205 at $6.8\sigma$ significance in the {\sc{CLEAN}}ed SB maps. The radio-source environment is not expected to pose a problem to the measured SZ signal: no sources were detected in the LB data and the 10 NVSS sources are not very bright $\lesssim 10$\,mJy and mostly $\gtrsim 5\arcmin\ $ away from the map center. The SZ signal is extended in the NS direction. X-ray counts at the 2 photon level can be seen in the ROSAT image covering the CARMA-8 SZ contours with a peak of 3 counts towards the edge of the SZ contours. In the {\it{Planck}} Union catalog this system has an SNR of 5.7. \\
{\bf{P351:}}\\
One radio source was detected in the LB data $\approx 4\arcmin\ $ away from the map center with a flux of $3.3$\,mJy and thus should not be affecting the cluster decrement. The subtraction of this source worked well leaving a noise-like LB map. The LB-detected source was associated with a 51.7\,mJy NVSS source. The remaining 7 NVSS sources do not lie on significant positive-flux features in the LB data. The brightest of these sources had a flux of 127.9\,mJy and was located 10\arcmin away from the map center. The LB peak flux density at this location was of $<1$\,mJy. The peak flux density for this SZ signal was one of the smaller ones for the sample $-2.1$\,mJy/beam but the detection is significant at $5.6\sigma$. Some weak X-ray emission coincident with the CARMA SZ signal is seen in the ROSAT image; the X-ray peak, at $\approx 3$ counts, lies at the edge of the 2$\sigma$ SZ contour. P351 is not included in the Union catalog and has a low SNR (3.8$\sigma$) in the {\it{Planck}} data.\\

\section{Cataloged Clusters and Galaxy Overdensities in the Vicinity of our Observations}
\label{sec:lit}
\begin{onecolumn}
\begin{table}
\scriptsize
\caption{Registered cluster and cluster candidates that lie within the 4\arcmin of our CARMA-8 map centers. X-ray mass estimates within $r_{500}$---the radius at which the mean gas density is 500 times the critical density---($M_{\rm{X},500}$) have been taken from the MCXC catalog (Piffaretti et al. 2011). Distances are given with respect to the pointing centre of the CARMA-8 observations, unless otherwise stated. For details on the CARMA-8 data towards these clusters see Section \ref{sec:discussion}. $\rm{N}_{\rm{gals}}$, the richness estimator, for the WHL clusters is calculated in \protect\cite{wen2012}  within $r_{200}^{(\dag)}$ and for the other clusters in  \protect\cite{gal2003} within a radius of $1h^{-1}$\,Mpc at the cluster redshift. Redshifts are spectroscopic, unless labelled (phot), in which case they are photometric. Cluster candidates detected in the CARMA-8 data have their cluster ID highlighted in bold font.}
 \label{tab:clustersfov}
\begin{tabular}{lccccccc}
\hline \hline
Cluster ID & Known Cluster & Distance  & $z$ & $\rm{N}_{\rm{gals}}$ & $M_{\rm{X},500}$ & Comments\\
  & & (\arcsec) & & & $\times 10^{14}M_{\sun}$ & \\ \hline

P028 & Abell 2108$^{(a)}$& 105 & 0.09 & & 1.919 &  Low mass and $z$---some SZ flux could be resolved out\\
P028 & NSC J154002+175240$^{(b)}$ & 150 & 0.0789& 49.2 &  & Low counts---likely to be a low-mass system \\
            &                                           &               &     &         &   & Low $z$---some SZ flux could be resolved out\\
P028 & WHLJ154000.1+175609$^{(c)}$ & 177 & 0.39 (phot) & 52 &  & Low counts---likely to be a low-mass system   \\
         & & & & & & \\
  P049 & A1961$^{(a)}$ & 132 & 0.23 &  & 3.532 & Ought to be seen yet there is\\
             & WHLJ144431.8+311336$^{(c)}$ & 155 & 0.23 (phot) & 103 &  & no associated decrement in the C8 data \\
P049 & NSC J144432+311149$^{(b)}$ & 237 & 0.2334& 66.5 &  & Substantial primary-beam attenutation\\
         & & & & & & \\
P052 & WHLJ211849.1+003337$^{(c)}$ & 204 & 0.28 (phot) & 76 &  &\\
           &               & & & &  & \\
P057 & Abell 2131$^{(a)}$ & 240 &  &  & 69 & Substantial primary-beam attenutation.\\
            &                                 &                &   & & &Likely to be low mass system \\
            &                                 &                 &   & & &        and no X-ray detection.\\
P057 & WHLJ154833.7+360536$^{(c)}$ & 138 & 0.24 (phot) & 79 &  &\\
           &               & & & &  & \\
P121 & WHLJ130331.7+672638$^{(c)}$ & 60 & 0.21 (phot)  & 86 &  &\\
           &               & & & &  & \\
P134 & WHLJ115049.1+621948$^{(c)}$ & 237 & 0.35 (phot) & 73 &  & Substantial primary-beam attenutation \\
           & & & & & & \\ 
{\bf{P014}} & WHLJ160319.0+031645$^{(c)}$ & 153 & 0.22 (phot) &114 & & Not associated with a CARMA decrement. \\
& & & & & & Detected CARMA SZ signal 107.448\arcsec away.\\
& & & & & &\\
{\bf{P086}} & WHLJ151351.9+524960$^{(c)}$ & 199 & 0.68 (phot) &  11 & & Not coincident with the CARMA SZ decrement (152\arcsec away).\\
         & & & & & & likely to be low mass\\
           &               & & & &  & \\
{\bf{P097}}  & WHLJ145526.9+585030$^{(c)}$ & 124 & 0.33 (phot)  &  22 &  & Not coincident with the CARMA  SZ decrement (111\arcsec away).\\
           & & & & & & likely to be low mass\\
           & & & & & &\\
{\bf{P170}}  & WHLJ085058.7+483003$^{(c)}$ &  66 & 0.51 (phot) & 33 &  & Coincident with  the CARMA SZ detection (34\arcsec away)\\
          & & & & &  & \\
{\bf{P187}} & Abell 586$^{(a)}$ & 50 & 0.171 & &  5.197  & Coincident with CARMA SZ detection (15\arcsec away) \\ 
  & WHLJJ073220.3+313801$^{(c)}$ &48 & 0.18 (phot) & 145 &  &   \\  
{\bf{P190}} & WHLJ110608.5+333340$^{(c)}$ & 55 & 0.49 (phot) & 73 &  &  Coincident with the SZ CARMA detection (21\arcsec away )\\
          & & & & & & \\
{\bf{P205}} & WHLJ113808.9+275431$^{(c)}$ & 69 & 0.34 (phot) & 12 &   & Coincident with the SZ CARMA detection (15\arcsec away)\\ \hline

 \end{tabular}
  \begin{tablenotes}
 \small
 \item (a) Reference: \cite{abell1989}.
 \item (b) Reference: \cite{gal2003}. 
 \item (c) Reference: \cite{wen2012}.
 \item ($\dag$) Typical $r_{200}$ in Wen et al. (2012) correspond to scales $\approx 4-8\arcmin$.
  \end{tablenotes}
\end{table}
\end{onecolumn}

\section{WISE multi-color plots}
\label{app:multicol}
\begin{figure}
\begin{center}
\centerline{\includegraphics[width=12.0cm,clip=,angle=0.]{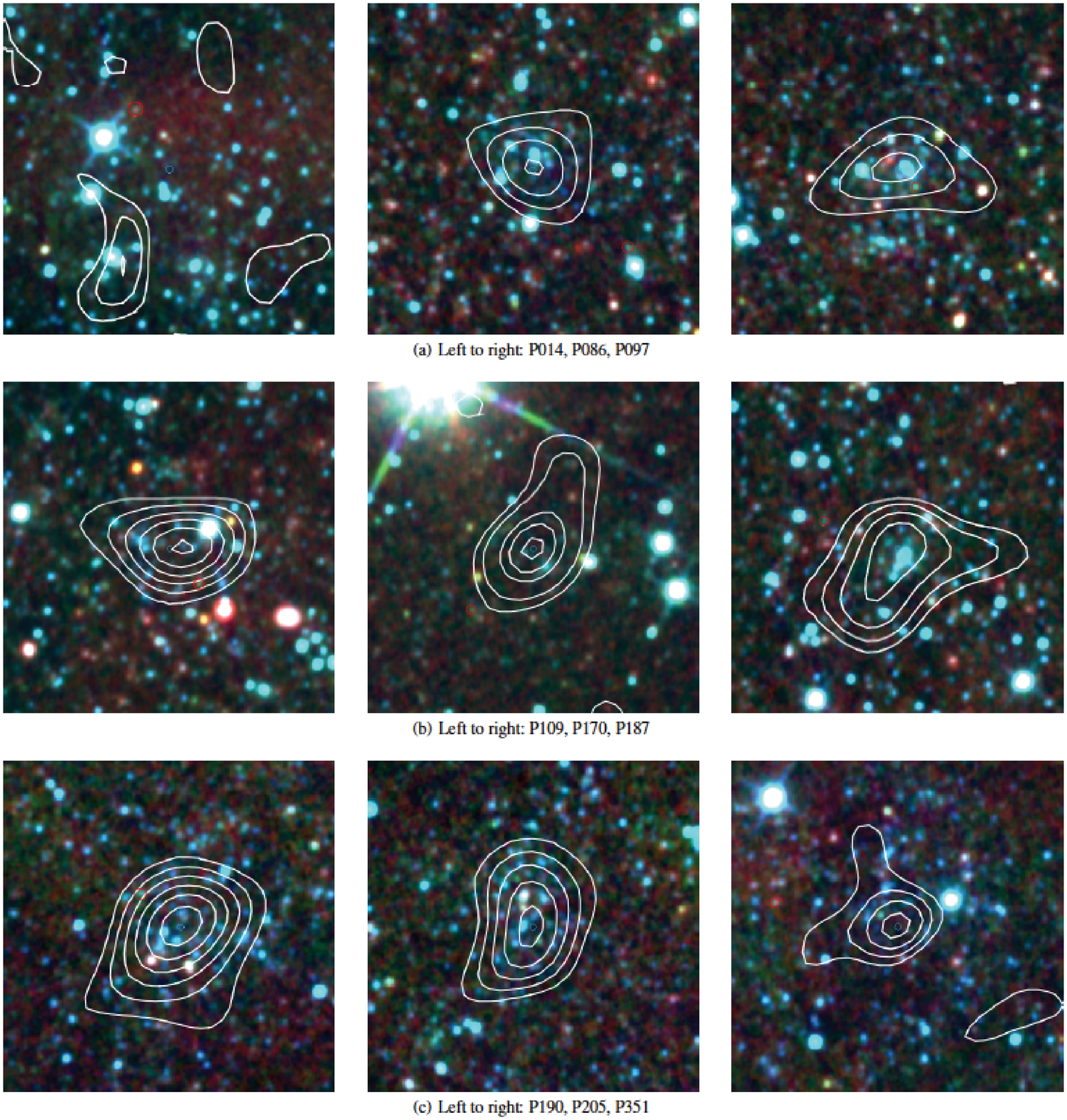}}
\caption{WISE multi-color plots with the CARMA-8 negative contours overlaid. W1 [3.4 microns] is shown in blue , W2 [4.6 microns] in green and W3 [12 microns] in red. The red circle indicates the position of the putative (red) BCG and the blue circle is at the pointing centre of the CARMA observations.}
\label{fig:wiseccimdet}
\end{center}
\end{figure}
\newpage

\section{ROSAT X-ray images}
\label{sec:rosat}
\begin{figure*}
\begin{center}
\centerline{\includegraphics[width=12.0cm,clip=,angle=0.]{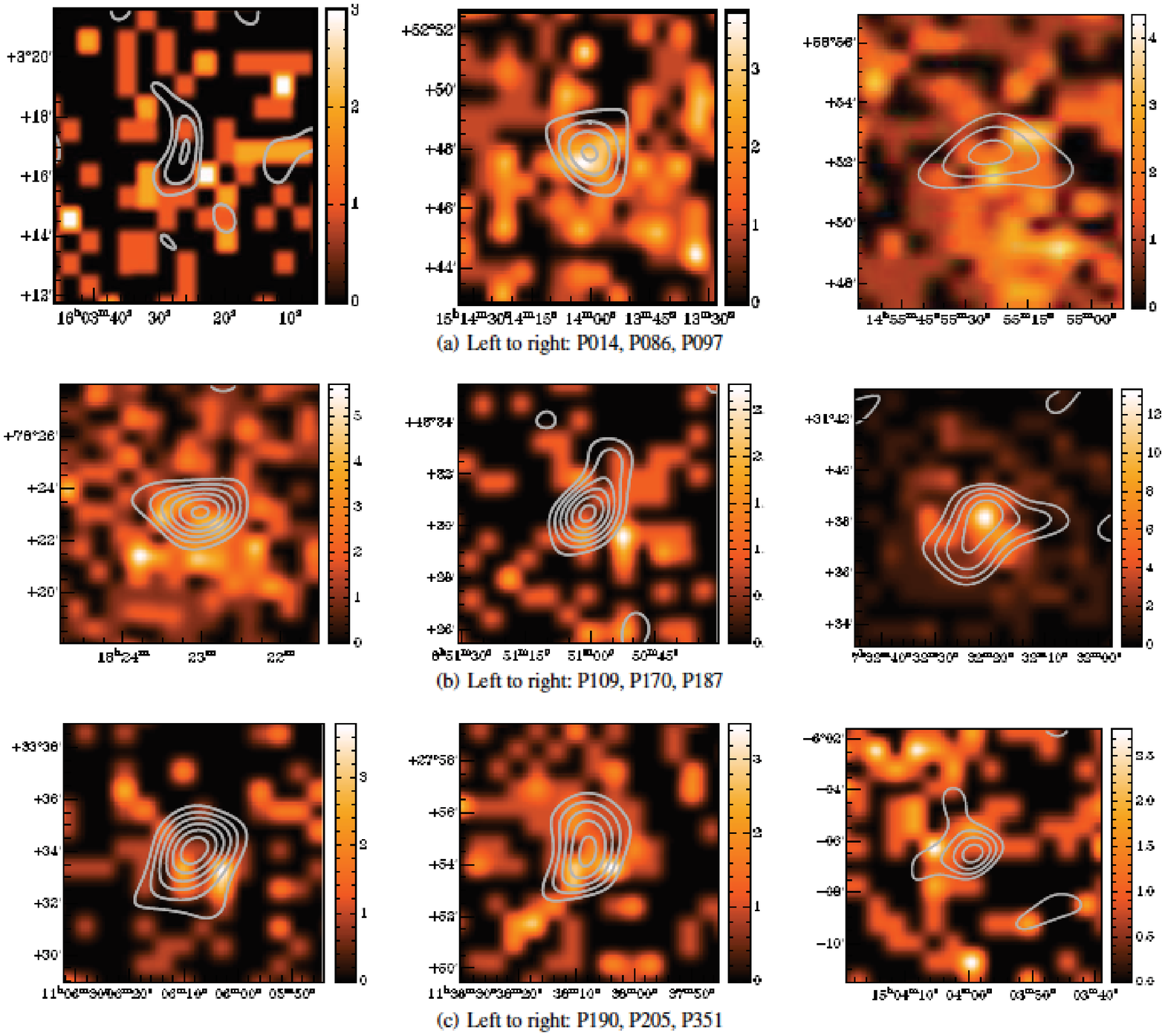}}
\caption{ROSAT images for CARMA-detections smoothed by a Gaussian function with a FWHM of 5$\arcmin$. The images are centered at the position of the CARMA SZ decrement and are $10\arcmin \times 10\arcmin\ $. The fit files were taken from SkyView \citep{McGlynn1998}.}
\label{fig:rosatmapsdet}
\end{center}
\end{figure*}
\newpage

\section{Wise Color-magnitude Plots}
\label{sec:wisecc}
 \setcounter{subfigure}{0}
\begin{figure*}
\begin{center}
\setlength{\tabcolsep}{2mm}
\subfigure[Left to right:  P014, P086, P097]{
\centering
\includegraphics[width=5.0cm,clip=true,trim= 10 10 10 10 ,angle=0.]{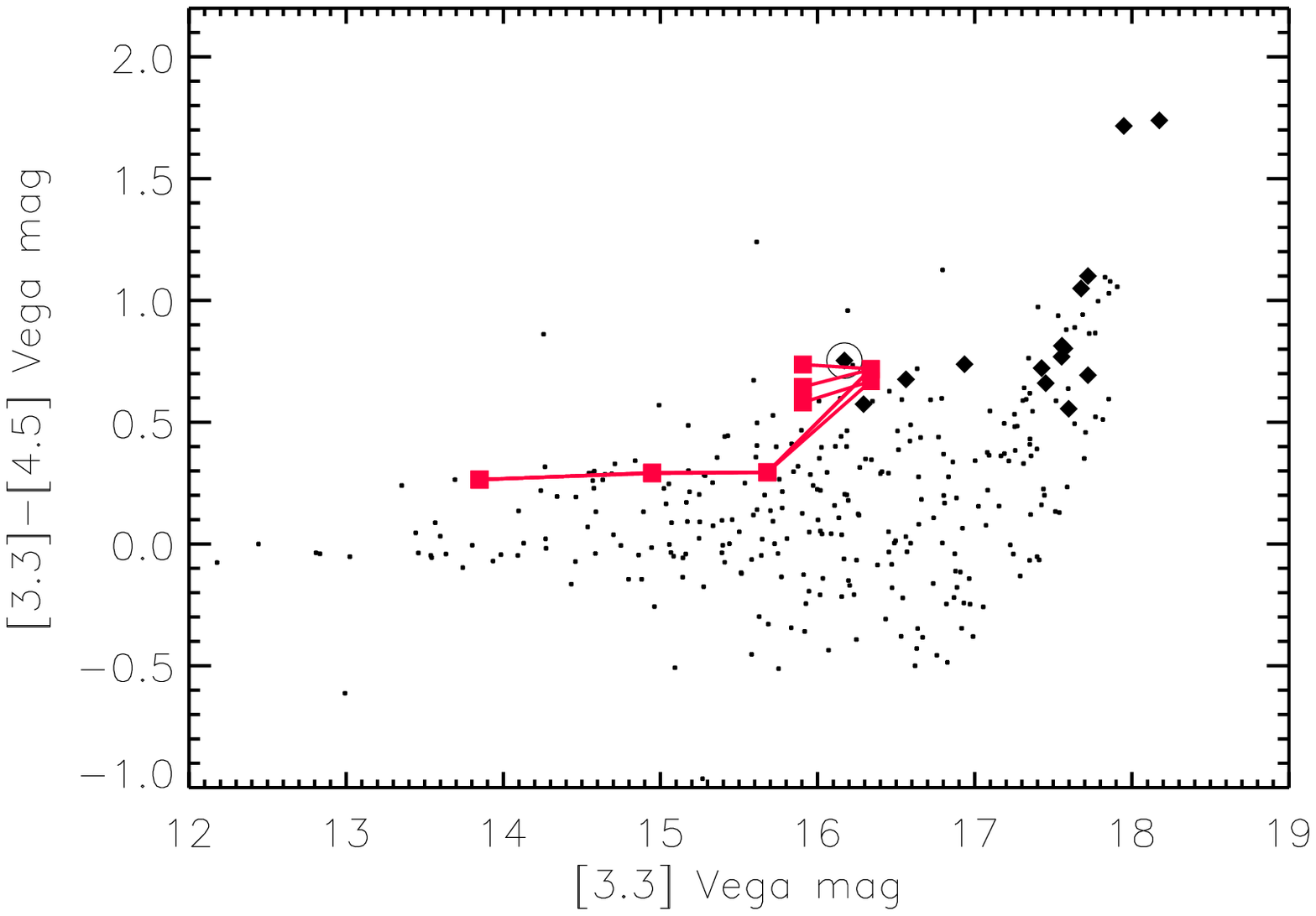}\qquad\includegraphics[width=5.0cm,clip=true,trim= 10 10 10 10,angle=0.]{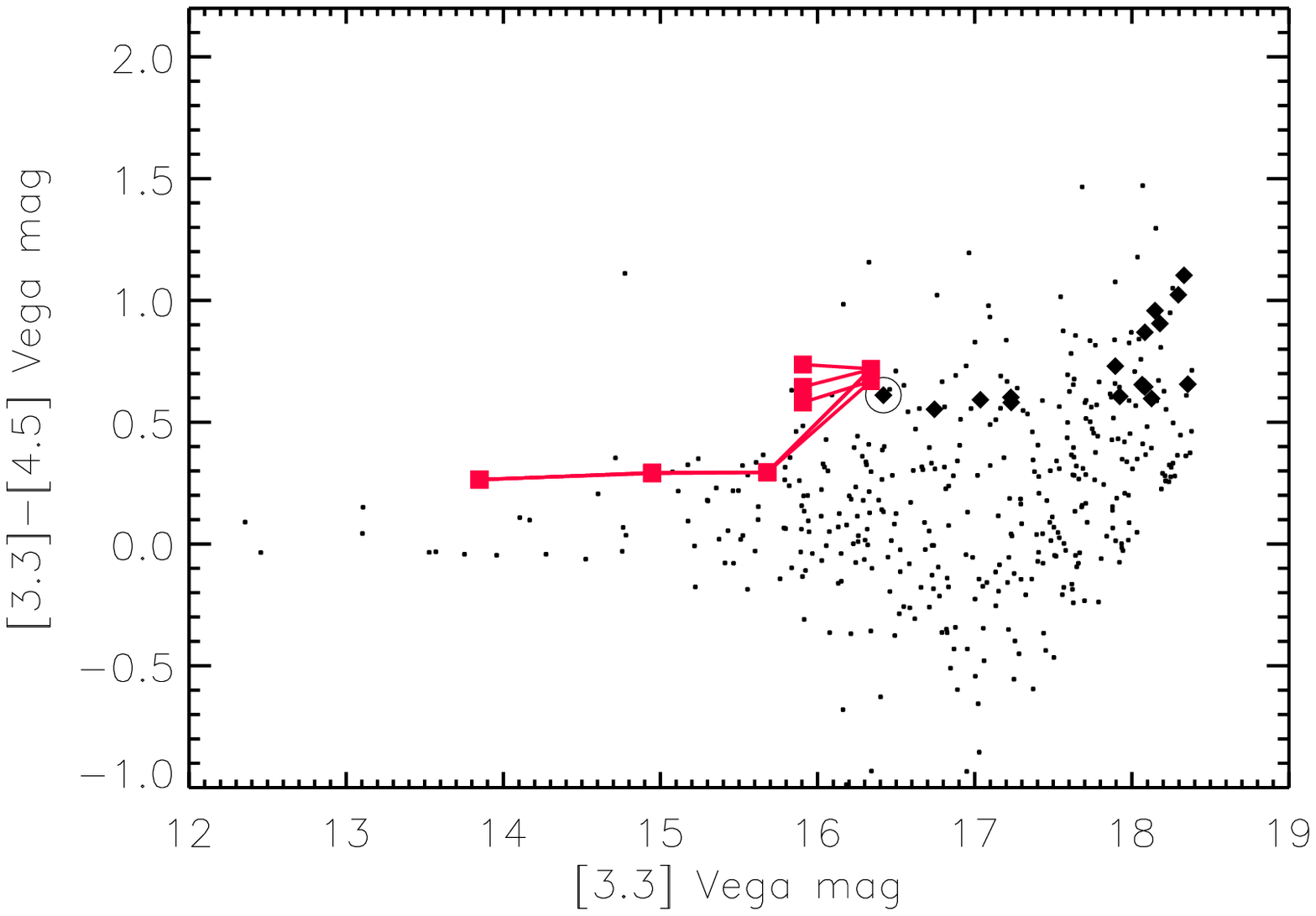}\qquad\includegraphics[width=5.0cm,clip=true,trim= 10 10 10 10,angle=0.]{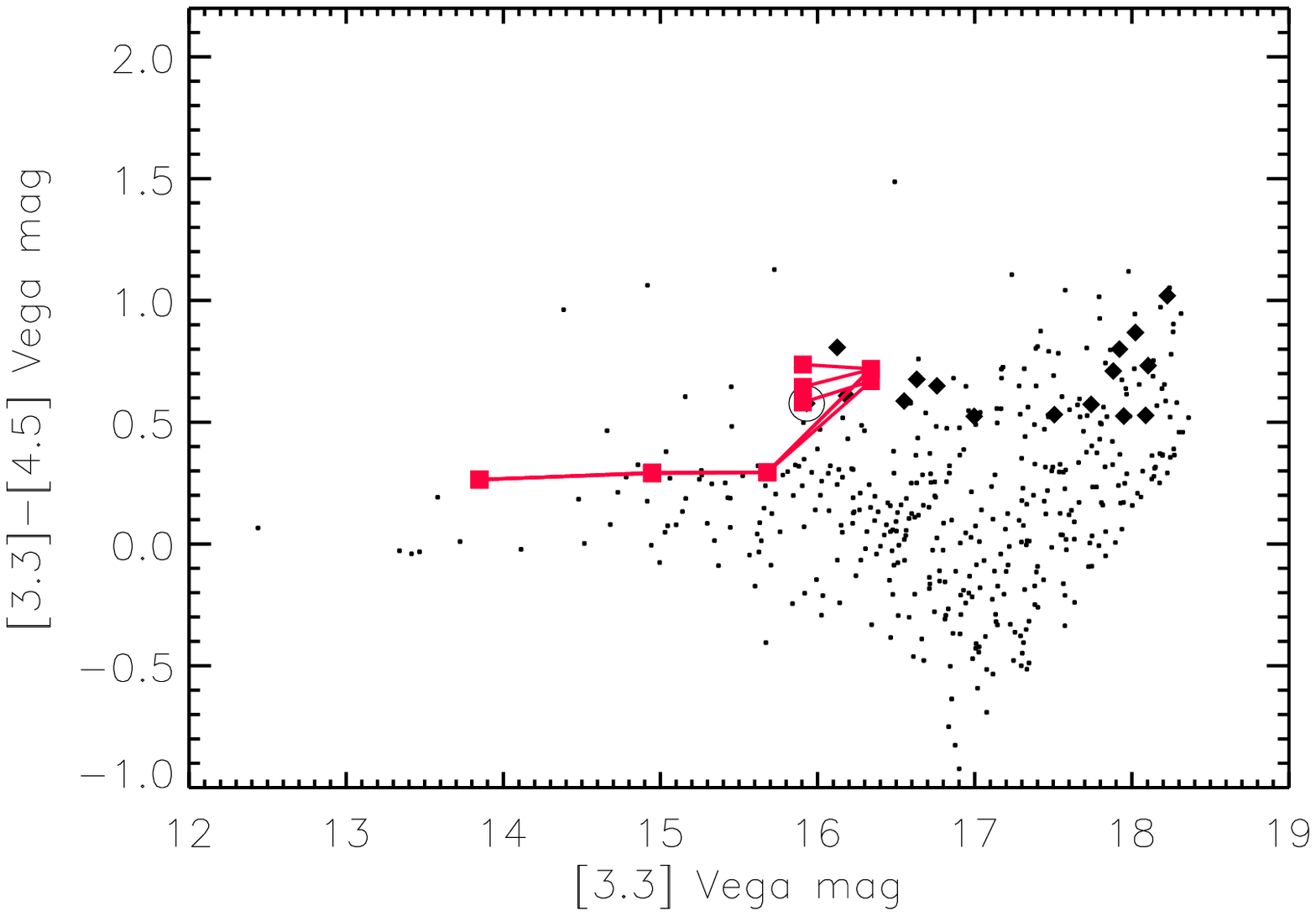}}\\
\subfigure[Left to right:   P109, P170, P187]{
\centering
\includegraphics[width=5.0cm,clip=true,trim= 10 10 10 10 ,angle=0.]{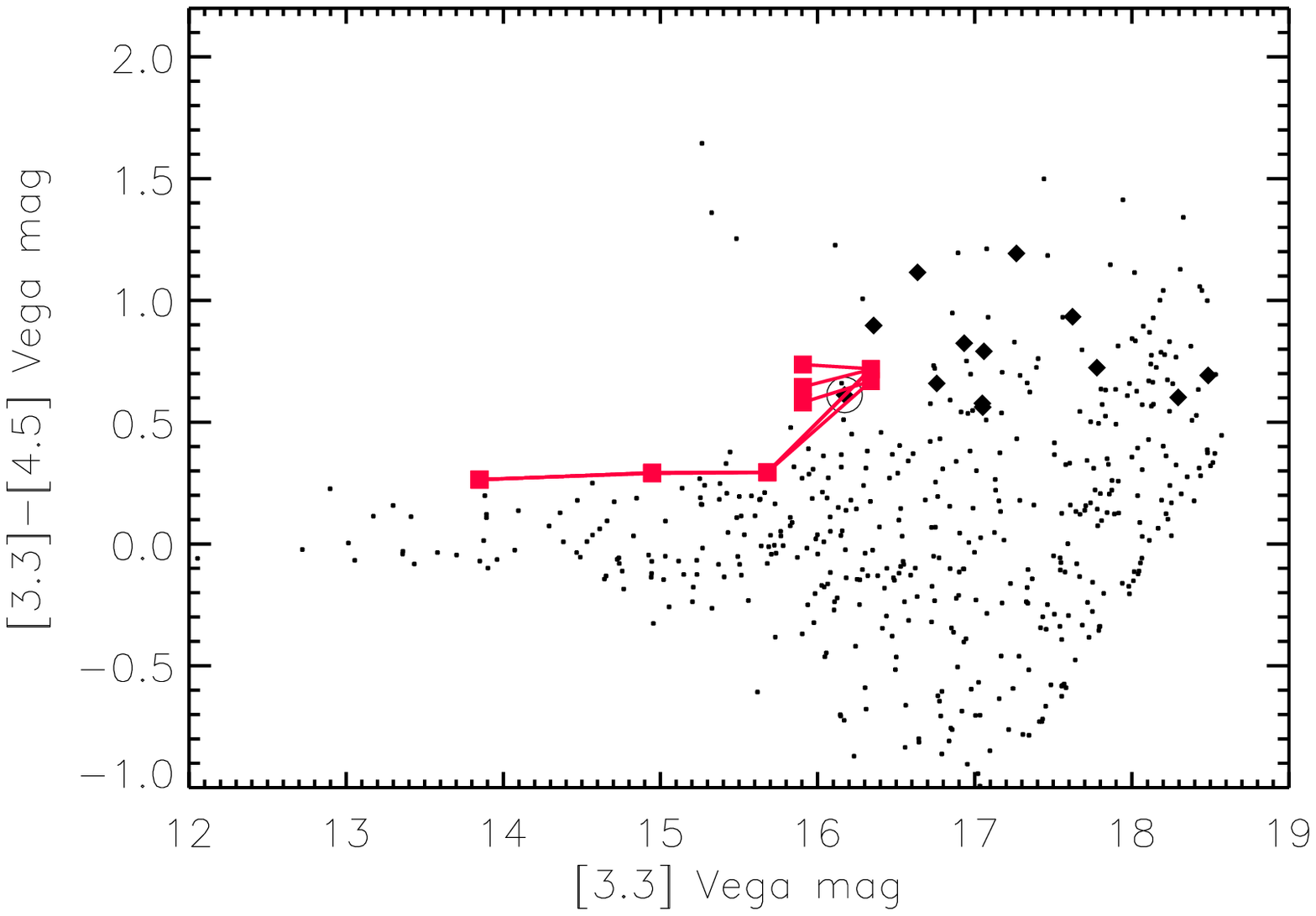}\qquad\includegraphics[width=5.0cm,clip=true,trim= 10 10 10 10,angle=0.]{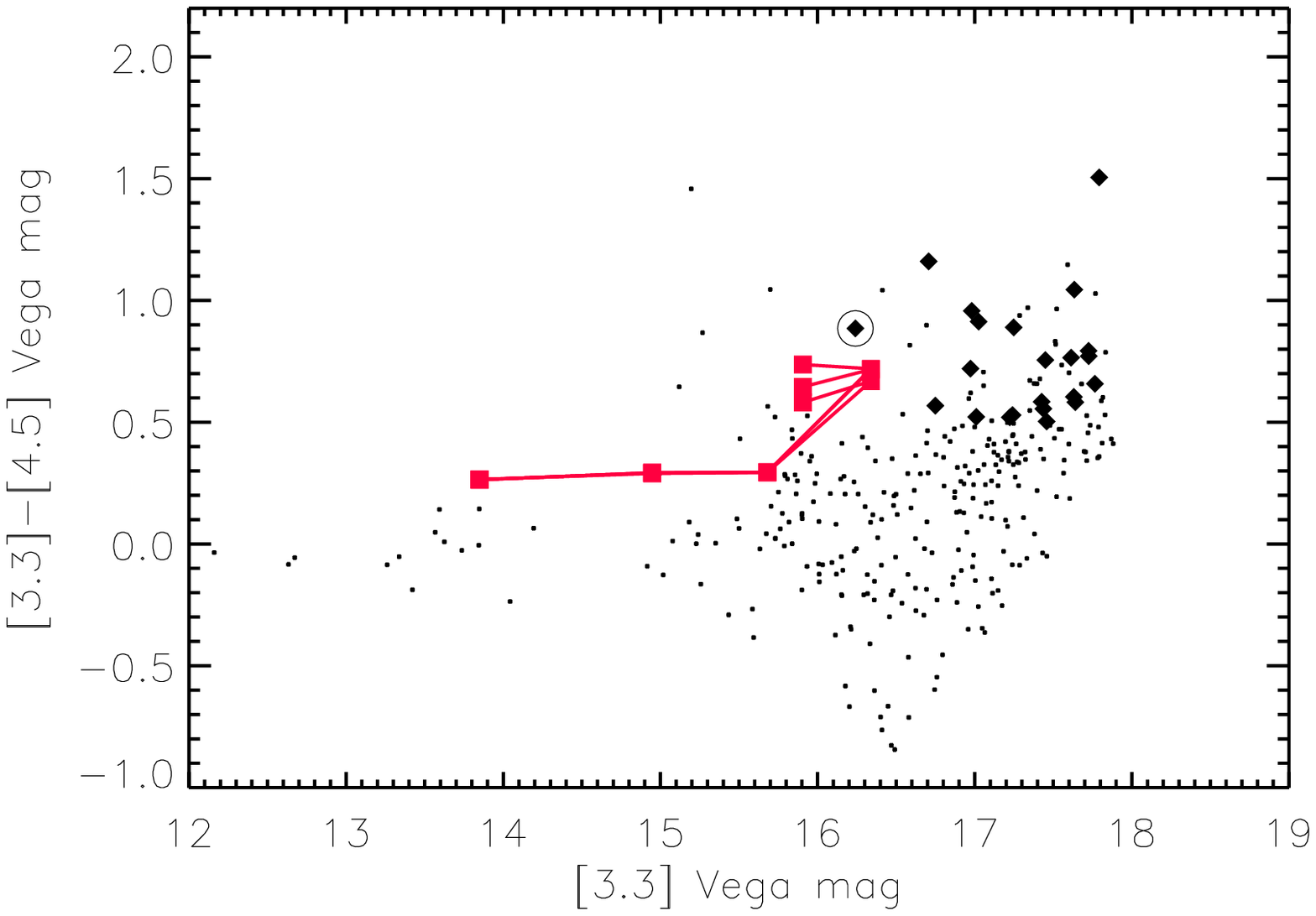}\qquad\includegraphics[width=5.0cm,clip=true,trim=10 10 10 10,angle=0.]{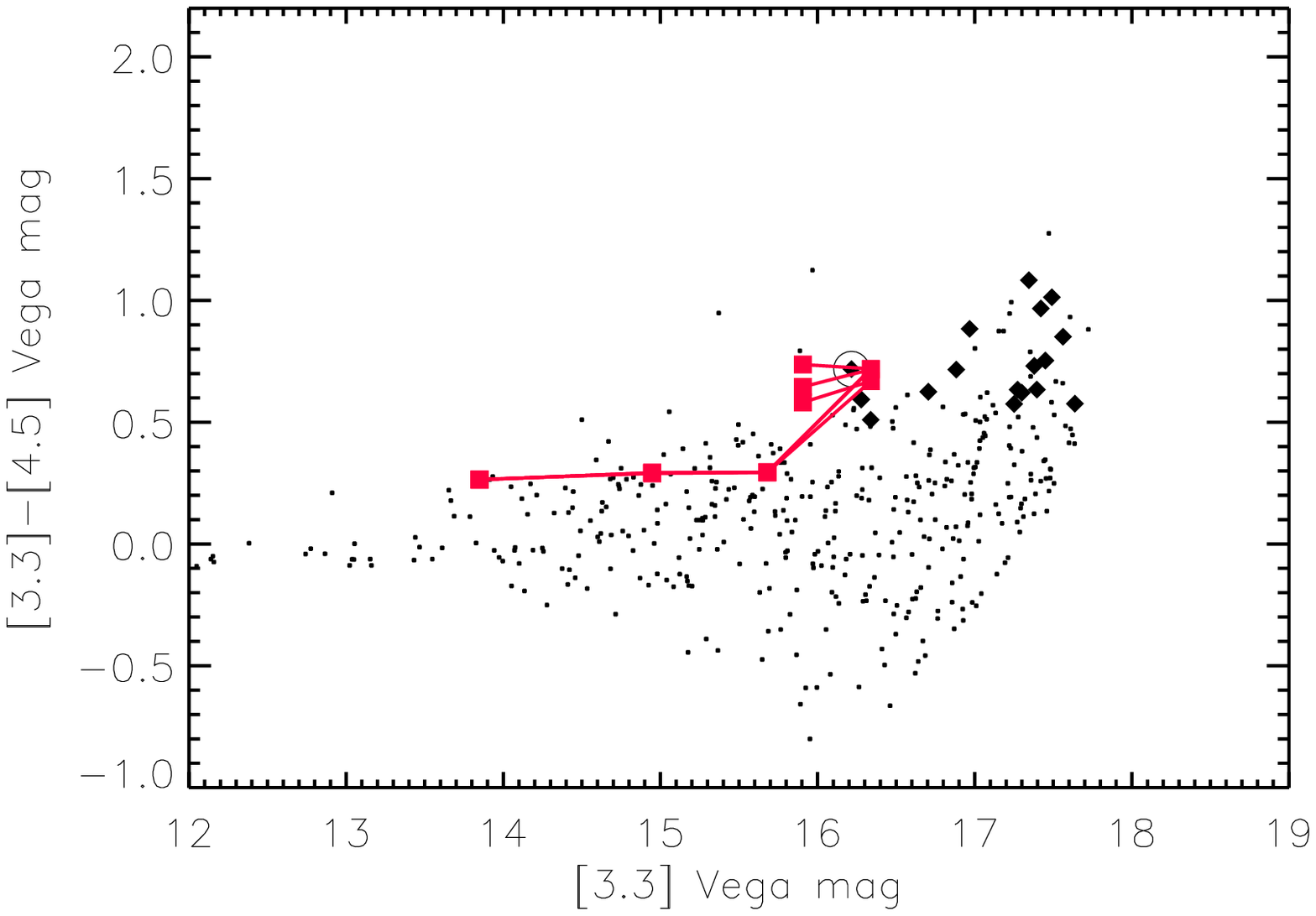}}\\
\subfigure[Left to right:   P190, P205, P351]{
\centering
\includegraphics[width=5.0cm,clip=true,trim= 10 10 10 10 ,angle=0.]{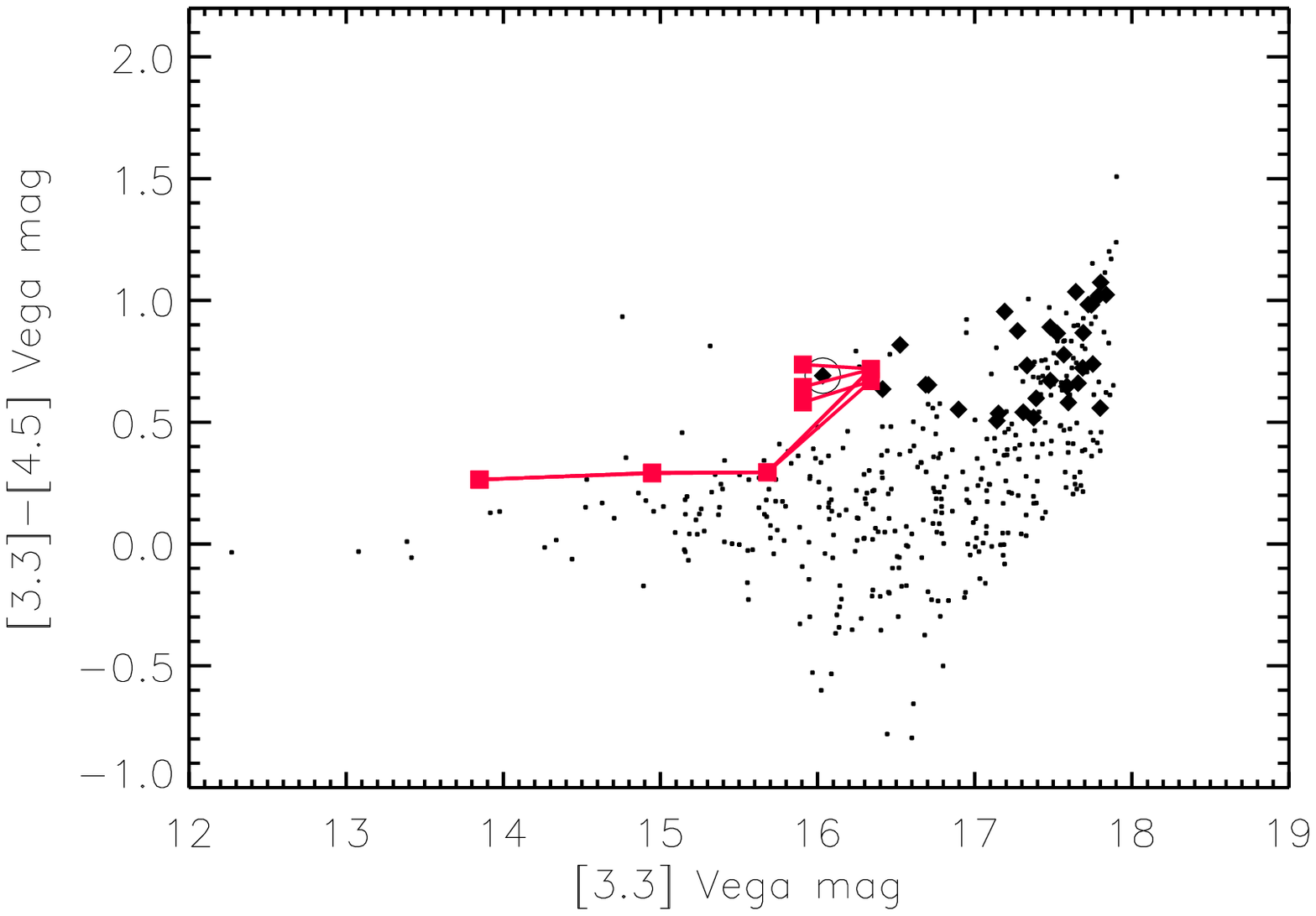}\qquad\includegraphics[width=5.0cm,clip=true,trim= 10 10 10 10,angle=0.]{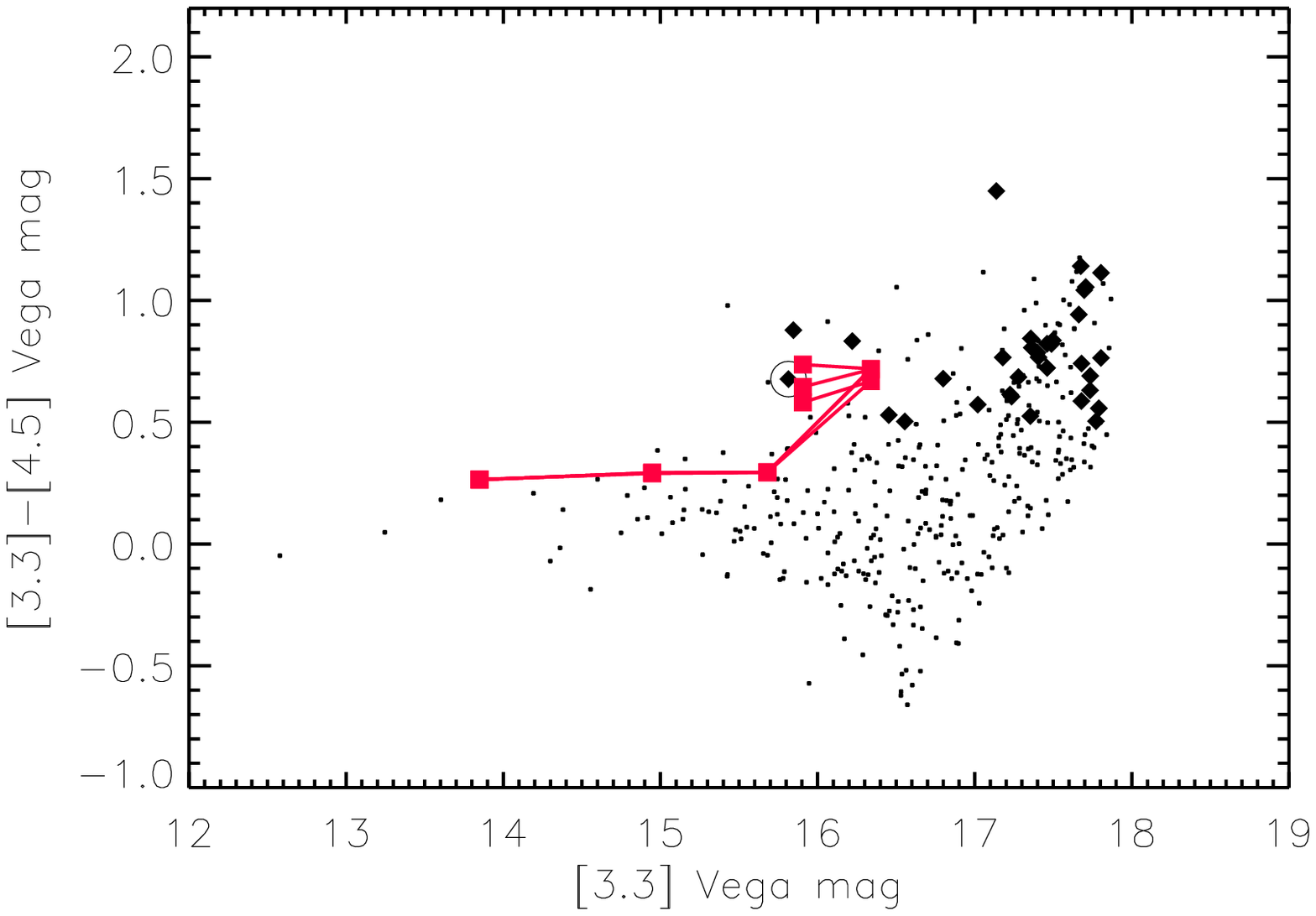}\qquad\includegraphics[width=5.0cm,clip=true,trim=10 10 10 10,angle=0.]{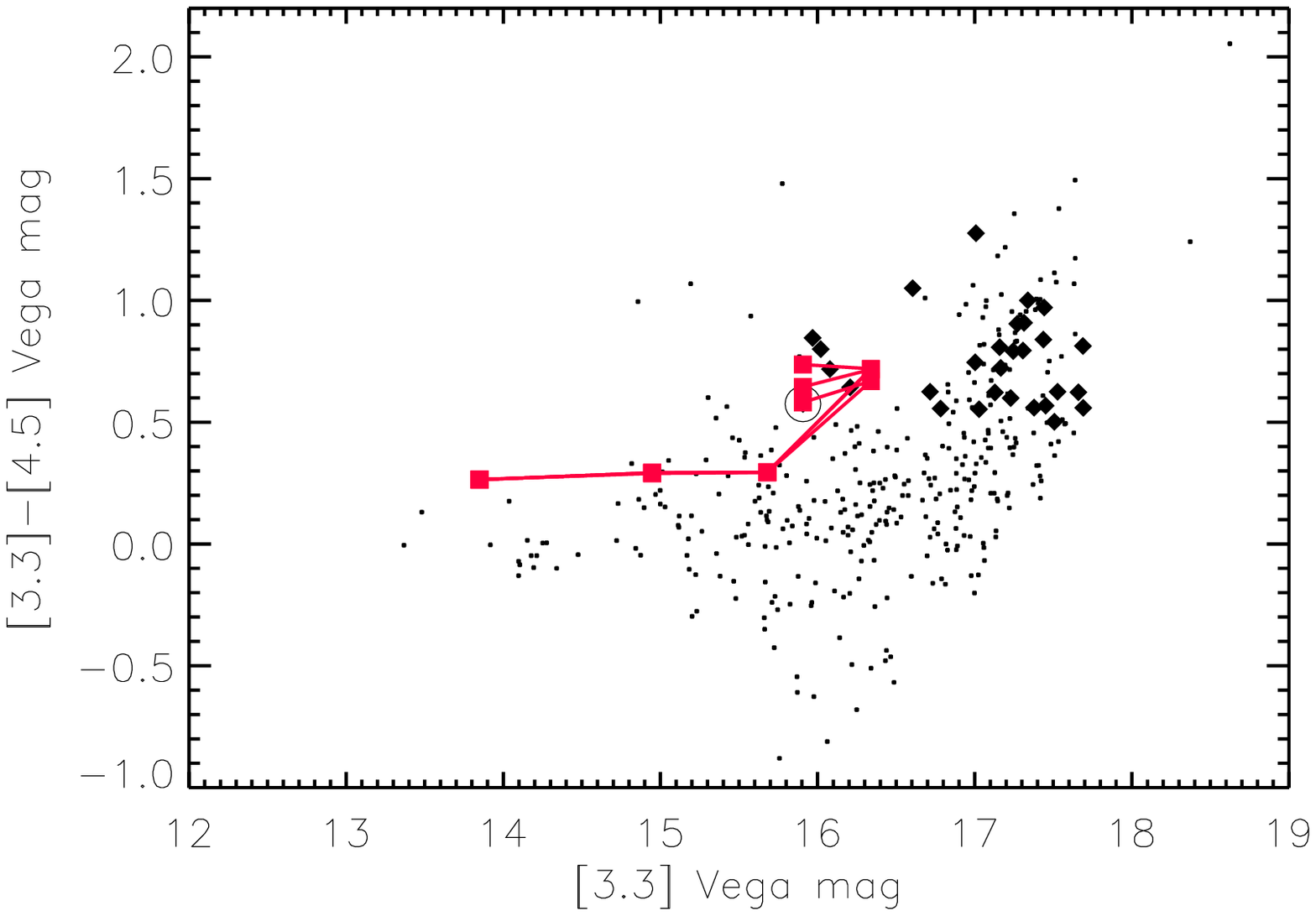}}\\
\caption{ The color-magnitude distribution of WISE sources
at the location of the {\sc{CARMA}} SZ-detected galaxy clusters. The dots
show the color-magnitude relation of the objects corresponding to the
overdensity i.e. within 4.75\arcmin of the SZ centroid. The solid black symbols show
the red ($[3.4]-[4.6] > -0.1$ AB mag) objects which are within 2.5\arcmin of the CARMA SZ centroid.
The circled object is the brightest galaxy among the red objects and the
putative bright cluster galaxy. Also shown are tracks for a passively
evolving starburst of mass $1\times 10^{12}$ M$_{\sun}$, and an e-folding timescale of
star-formation
of 100 Myr with the square symbols marked at redshifts of 3, 2, 1, 0.5 and 0.25.
The formation redshift of the burst was 3, 5 or 7. Between redshifts of 1 and 2, the
tracks
become bluer from the k-correction due to the shape of the stellar SED
around the vicinity of the 1.6$\mu$m bump. Galaxies to the top
right of these tracks are likely lower mass starbursts with moderate extinction.}
\label{fig:wiseccdet}
\end{center}
\end{figure*}

\setcounter{subfigure}{0}
\begin{figure*}
\begin{center}
\setlength{\tabcolsep}{2mm}
\subfigure[Left to right: P028, P031, P049, P052] {
\centering
\includegraphics[width=4.0cm,clip=true,trim=10 10 10 10 ,angle=0.]{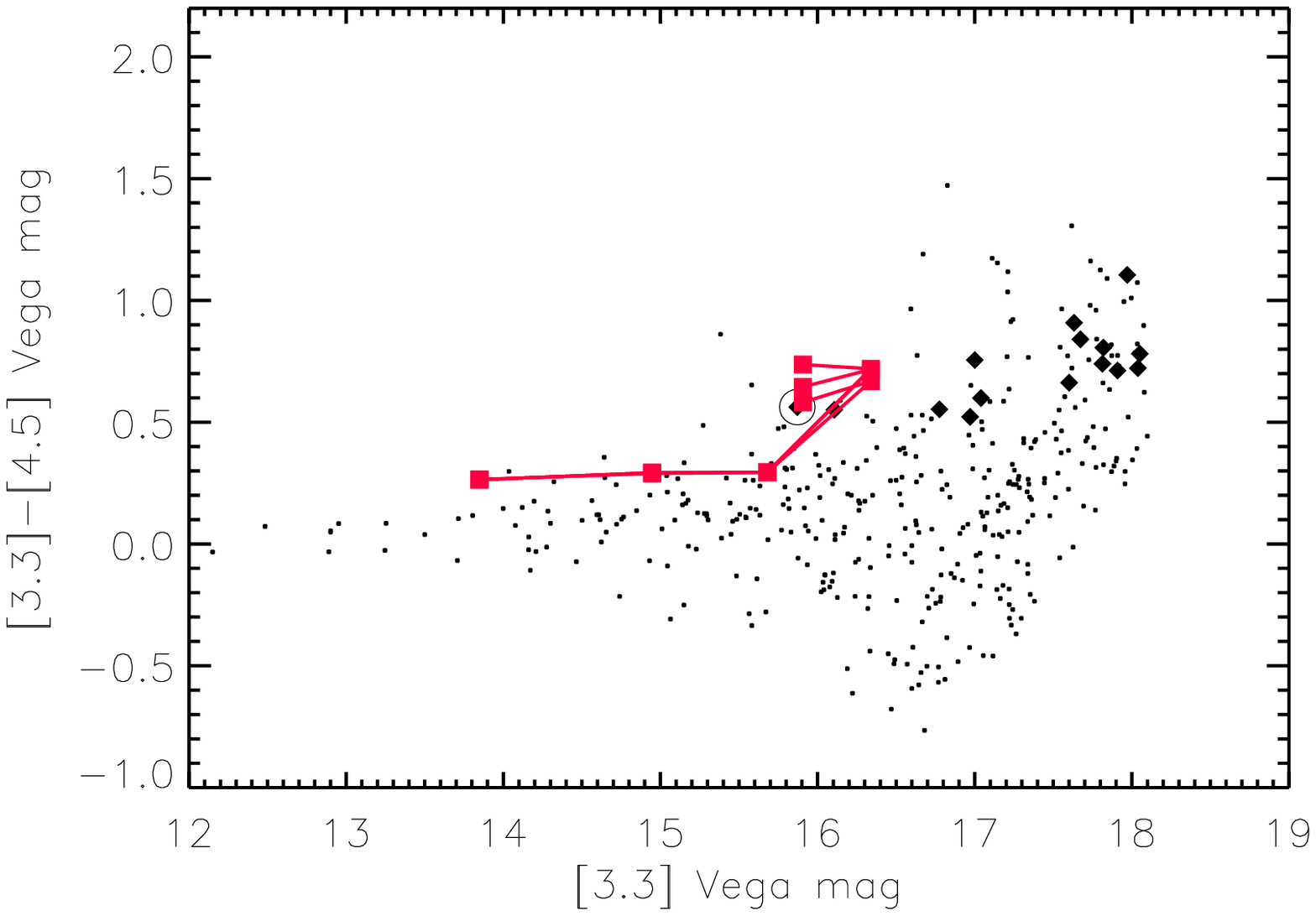}\qquad\includegraphics[width=4.0cm,clip=true,trim=10 10 10 10,angle=0.]{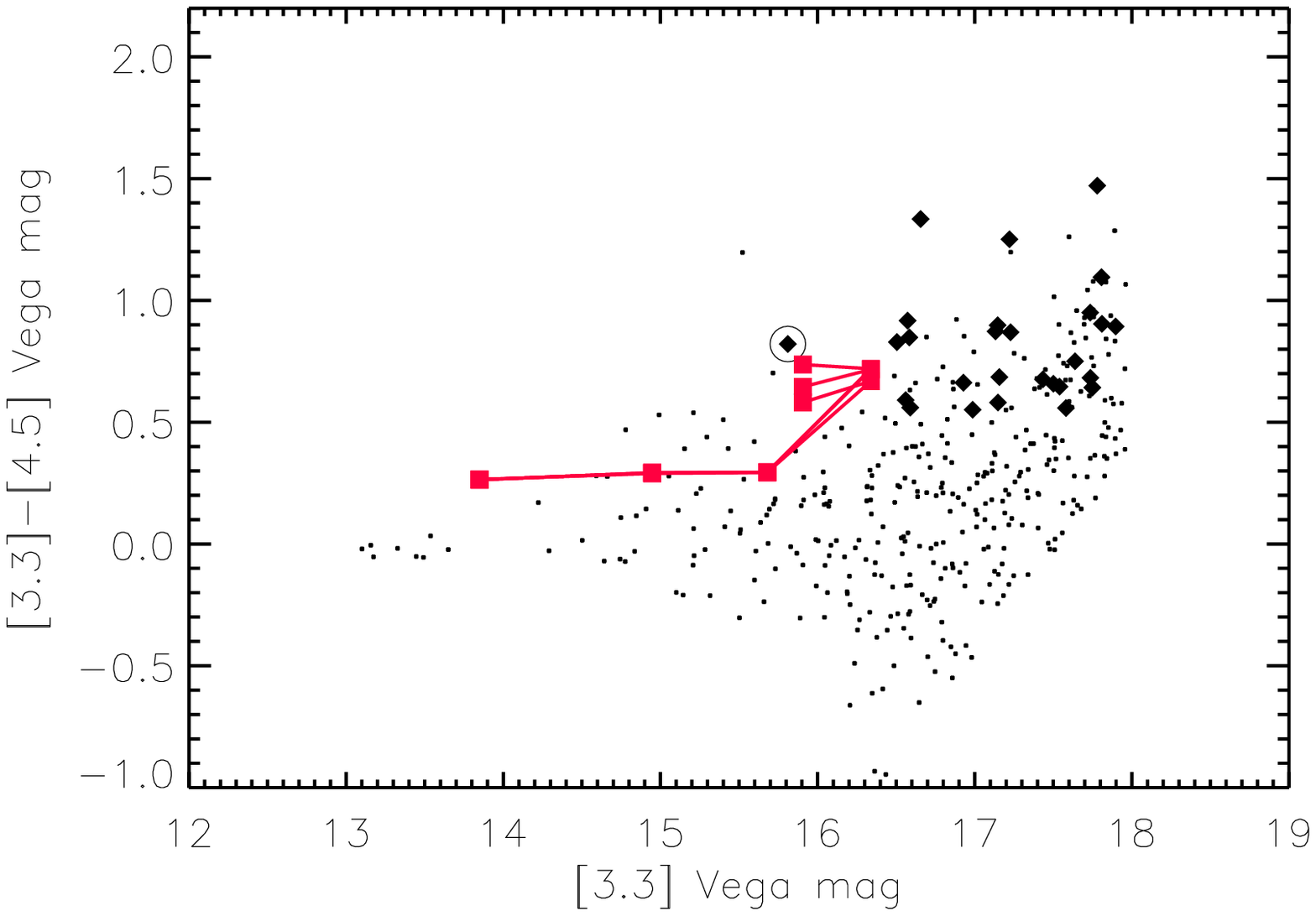}\qquad
\includegraphics[width=4.0cm,clip=true,trim=10 10 10 10,angle=0.]{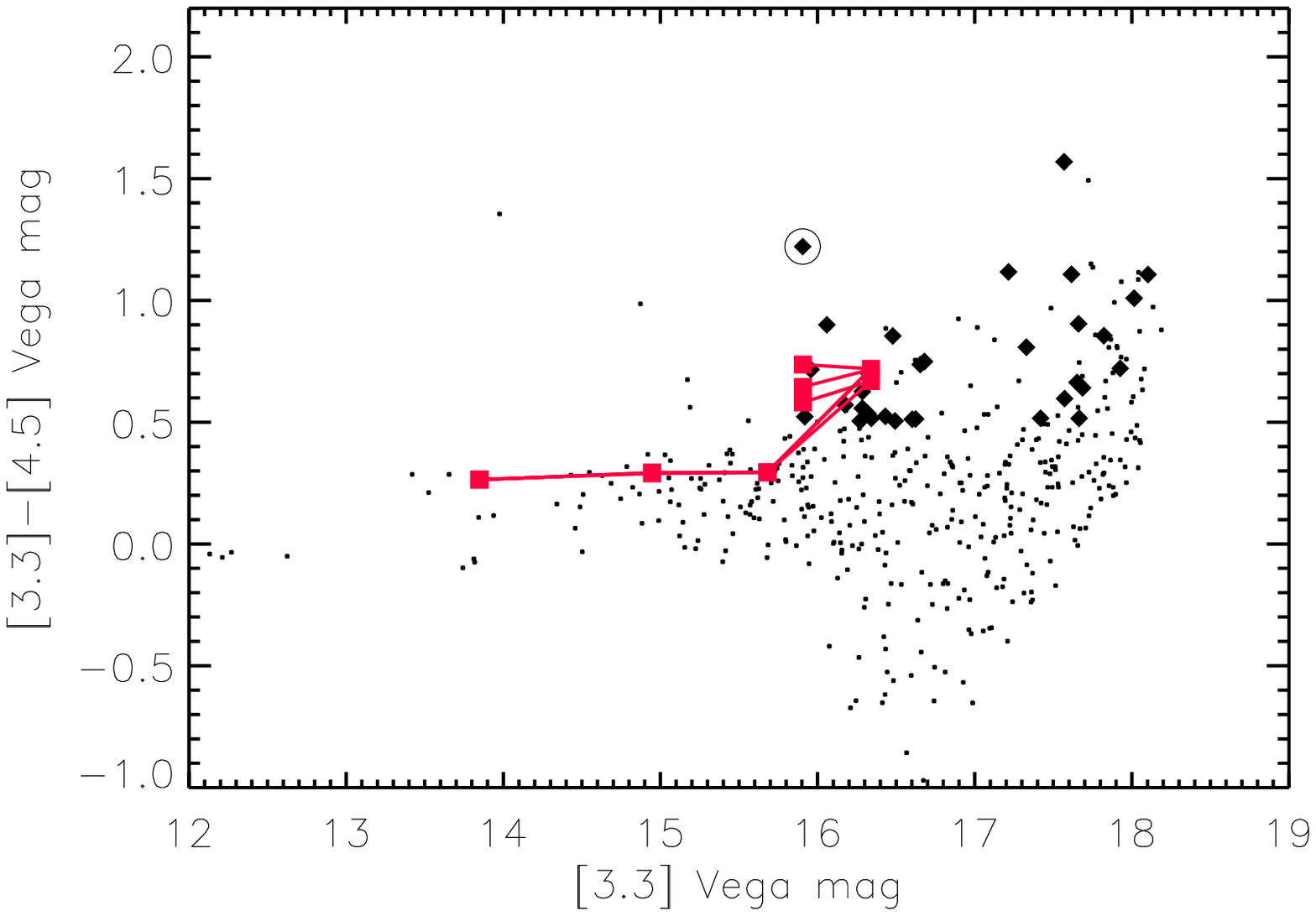}\qquad
\includegraphics[width=4.0cm,clip=true,trim=10 10 10 10,angle=0.]{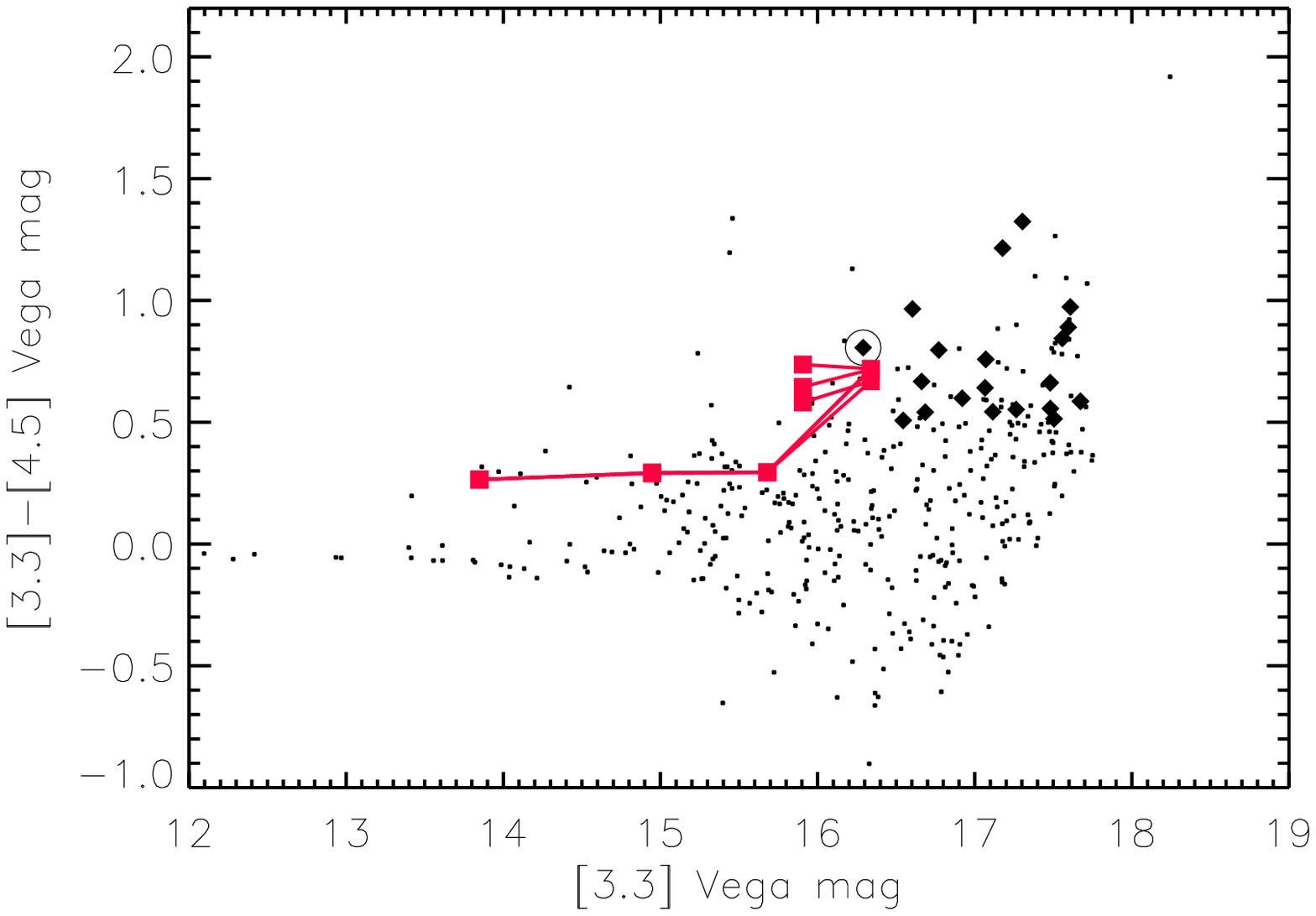}}\\
\subfigure[Left to right: P057, P090, P121, P134 ]{
\centering
\includegraphics[width=4.0cm,clip=true,trim=10 10 10 10 ,angle=0.]{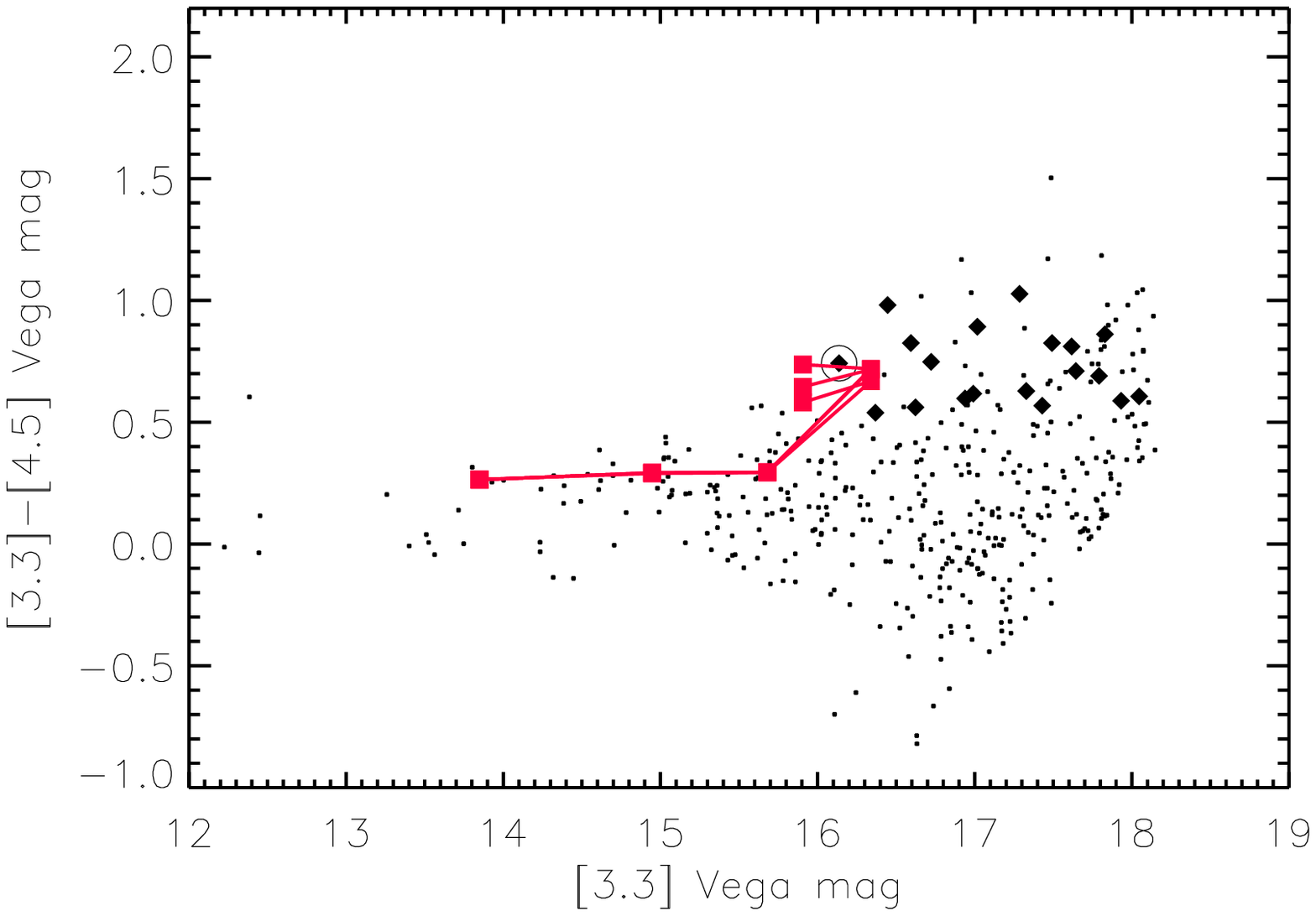}\qquad\includegraphics[width=4.0cm,clip=true,trim=10 10 10 10,angle=0.]{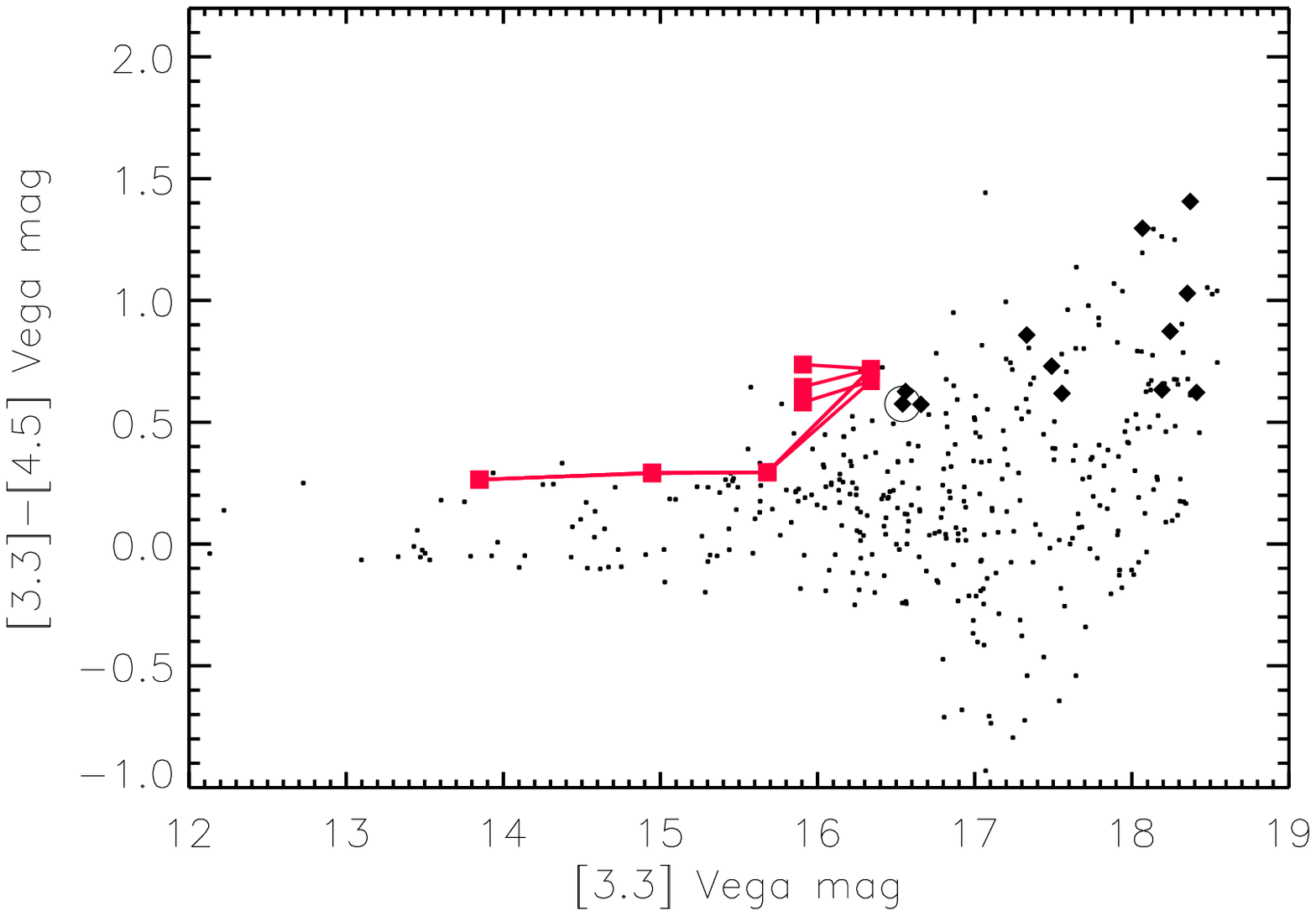}\qquad
\includegraphics[width=4.0cm,clip=true,trim=10 10 10 10,angle=0.]{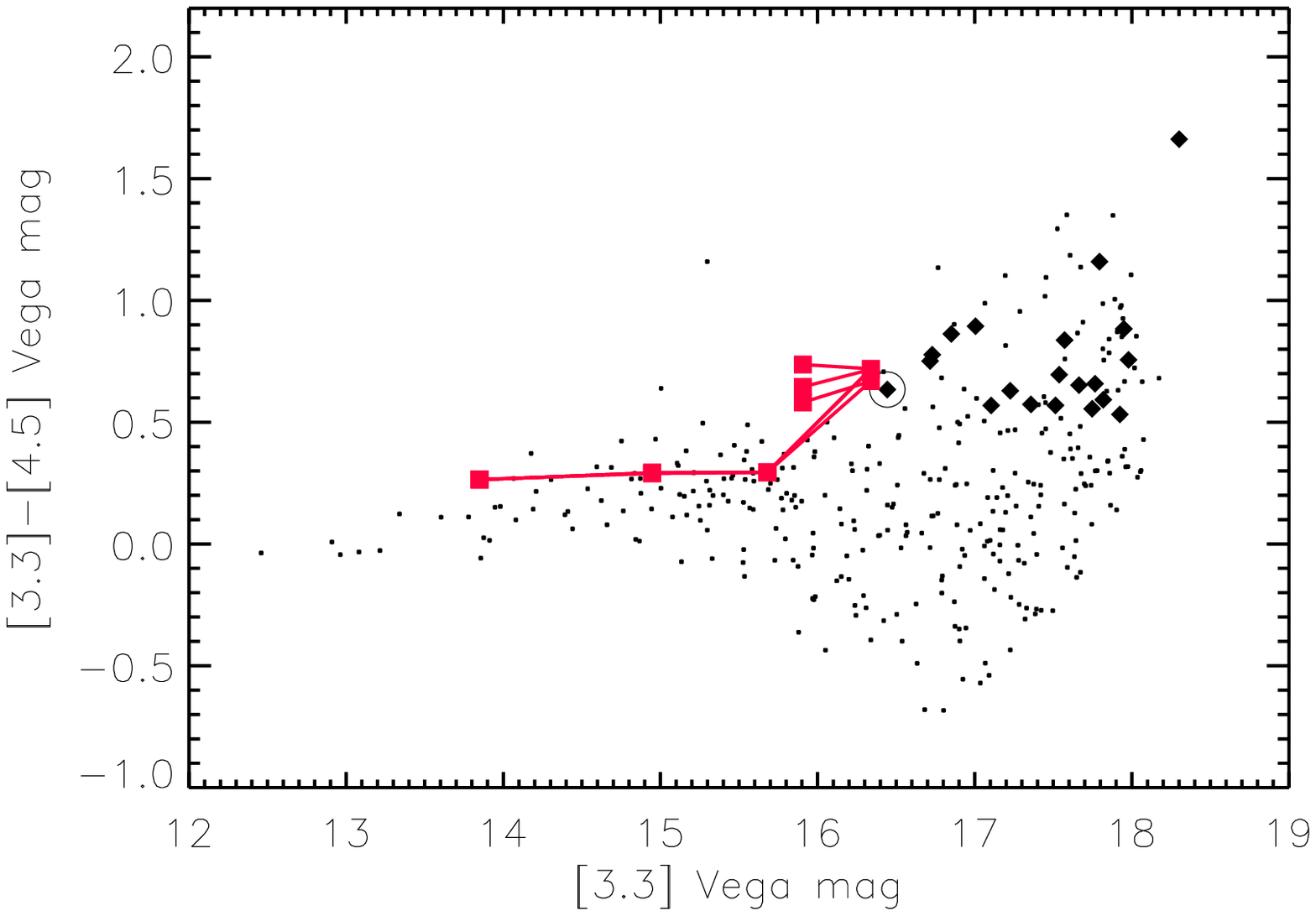}\qquad
\includegraphics[width=4.0cm,clip=true,trim=10 10 10 10,angle=0.]{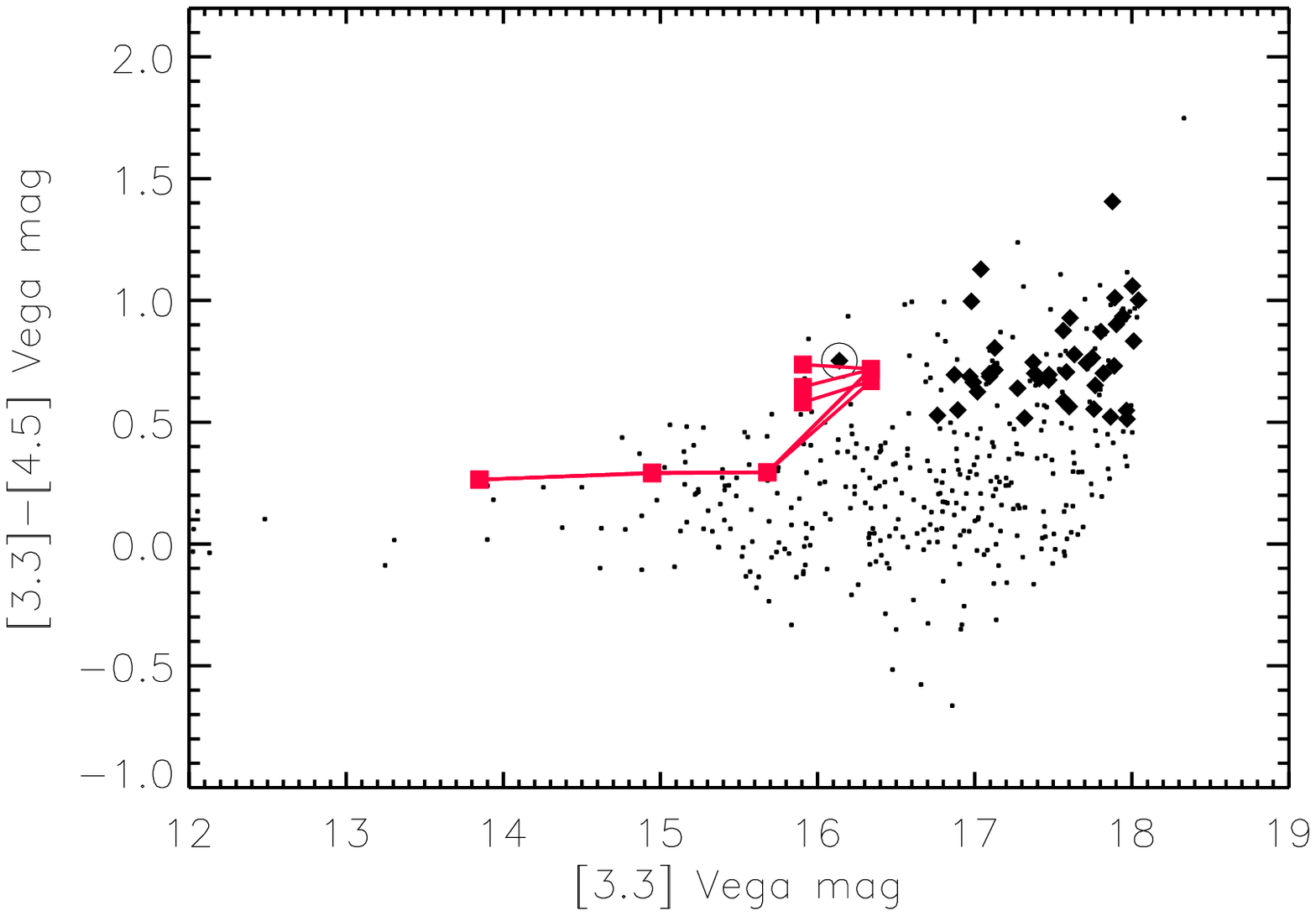}}\\
\subfigure[Left to right:   P138, P264]{
\centering
\includegraphics[width=4.0cm,clip=true,trim=10 10 10 10 ,angle=0.]{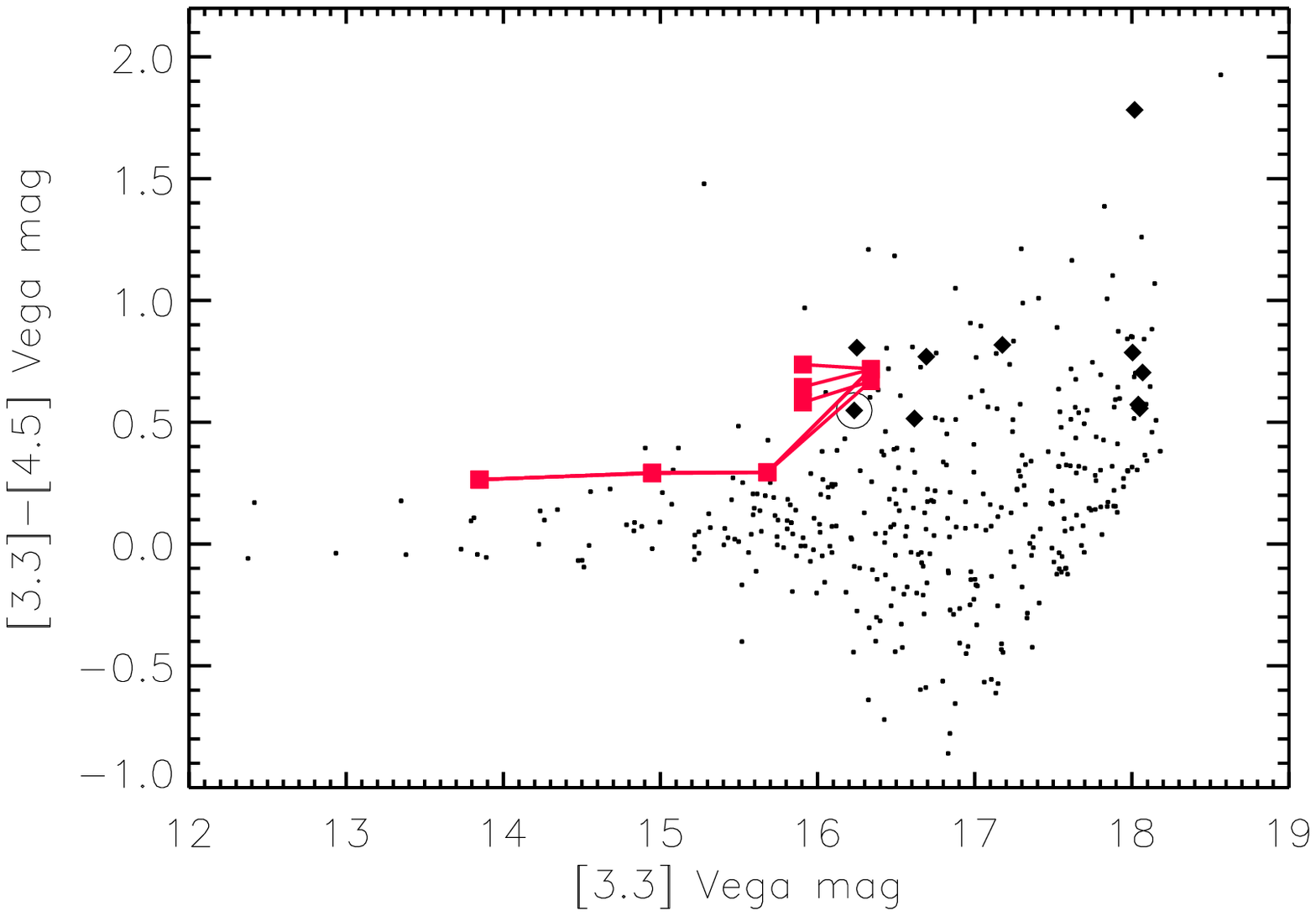}\qquad\includegraphics[width=4.0cm,clip=true,trim=10 10 10 10,angle=0.]{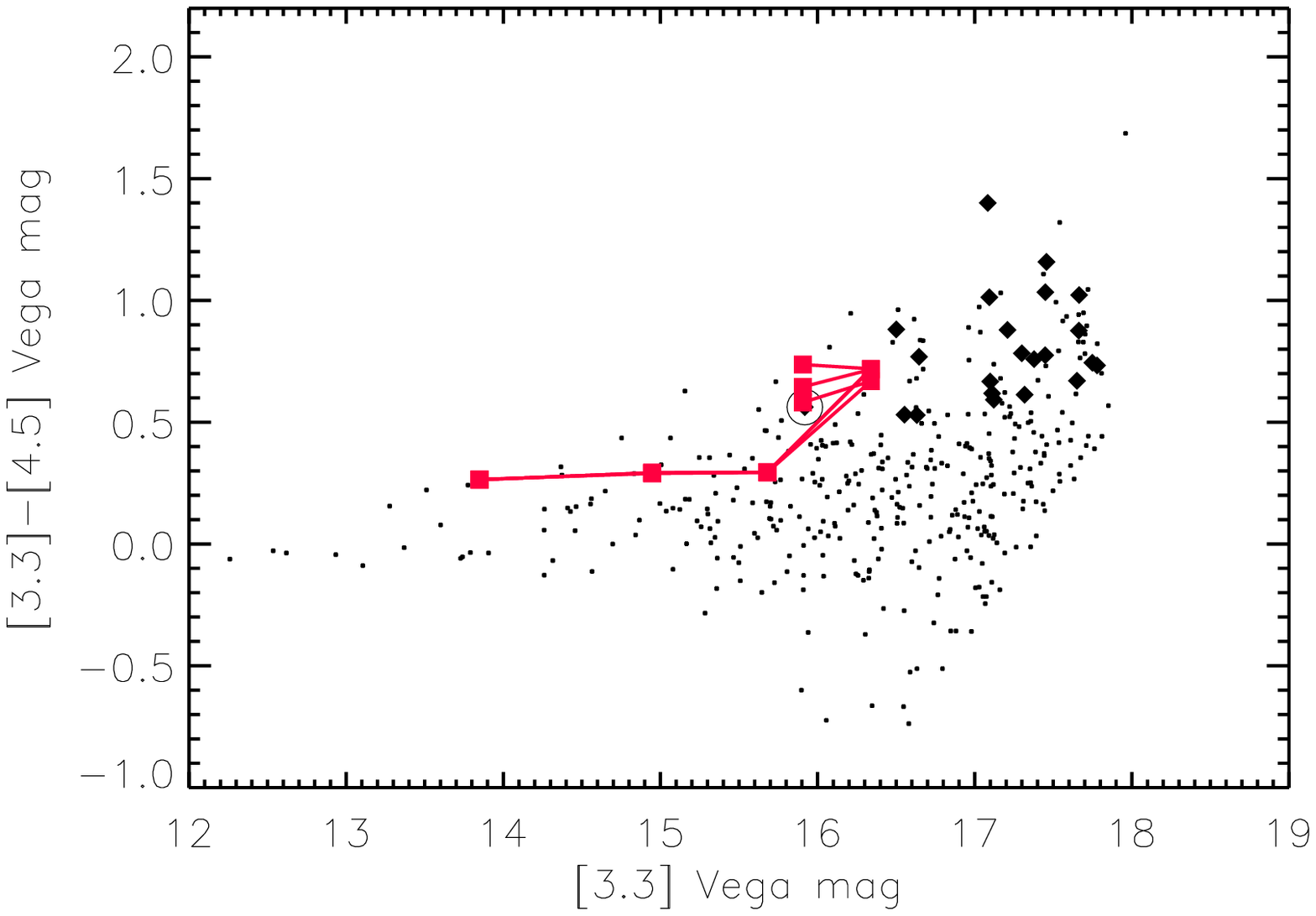}}\\
\caption{Colour-magnitude plots like the ones above but this time for cluster candidates without a CARMA-8 detection and centered at the {\it{Planck}} cluster position.}
\label{fig:wiseccndet}
\end{center}
\end{figure*}
\clearpage

\section{SDSS photometric redshifts}
\label{sec:photoz}
\clearpage
\begin{figure}
\begin{center}
\centerline{\includegraphics[width=12.0cm,clip=,angle=0.]{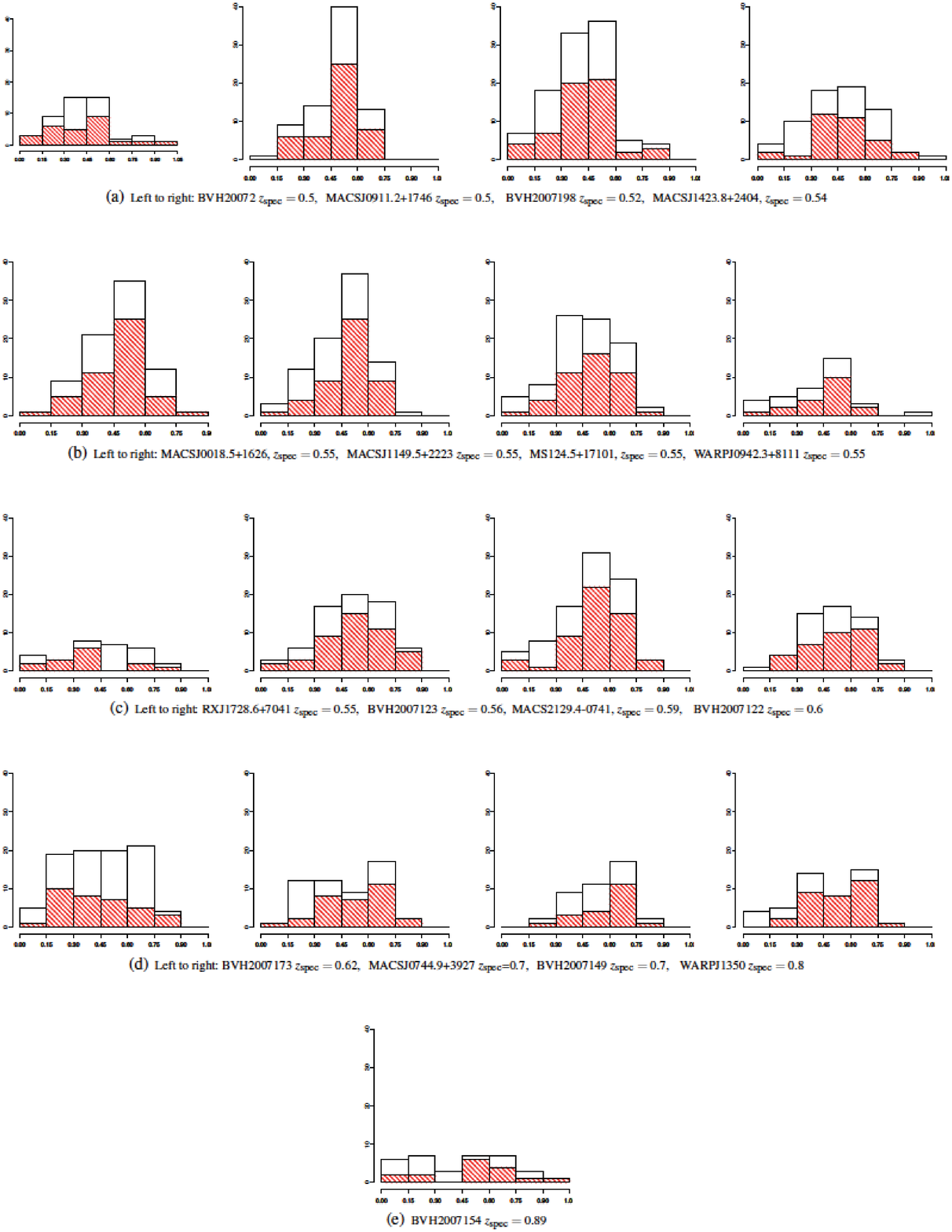}}
\caption{Histograms of the photometric redshifts of SDSS objects lying within 1.5\arcmin (white) and 1.0\arcmin(red, diagonal lines) of the MCXC cluster position (Piffaretti et al. 2011). Each plot is labelled with the cluster name and the spectroscopic redshift estimates given in the MCXC catalog. The $N_{\rm{gals}}$ derived from this work need not match those from Wen et al. (2012) for various reasons, including the search radius, which, for Wen et al., is typically $\approx 4-8\arcmin\ $ and the filters applied by Wen et al for quality purposes e.g., they require small photometric errors.}
\label{fig:photozMCXC}
\end{center}
\end{figure}

\begin{figure*}
\begin{center}
\centerline{\includegraphics[width=12.0cm,clip=,angle=0.]{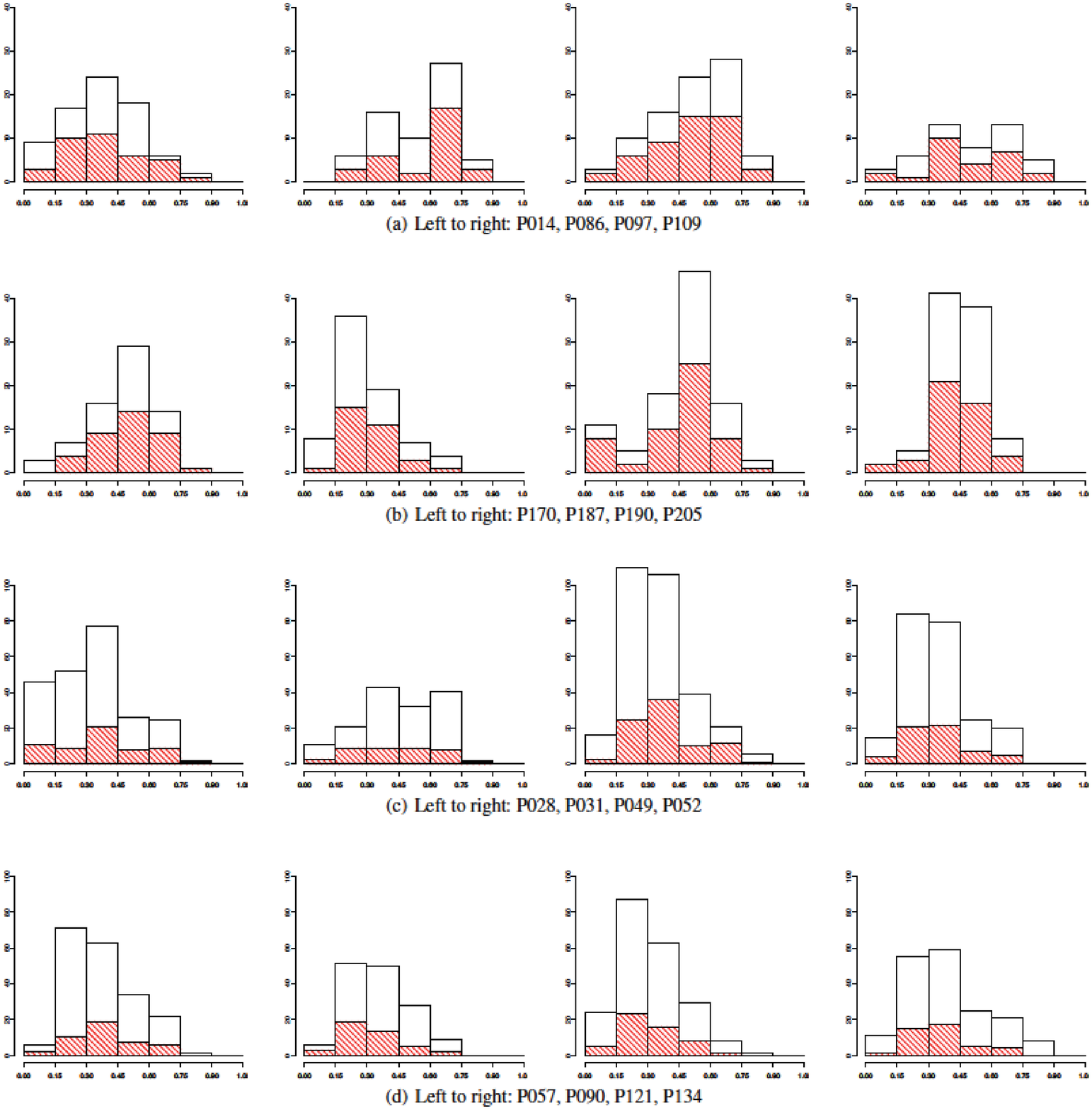}}
\caption{Histograms of the photometric redshifts of SDSS objects lying within 1.5\arcmin (white) and 1.0\arcmin(red, diagonal lines) of the CARMA-derived cluster position for CARMA detected systems (first two rows) and within 3.0\arcmin (white) and 1.5\arcmin(red, diagonal lines) of the {\it{Planck}} position for cluster candidates without an SZ CARMA detection (last two rows). Each plot is labelled with the shorthand ID name. The data suggest that many of the CARMA-8 non-detections are likely at z$\sim$0.3.}
\label{fig:photozP}
\end{center}
\end{figure*}


\clearpage


\begin{thebibliography}{}
\setlength{\labelwidth}{0pt}

\bibitem[\protect\citeauthoryear{Abell, Corwin, \& Olowin}{1989}]{abell1989} Abell G.~O., Corwin H.~G., Jr., Olowin R.~P., 1989, ApJS, 70, 1 

\bibitem[\protect\citeauthoryear{Allen et al.}{2011}]{allen2011} Allen, S.~W., Evrard, A.~E., \& Mantz, A.~B.\ 2011, \araa, 49, 409 

\bibitem[\protect\citeauthoryear{AMI Consortium: Hurley-Walker et al.}{2011}]{hurley2011} AMI Consortium, Hurley-Walker, N., Brown, M.~L., et al.\ 2011, \mnras, 414, L75 

\bibitem[\protect\citeauthoryear{AMI Consortium: Rodr{\'i}guez-Gonz{\'a}lvez et al.}{2012}]{carmen2012} AMI Consortium: Rodr{\'i}guez-Gonz{\'a}lvez et al., 2012, MNRAS, 425, 162 

\bibitem[\protect\citeauthoryear{Assef et al.}{2013}]{assef2013} Assef, R.~J., Stern, D., Kochanek, C.~S., et al.\ 2013, \apj, 772, 26 

\bibitem[\protect\citeauthoryear{Bahcall \& Fan}{1998}]{bahcall1998} Bahcall N.~A., Fan X., 1998, ApJ, 504, 1 

\bibitem[\protect\citeauthoryear{Bardeen et al.}{1986}]{bardeen1986} Bardeen, J.~M., Bond, J.~R., Kaiser, N., \& Szalay, A.~S.\ 1986, \apj, 304, 15 

\bibitem[\protect\citeauthoryear{Bonaldi et al.}{2007}]{bonaldi2007} Bonaldi, A., Tormen, G., Dolag, K., \& Moscardini, L.\ 2007, \mnras, 378, 1248 

\bibitem[\protect\citeauthoryear{Bond et al.}{1996}]{bond1996} Bond, J.~R., Kofman, L., \& Pogosyan, D.\ 1996, \nat, 380, 603 

\bibitem[\protect\citeauthoryear{Bond \& Myers}{1996}]{myers1996} Bond, J.~R., \& Myers, S.~T.\ 1996, \apjs, 103, 1 

\bibitem[\protect\citeauthoryear{Brodwin et al.}{2012}]{brodwin2012} Brodwin M., et al., 2012, ApJ, 753, 162 

\bibitem[\protect\citeauthoryear{Carlstrom et al.}{2002}]{carlstrom2002} Carlstrom, J.~E., Holder, G.~P., \& Reese, E.~D.\ 2002, \araa, 40, 643 

\bibitem[\protect\citeauthoryear{Carvalho, Rocha, \& Hobson}{2009}]{carvalho2009} Carvalho P., Rocha G., Hobson M.~P., 2009, MNRAS, 393, 681 

\bibitem[\protect\citeauthoryear{Carvalho et al.}{2012}]{carvalho2012} Carvalho P., Rocha G., Hobson M.~P., Lasenby A., 2012, MNRAS, 427, 1384 

\bibitem[\protect\citeauthoryear{Gal et al.}{2003}]{gal2003} Gal, R.~R., de Carvalho, R.~R., Lopes, P.~A.~A., et al.\ 2003, \aj, 125, 2064 

\bibitem[\protect\citeauthoryear{Galametz et al.}{2012}]{galametz2012} Galametz A., et al., 2012, ApJ, 749, 169

\bibitem[\protect\citeauthoryear{Gettings et al.}{2012}]{gettings2012} Gettings, D.~P., Gonzalez, A.~H., Stanford, S.~A., et al.\ 2012, \apjl, 759, L23  

\bibitem[\protect\citeauthoryear{Hasselfield et al.}{2013}]{hasselfield2013} Hasselfield M., et al., 2013, JCAP, 7, 8 

\bibitem[\protect\citeauthoryear{Herranz et al.}{2002}]{herranz2002} Herranz D., Sanz J.~L., Hobson M.~P., Barreiro R.~B., Diego J.~M., Mart{\'{\i}}nez-Gonz{\'a}lez E., Lasenby A.~N., 2002, MNRAS, 336, 1057 


\bibitem[\protect\citeauthoryear{H\"{o}gbom et al.}{1974}]{hogbom1974} H\"{o}gbom~J. A., 1974, Astron. Astrophys. Suppl, 197

\bibitem[\protect\citeauthoryear{Kay et al.}{2012}]{kay2012} Kay, S.~T., Peel, M.~W., Short, C.~J., et al.\ 2012, \mnras, 422, 1999 






\bibitem[\protect\citeauthoryear{Marriage et al.}{2011}]{marriage2011} Marriage T.~A., et al., 2011, ApJ, 737, 61 



\bibitem[\protect\citeauthoryear{McGlynn, Scollick, \& White}{1998}]{McGlynn1998} McGlynn T., Scollick K., White N., 1998, IAUS, 179, 465 


\bibitem[\protect\citeauthoryear{Melin, Bartlett, \& Delabrouille}{2006}]{melin2006} Melin J.-B., Bartlett J.~G., Delabrouille J., 2006, A\&A, 459, 341 


\bibitem[\protect\citeauthoryear{Montero-Dorta \& Prada}{2009}]{montero2009} Montero-Dorta A.~D., Prada F., 2009, MNRAS, 399, 1106 

\bibitem[\protect\citeauthoryear{Motl et al.}{2005}]{motl2005} Motl, P.~M., Hallman, E.~J., Burns, J.~O., \& Norman, M.~L.\ 2005, \apjl, 623, L63 



\bibitem[\protect\citeauthoryear{Muchovej et al.}{2007}]{muchovej2007} Muchovej, S., Mroczkowski, T., Carlstrom, J.~E., et al.\ 2007, \apj, 663, 708 

\bibitem[\protect\citeauthoryear{Muchovej et al.}{2010}]{muchovej2010} Muchovej S., et al., 2010, ApJ, 716, 521 







\bibitem[\protect\citeauthoryear{Muzzin et al.}{2013}]{muzzin2013} Muzzin, A., Wilson, G., Demarco, R., et al.\ 2013, \apj, 767, 39 

\bibitem[\protect\citeauthoryear{Papovich}{2008}]{papovich2008} Papovich C., 2008, ApJ, 676, 206 

\bibitem[\protect\citeauthoryear{Perrott}{2014}]{perrott2014} Perrott, Y.~C., Olamaie, M., Rumsey, C., et al.\ 2014, arXiv:1405.5013 

\bibitem[\protect\citeauthoryear{Piffaretti et al.}{2011}]{pifa2010} Piffaretti R., Arnaud M., Pratt G.~W., Pointecouteau E., Melin J.-B., 2011, A\&A, 534, A109 


\bibitem[\protect\citeauthoryear{Planck Collaboration et al.}{2011 I}]{planck1} {\it{Planck}} Collaboration I, et al., 2011, A\&A, 536, A1 

\bibitem[\protect\citeauthoryear{Planck Collaboration et al.}{2011 VIII}]{ESZ} ${\it{Planck}}$ Collaboration VIII, 2011, A\&A, 536, A8 

\bibitem[\protect\citeauthoryear{Planck Collaboration et al.}{2013 II}]{PlanckAMI} ${\it{Planck}}$ Collaboration, Intermediate Results II, 2013, A\&A, 550, A128 

\bibitem[\protect\citeauthoryear{Planck Collaboration et al.}{2013 XXIX}]{Planck2013} ${\it{Planck}}$ Collaboration, et al., 2013 XXIX, arXiv, arXiv:1303.5089 

\bibitem[\protect\citeauthoryear{Planck Collaboration et al.}{2013}]{Planckcosmo} Planck Collaboration, et al.\ 2013 XX, arXiv:1303.5080 

\bibitem[\protect\citeauthoryear{Reichardt et al.}{2013}]{reichardt2013} Reichardt C.~L., et al., 2013, ApJ, 763, 127 

\bibitem[\protect\citeauthoryear{Rettura et al.}{2014}] {rettura2014} Rettura, A., Martinez-Manso, J., Stern, D., et al.\ 2014, arXiv:1404.0023 

\bibitem[\protect\citeauthoryear{Rodr{\'i}guez-Gonz{\'a}lvez et al.}{2014}]{rodriguez2014b} Rodr{\'i}guez-Gonz{\'a}lvez et al., in prep 


\bibitem[\protect\citeauthoryear{Sayers et al.}{2012}]{sayers2012} Sayers, J., Czakon, N.~G., Bridge, C., et al.\ 2012, \apjl, 749, L15 


\bibitem[\protect\citeauthoryear{Shepherd}{1997}]{shepherd1997} Shepherd M.~C., 1997, ASPC, 125, 77 


\bibitem[\protect\citeauthoryear{Stern et al.}{2012}]{stern2012} Stern, D., Assef, R.~J., Benford, D.~J., et al.\ 2012, \apj, 753, 30 

\bibitem[\protect\citeauthoryear{Sunyaev \& Zel'dovich}{1972}]{Sunyaev_1972} Sunyaev R.~A., Zel'dovich Y.~B., 1972, CoASP, 4, 173

\bibitem[\protect\citeauthoryear{Tauber et al.}{2010}]{tauber2010} Tauber J.~A., et al., 2010, A\&A, 520, A1 

\bibitem[\protect\citeauthoryear{Wen et al.}{2012}]{wen2012} Wen, Z.~L., Han, J.~L., \& Liu, F.~S.\ 2012, \apjs, 199, 34 

\bibitem[\protect\citeauthoryear{Wenger et al.}{2000}]{wenger2000} Wenger M., et al., 2000, A\&AS, 143, 9 

\bibitem[\protect\citeauthoryear{Williamson et al.}{2011}]{williamson2011} Williamson R., et al., 2011, ApJ, 738, 139 

\bibitem[\protect\citeauthoryear{Wright et al.}{2010}]{wright2010} Wright, E.~L., Eisenhardt, P.~R.~M., Mainzer, A.~K., et al. 2010, \aj, 140, 1868 

\bibitem[\protect\citeauthoryear{Viana \& Liddle}{1996}]{viana1996} Viana P.~T.~P., Liddle A.~R., 1996, MNRAS, 281, 323 

\bibitem[\protect\citeauthoryear{Voit}{2005}]{voit2005} Voit, G.~M.\ 2005, Reviews of Modern Physics, 77, 207 



\bibitem[\protect\citeauthoryear{Zwart et al.}{2008}]{zwart2008} Zwart J.~T.~L., et al., 2008, MNRAS, 391, 1545 

\end{thebibliography}
\end{document}